\documentclass[%
reprint,
notitlepage,
superscriptaddress,
nofootinbib,
amsmath,amssymb,
aps,
prd,
]{revtex4-2}

\usepackage{graphicx}
\usepackage{subfig}
\usepackage[export]{adjustbox}
\usepackage[bottom]{footmisc}
\usepackage{tablefootnote}
\graphicspath{{.}}

\usepackage{dcolumn}% Align table columns on decimal point
\usepackage{bm}% bold math
\usepackage[mathlines]{lineno}% Enable numbering of text and display math
\usepackage[dvipsnames,table,xcdraw]{xcolor}
\usepackage{multirow}
\usepackage[hidelinks]{hyperref}
\usepackage{orcidlink}

\usepackage[utf8]{inputenc}
\usepackage[T1]{fontenc}
\usepackage{slashed}

\usepackage{enumerate}

\usepackage{mathalfa,amssymb,amsthm}

\usepackage{tensor}
\usepackage{wasysym}

\newcommand{\damtp}{Department of Applied Mathematics and Theoretical Physics, 
Centre for Mathematical Sciences, University of Cambridge, 
Wilberforce Road, Cambridge CB3 0WA, United Kingdom}
\newcommand{\kavli}{Kavli Institute for Cosmology Cambridge, 
Madingley Road CB3 0HA, Cambridge, UK}

\begin{document}

\title{Measuring the ringdown scalar polarization of gravitational
waves in Einstein-scalar-Gauss-Bonnet gravity}

%-----------------------------------------------------------------------------
\author{Tamara Evstafyeva \orcidlink{0000-0002-2818-701X}}
\email{te307@cam.ac.uk}
\affiliation{\damtp}
\author{Michalis Agathos \orcidlink{0000-0002-9072-1121}}
\email{magathos@damtp.cam.ac.uk}
\affiliation{\kavli}%
\affiliation{\damtp}
\author{Justin L. Ripley \orcidlink{0000-0001-7192-0021}}
\email{ripley@illinois.edu}
\affiliation{%
Illinois Center for Advanced Studies of the Universe \& Department of Physics,
University of Illinois at Urbana-Champaign, Urbana, Illinois 61801, USA
}%
\affiliation{\damtp}%
%-----------------------------------------------------------------------------

\date{\today}% It is always \today, today,
   %  but any date may be explicitly specified

\begin{abstract}
   We model the scalar waves produced during the ringdown stage of
   binary black hole coalescence in Einstein-scalar-Gauss-Bonnet (EsGB)
   gravity, 
   using numerical relativity simulations of the theory in 
   the decoupling limit.
   Through a conformal coupling of the scalar field to the metric in the
   matter-field action, 
   we show that the gravitational waves in this theory can have 
   a scalar polarization.
   We model the scalar quasinormal modes of the ringdown signal in 
   EsGB gravity, and quantify the extent to which current and future 
   gravitational wave detectors could observe the spectrum of scalar 
   radiation emitted during the ringdown phase of binary black hole coalescence.
   We find that within the limits of the theory's coupling parameters 
   set by current theoretical and observational constraints, 
   the scalar ringdown signal from black hole remnants in the 
   $10^1-10^3 \, M_{\odot}$ mass range is expected to be well below 
   the detectability threshold with the current network of 
   gravitational-wave detectors (LIGO-Virgo-KAGRA), 
   but is potentially measurable with next-generation detectors 
   such as the Einstein Telescope. 
\end{abstract}

\maketitle

%==============================================================================
%\allowdisplaybreaks
%\tableofcontents
%==============================================================================
\section{Introduction}\label{sec:introduction}
General relativity (GR) has passed all observational tests to
date~\cite{Will:2014kxa,Ishak:2018his} and, as of 2015, observations of gravitational wave signals in LIGO~\cite{TheLIGOScientific:2014jea} and Virgo~\cite{TheVirgo:2014hva} data have been consistent with its predictions~\cite{TheLIGOScientific:2016src,LIGOScientific:2019fpa,LIGOScientific:2018dkp,LIGOScientific:2020tif,LIGOScientific:2021sio,Cardoso:2019rvt}.
Nevertheless, there has been an increasing amount of work that tests
the predictions of non-GR theories of gravity 
(\emph{modified theories of gravity})
with gravitational wave measurements of binary black hole (BH) and neutron
star (NS) mergers 
\cite{Yunes:2016jcc,Baker:2017hug,
   Lyu:2022gdr,
   Perkins:2021mhb,
   Okounkova:2021xjv}.
These \emph{model-dependent} tests of GR have allowed for remarkably strong
constraints to be placed on a wide array of modified gravity theories.

Here we focus on the gravitational wave implications for black hole binaries
in Einstein-scalar-Gauss-Bonnet (EsGB) gravity.
EsGB gravity can be motivated from effective-field-theory-styled arguments~\cite{Weinberg:2008hq,Kovacs:2020pns}, and
has been shown to appear in the low-energy limits of heterotic
string theories~\cite{Gross:1986mw, Zwiebach:1985uq, Cano:2021rey}.
What makes EsGB interesting from a phenomenological point of view is that
black holes in the theory can have scalar
hair~\cite{Kanti:1995vq, Sotiriou:2014pfa} (for a review, see~\cite{Herdeiro:2015waa}).
Because of this, binary black hole (BBH) systems can emit 
scalar radiation~\cite{Yagi:2011xp}.
From a theoretical point of view, EsGB gravity is a Horndeski theory~\cite{Kobayashi_2019}
and, as such, has second order equations of motion,
circumventing the Ostrogradsky instability.
In addition, the theory has been shown to have a well-posed initial value 
problem in the weak coupling limit~\cite{Kovacs:2020pns} which allows for
numerical relativity simulations of the full 
theory~\cite{East_2021,AresteSalo:2022hua,East:2022rqi,Ripley:2022cdh}.~\footnote{It is,
however, not guaranteed that dynamical spacetimes will preserve the 
weak coupling of their initial data during their evolution, and 
local pathologies may form~\cite{Ripley:2019hxt,Ripley:2019irj,R:2022hlf}.}

In this paper, we focus on the leading order effects of EsGB on the
gravitational waveform emitted during the ringdown stage of BBH mergers.
It is expected that the strongest constraints on most modified theories of gravity
will come from the inspiral,
due to the cumulative dephasing of the signal from the GR predicted waveform, over a large number of observed cycles~\cite{Yunes:2013dva,Berti:2018cxi}.
Nevertheless, correctly modeling the ringdown is crucial to understanding the
properties of the remnant black hole formed from binary coalescence,
and to performing consistency tests with theory-specific estimates of the final black
hole properties inferred from the inspiral and merger~\cite{Berti:2018vdi}.
While the dephasing of the inspiral can be a common feature of many different theories of modified gravity and the theory-specific details can be difficult to model at high accuracy (see~\cite{Shiralilou:2020gah,Lyu:2022gdr} for perturbative approaches in EsGB), a spectroscopic analysis of the ringdown signal may reveal characteristic features that will help narrow down the origin of an observed deviation from GR\@.

Working in the decoupling limit of EsGB gravity,
we extract first-order tensorial and scalar waveforms
and focus on the ringdown portion of the signal. 
We study the quasinormal mode (QNM) spectrum
of the resulting scalar and tensor waveforms and numerically
compute the amplitudes of the excited modes. 
Unlike most previous studies on EsGB gravity, 
we study the impact of the radiated scalar field on 
the \emph{scalar polarization} 
measured by the gravitational wave detectors, 
through the presence of a nonminimal scalar 
field coupling to the metric tensor in the matter field action
(without such a coupling, gravitational waves would not have a scalar polarization component in EsGB gravity~\cite{Wagle:2019polarisation}).
While there has been work on model-independent tests of
the polarization of gravitational waveforms~\cite{Isi:2015cva,Isi:2017equ,Isi:2017fbj,Isi:2022mbx},
we present the first model-dependent analysis
of the feasibility of measuring a nontensor polarization from black hole
binaries with current and future gravitational wave detectors
(we note though that Ref.~\cite{Isi:2018miq} considers model-dependent
tests with stochastic gravitational wave backgrounds).
Ultimately, we find the scalar polarization---given current
constraints on the EsGB gravity and weak-field tests of general 
relativity---is unobservable with current gravitational wave detectors, 
although there is a higher chance that it could be
measured with next-generation detectors.
\raggedbottom
\par{The paper is organized as follows.
In Sec.~\ref{sec:form-4-deriv} 
we present the theoretical framework of EsGB gravity in the Einstein frame.
In Sec.~\ref{sec:polarization}, we review 
how to extract the polarization content of gravitational waves (GWs), 
and show how the conformal scalar field coupling can give rise to a scalar
polarization in gravitational waves. 
Next, we review and justify the application of the decoupling limit 
for solving the equations of motion of EsGB gravity in 
Sec.~\ref{sec:order_reduction_scheme}.
For reference, we present a summary of current observational and 
theoretical constraints on the couplings of the theory we consider
in Sec.~\ref{sec:constraints}.
In Secs.~\ref{sec:num-evolution}--\ref{sec:extraction}
we overview our numerical set-up and the diagnostics we used to extract 
physical quantities from our simulations. 
In Sec.~\ref{sec:numerical_results}, 
we present the scalar waveforms measured from our numerical relativity 
(NR) simulations.
We analyse the QNM spectrum of our resulting scalar waveforms in 
Sec.~\ref{sec:ringdown} and compute their signal-to-noise ratio (SNR) in
Sec.~\ref{sec:measurability} to assess the measurability of the scalar signal.
We then move on to a full Bayesian data analysis of our 
simulated signals by performing NR injections of
the scalar waveform into simulated current and 
future-generation detector noise using a tailored ringdown pipeline in
Sec.~\ref{sec:injection}.
We finish with concluding remarks in Sec.~\ref{sec:conclusion}.}

Our notation and conventions are as follows. 
The metric signature is $(-+++)$ and we use $M$ to denote the Arnowitt-Deser-Misner (ADM) mass. 
We use lower case Greek indices to index spacetime components 
(indexed $0,1,2,3$, with $0$ being the timelike index) 
and lower case Latin indices to index spatial components.  
The Riemann tensor is
$R^{\alpha}{}_{\mu\beta\nu}=\partial_{\beta}\Gamma^{\alpha}_{\mu\nu}-\cdots$.

%==============================================================================
\section{Theoretical background}
\label{sec:theory}

\subsection{Formulation of four-derivative scalar-tensor ($4\partial$ST) gravity}
\label{sec:form-4-deriv}

The general action for Einstein-scalar-Gauss-Bonnet gravity is
\begin{align}
\label{eq:fullaction_einstein}
   S
   =
   \frac{c^4}{16 \pi G} & \int d^{4}x\sqrt{-g}
   \left[
      R 
      + 
      X  
      - 
      V\left(\varphi\right)
      + 
      \bar{\alpha}\left(\varphi\right)X^2 \right. \nonumber \\
      & \left. +
      \alpha_{\rm{GB}}\beta(\varphi) \mathcal{R}_{GB} 
   \right] 
    + 
   S_{M}[\Psi, A^2(\varphi) g_{\mu \nu}] 
   ,
\end{align}
where $V(\varphi)$ is the scalar field potential, 
$\bar{\alpha}(\varphi)$, $\beta(\varphi)$ are arbitrary coupling functions 
of the scalar field, 
$X \equiv - \frac{1}{2} (\nabla \varphi)^2$,
$S_M$ is the action functional for the matter fields 
(which we denote schematically with $\Psi$),
$A^2(\varphi)$ is the conformal coupling of matter to the metric,
and $\mathcal{R}_{\rm{GB}}$ is the Gauss-Bonnet scalar
\begin{align}
   \mathcal{R}_{GB}
   \equiv
   \frac{1}{4}
   \epsilon^{\mu\nu\alpha\beta}
   \epsilon^{\rho\sigma\gamma\delta}
   R_{\mu\nu\rho\sigma} R_{\alpha\beta\gamma\delta}
   .
\end{align}
If $A\neq0$, the scalar field couples directly to matter. 
This coupling generally leads to violations of the weak-equivalence principle 
(barring the presence of \emph{screening} mechanisms~\cite{Joyce:2014kja}), 
as the effective gravitational field felt by matter fields will depend on
the value of the scalar field $\varphi$. 
Weak field tests of gravity have strongly constrained violations of the weak-equivalence
principle, which can be translated 
to constraints on the strength of the coupling $A$~\cite{Will:2014kxa}.
In geometric units, the coupling parameter $\alpha_{\rm{GB}}$ has dimensions 
length squared, $L^{2}$. 
It is then natural to expect strong constraints on the scalar
Gauss-Bonnet coupling will come from regions of high spacetime curvature
\cite{Yunes:2013dva,Berti:2018cxi,Berti_2015}.

In this work we examine a subset of theories for which $V=\bar{\alpha}=0$. 
We assume the beyond-GR corrections give only relatively small deviations
to the background GR solution. 
Therefore, we impose that $|\varphi|, |\alpha_{\rm{GB}}|, |\beta| \ll 1$ 
and to leading order we have 
\begin{eqnarray}
   \label{eq:actionfinal}
   S =
   \frac{c^4}{16 \pi G}\int d^4x & \sqrt{-g} \left[
      R 
      - 
      \frac{1}{2}(\nabla \varphi)^2  
      + 
      \beta_{0} \varphi \mathcal{R}_{\rm{GB}} 
   \right] \nonumber \\
   & +
   S_M\left[
      \Psi
      ,
      \left(
         1 
         +
         a_0\varphi
      \right)
      g_{\mu\nu}
   \right]
   ,
\end{eqnarray} 
where we have Taylor expanded $\beta$ and $A$ in $\varphi$,
set $A(0)=1$, $A'(0)\equiv a_0/2$, and $ \beta_{0} \equiv \alpha_{\rm{GB}} \beta'(0)$. 
We have discarded the term $\beta(0)$ as the Gauss-Bonnet scalar
is a total derivative in four spacetime dimensions, so it does not
contribute to the equations of motion.
%
%==============================================================================
\subsection{\label{sec:polarization} Polarization content}
In alternative metric theories of gravity, there may exist up to 
six polarizations associated with additional degrees of freedom: 
tensorial (plus and cross), vector ($x$ and $y$) and scalar 
(breathing and longitudinal)~\cite{1973PhRvL..30..884E}.
Because of their symmetries, 
the breathing and longitudinal modes are fully degenerate to networks of 
quadrupolar antennas~\cite{Chatziioannou:2012rf},
so we will refer to them jointly as the scalar polarization. 
The scalar polarization state is generally a 
mixture of the transverse breathing 
and longitudinal polarizations and is excited by a massive scalar field. 
In the massless limit, the longitudinal polarization disappears, 
while the breathing one persists~\cite{Hou:2018hordenski}.
Since the scalar field we consider is massless, 
we only study the breathing polarization in this work.

\begin{figure*}[hbt!]
\includegraphics[width=.53\linewidth]{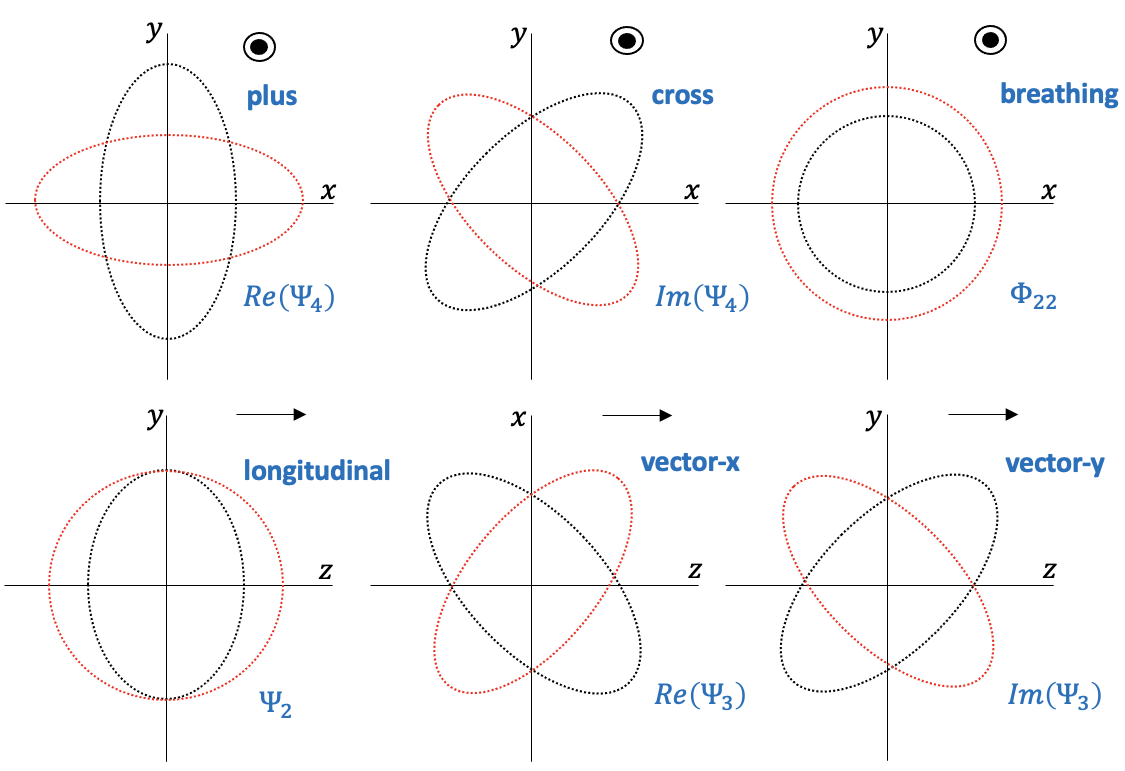}
   \caption{The classification of the six polarizations encoded in the four Newman-Penrose scalars. The shapes represent displacement that each mode induces on a sphere of test particles. The wave is assumed to propagate in the $z$-direction indicated by the icons/arrows in the right corners.}
\label{polarizations_E2}
\end{figure*}

The six polarization states are encoded in six ``electric'' 
components of the Riemann tensor, $R_{i0j0}$ ($i,j$ spatial),
which can be written in terms of the following
Newman-Penrose scalars~\cite{1973PhRvL..30..884E}
\begin{align}
   \Psi_2 &= \frac{1}{6} R_{z0z0}, \\
   \Psi_3 &= \frac{1}{2} \left(-R_{x0z0} + i R_{y0z0} \right),\\
   \Psi_4 &= R_{y0y0} - R_{x0x0} + 2iR_{x0y0}, \\
   \Phi_{22} &= - (R_{x0x0} + R_{y0y0}).
\end{align}
In Fig.~\ref{polarizations_E2} we illustrate the form of the
different polarization states.
Perturbing the metric and scalar field, the metric $\hat{g}_{\mu\nu} = (1+a_0 \varphi) g_{\mu \nu}$ that couples to the
matter fields (such as a gravitational wave detector) in the wave zone is 
\begin{align}
   \label{eq:wave_zone_expansion}
   \hat{g}_{\mu\nu}
   = 
   \eta_{\mu\nu}
   +
   h_{\mu\nu}
   +
   a_0\eta_{\mu\nu} \varphi
   +
   \cdots
   ,
\end{align}
where $\eta_{\mu\nu}$ is the Minkowski metric.
The tensorial perturbation $h_{\mu\nu}$ gives the
plus and cross polarizations, 
while the coupling constant $a_0$, along with a nonzero
value of $\varphi$, gives a new scalar 
breathing-mode polarization to gravitational 
waves\footnote{We note that if $a_0=0$,
then gravitational waves have no scalar polarization
in EsGB gravity~\cite{Wagle:2019polarisation}.}
\cite{PhysRevD.10.2374,Isi:2018miq}.

%==============================================================================
\subsection{\label{sec:order_reduction_scheme}Perturbative equations of motion}
\label{sec:perturbative}
Unless otherwise stated, in this section we work in geometric units $G=c=1$.
Our discussion roughly follows that of~\cite{Okounkova:2017yby,Witek:2018dmd}.
We start with the full equations of motion derived from action~\eqref{eq:actionfinal}
\begin{align}
\label{eq:eoms_grav}
   G_{\mu \nu} 
   &
   = \frac{1}{2} T_{\mu \nu} - \beta_{0} C^{\rm{GB}}_{\mu \nu},  \\ 
\label{eq:eoms_scalar}   
   \Box \varphi  
   &= 
   - \beta_{0} \mathcal{R}_{\rm{GB}} 
   ,
\end{align}
where
\begin{align}
   T_{\mu \nu} 
   &= 
   \nabla_{\mu} \varphi \nabla_{\nu} \varphi 
   - 
   \frac{1}{2} g_{\mu \nu} \nabla_{\rho} \nabla^{\rho} \varphi
   , \\
   C^{\rm{GB}}_{\mu \nu} 
   &= 
   2g_{\rho(\mu}g_{\nu)\alpha} \epsilon^{\kappa \alpha \tau \eta} 
   \nabla_{\lambda} 
   (^*\tensor{R}{^{\rho}^{\lambda}_{\tau}_{\eta}}  \nabla_{\kappa} \varphi).
\end{align}
We expand the metric and the scalar field as
\begin{align} 
   \label{eq:phiepansion}
   \varphi 
   &= 
   \sum_{k=0}^{\infty}{\frac{1}{k!} \epsilon^k \varphi^{(k)}} 
   ,\\
   \label{eq:metricexpansion}
   g_{\mu \nu} 
   &= 
   g_{\mu \nu}^{(0)} 
   +  
   \sum_{k=1}^{\infty}{\frac{1}{k!} \epsilon^k h_{\mu \nu}^{(k)}}
   .
\end{align}
where $\epsilon$ is a book-keeping parameter for our perturbative expansion. 
By considering the equations of motion order by order in $\epsilon$,
one finds that up to $\mathcal{O}(\epsilon)$~\cite{Witek:2018dmd}
\begin{align} 
   &
   \varphi^{(0)} 
   = 
   0, 
   &
   g_{\mu \nu}^{(0)} 
   = 
   g_{\mu \nu}^{\text{GR}}
   , 
   \label{eq:sol1} 
   \\
   & 
   \varphi^{(1)} 
   = 
   \varphi^{(1)}, 
   &
   h_{\mu \nu}^{(1)} 
   = 0
   . 
   \label{eq:sol2}
\end{align}
We note that the zeroth order equations of motion are the Einstein field equations 
minimally coupled to a massless scalar field.
The choice of $\varphi^{(0)}=0$ in the zeroth order solution~\eqref{eq:sol1}
is motivated by the fact that asymptotically flat black holes cannot carry scalar 
hair~\cite{cmp/1103857885},
which if initially present would be radiated away at late times, and that
a cosmological value of $\phi$ for EsGB gravity has been constrained by
measurements of the speed of gravitational waves~\cite{LIGOScientific:2017zic,Baker:2017hug}.
Using Eqs.~\eqref{eq:sol1} and~\eqref{eq:sol2},
the equations of motion to $\mathcal{O}(\epsilon)$ reduce to
\begin{align}
\label{eq:eomfinal1st}
   G_{\mu\nu}^{(0)}
   &= 0, \\
\label{eq:eomfinal2nd}
   \Box^{(0)} \varphi^{(1)} 
   &= 
   - \mathcal{R}_{GB}^{(0)} \beta_{0}.
\end{align}
We note that there is no backreaction of the scalar field on the 
background geometry, however at $\mathcal{O}(\epsilon^2)$ 
the metric will be corrected by the backreaction of the scalar field 
$\varphi^{(0)}$ on the spacetime 
\cite{Witek:2018dmd,Okounkova:2020rqw,Corman:2022xqg}. 

\begin{figure*}[hbt!]
\includegraphics[width=0.5\linewidth]{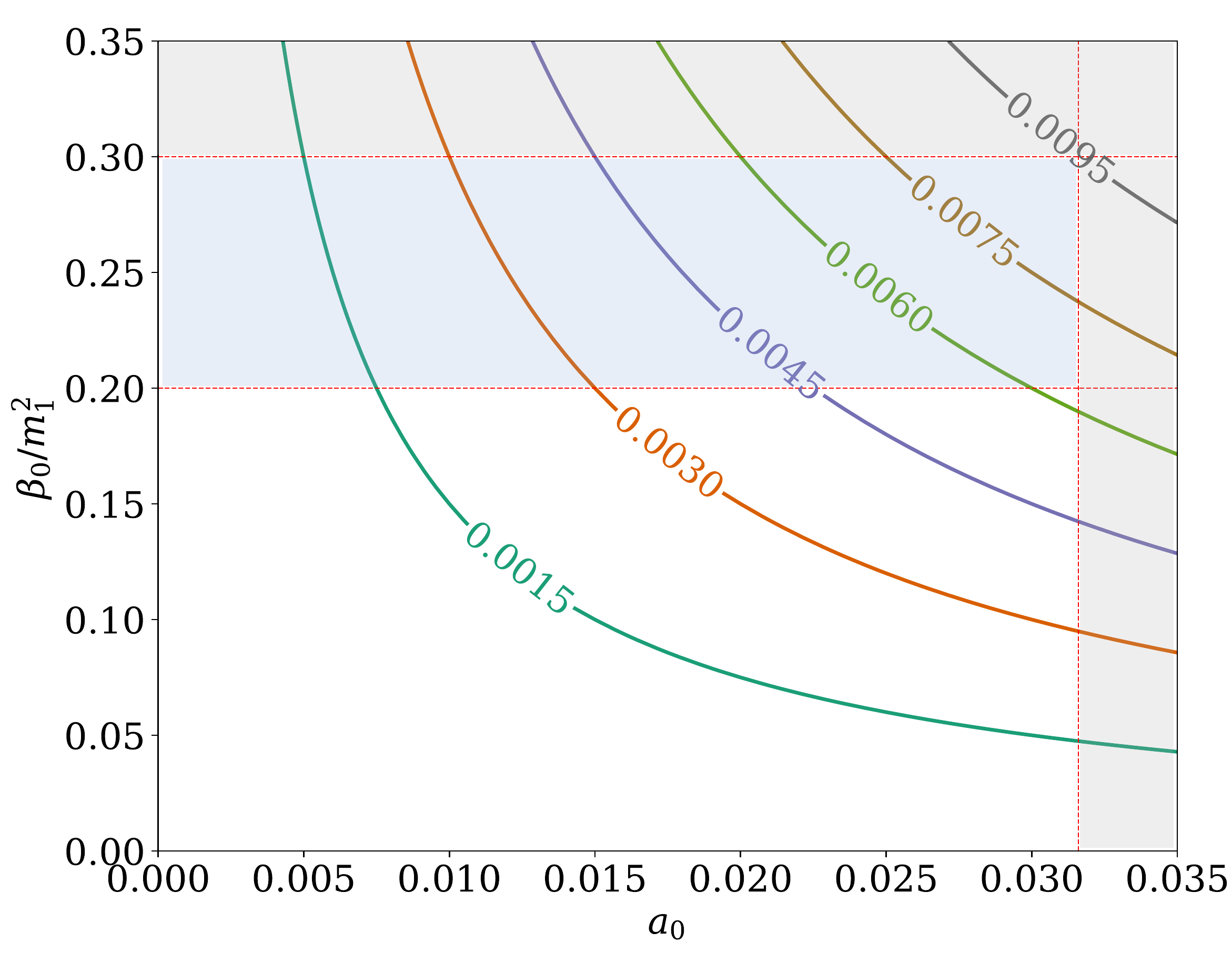}
\caption{Range of validity of our EsGB theory, 
   taking into account theoretical constraints on the dimensionless coupling 
   $\beta_0 / m_1^2$ and experimental constraint on the conformal factor $a_0$. 
   The vertical dashed line separates the maximally allowed value of $a_0$
    (to the left) from the observationally ruled out values (shaded gray region
    to the right). The two horizontal lines indicate the two theoretical constraints
    on the value of $\beta_0/m_1^2$. The upper line at
    $\beta_0/m_1^2=0.3$ represents the maximum coupling allowed before 
    naked singularities appear in static solutions to the theory, while the line at
    $\beta_0/m_1^2=0.2$ represents the upper bound before the theory suffers
    a breakdown in hyperbolicity. See Secs.~\ref{sec:perturbative} and  ~\ref{sec:constraints} for more discussion.}
\label{fig:validity}
\end{figure*}

We reintroduce dimensions by 
choosing the characteristic length scale of the system 
to be $L$, and introduce a dimensionless coordinate $\hat{x} = L x$. 
By redefining $x \to \hat{x}$, we acquire a factor of $L^{-2}$ 
in front of the wave operator and a factor of $L^{-4}$ 
in front of the Gauss-Bonnet scalar, leading to
\begin{equation}
   \hat{\Box}^{(0)} \varphi^{(1)} 
   =  
   -
   \beta_{0} L^{-2} \hat{\mathcal{R}}_{GB}^{(0)} 
   .
\end{equation}
We then may rescale the scalar field
\begin{equation} 
   \label{eq:dimfulscalar}
   \varphi^{(1)} 
   = 
   \left(\frac{\beta_{0}}{L^2} \right) \hat{\varphi}
   ,
\end{equation}
so that
\begin{equation}
   \hat{\Box}^{(0)} \hat{\varphi} 
   = 
   - 
   \hat{\mathcal{R}}_{GB}^{(0)}
   .
\end{equation}
This implies that a solution with a given $\beta_{0}/L^2$ 
parametrizes a family of solutions with different $L$ and $\beta_0$.
Further, Eq.~\eqref{eq:dimfulscalar} naturally introduces the choice for
the dimensionless parameter $\epsilon$, which we use to parametrize the weak coupling
of our solutions. 
Since we are interested in binary black hole systems and 
the corrections become larger for smaller masses, 
we choose $L=m_1$, 
which is the mass of the smaller black hole in our binary black hole configuration. 
We then define
\begin{align}
   \label{eq:definition_epsilon}
   \epsilon
   \equiv
   \frac{\beta_{0}}{m_1^2}
   \ll1
   .
\end{align}
By combining Eqs.~\eqref{eq:dimfulscalar} 
and~\eqref{eq:wave_zone_expansion}, 
we note that the scalar GW amplitude is then controlled by the couplings of 
our theory via $a_{0} \beta_{0}/m_1^{2}$. 
The dimensionless EsGB parameter $\epsilon$ 
plays an important role in EsGB gravity. 
Depending on the coupling $\beta_0$ chosen in the theory, 
a stationary BH solution of mass $m_1$ cannot exist above a certain threshold 
$\max\left(\beta_0/m_1^2\right) = \epsilon_{\rm{thr}} > \epsilon$. 
Such threshold is controlled by the regularity of the scalar field on the 
horizon and the finite radius singularity being hidden behind it for sufficiently 
small $\beta_{0}$. If the coupling becomes too large, 
a naked singularity emerges from within the horizon 
in solutions to the full theory. 
For instance, in the spherical-symmetric case of shift-symmetric EsGB, 
$\epsilon_{\rm{thr}} \sim 0.3$~\cite{Sotiriou:2014pfa}.
Hyperbolicity constraints on the theory suggest a somewhat more stringent
bound of $\epsilon_{\rm{thr}} \sim 0.2$ \cite{Ripley:2019aqj}.
In Fig.~\ref{fig:validity}, 
we outline schematically the range of validity of our EsGB theory in the 
parameter space of $a_0 \times \beta_0 / m_1^2$, 
taking into consideration the observational constraints on $a_0$ from 
Table~\ref{tab:constraints_gb} and theoretical constraints on $\beta_0/m_1^2$. 
The contour lines therefore represent possible scalar GW amplitude values.

%==============================================================================
\subsection{Current observational constraints}
\label{sec:constraints}
Here we discuss current experimental tests and observational bounds 
on the parameters $a_0$ and $\beta_0$. 
The EsGB coupling $\beta_0$  has been most strongly constrained through the 
analysis of inspiral and merger of binary systems of compact objects using GW data released by the LIGO-Virgo-KAGRA (LVK) collaboration~\cite{Lyu_2022}.
Weak-field tests of GR
also place a constraint on the coupling, 
but do not constrain it as significantly~\cite{Amendola:2007ni}, 
as the Gauss-Bonnet coupling does not lead to much scalarization 
of stellar solutions~\cite{Yagi:2011xp}.
This being said, as we are considering $a_0 \neq 0$, stars can scalarize in the theory we consider. In this work, we use the first constraint on $a_0$, obtained from the Viking relativity experiment on verification of signal retardation by solar gravity~\cite{1979ApJ...234L.219R}\footnote{We note however that the most stringent constraint of $a_0^2\lesssim 10^{-5}$ is known to be from the Cassini measurements of the Shapiro time delay~\cite{Bertotti_2022,Will:2014kxa}. Here we use a weaker constraint from the Viking experiment to see how well in principle the scalar polarization could be measured by present GW detectors.}, $a_0^2\lesssim 10^{-3}$.
Table~\ref{tab:constraints_gb} references the studies constraining the EsGB coupling\footnote{If one assumes that both objects in GW190814 are BHs, then an even more stringent constraint on the EsGB coupling can be found \cite{Wang:2021jfc}. However, in this work we take a conservative approach, by quoting the constraints from GW190814 that assume the secondary object is a neutron star.}
$\beta_0$ and the conformal coupling $a_0$. 
We additionally summarize theoretical constraints on the dimensionless coupling 
$\beta_0/m_1^2$ as discussed in Sec.~\ref{sec:order_reduction_scheme}.

\begin{table*} [hbt!]
\begin{tabular}{| c  c  c |}
    \hline
    \multicolumn{3}{|c|}{\textbf{Coupling} $\bm{\sqrt{\beta_{0}}}$} \\
    \hline
    \textbf{Study} & \textbf{References} &  \textbf{Constraint}\\ 
   \hline
   Solar system constraints & \cite{Amendola:2007ni} & $\mathcal{O}(10^7)$ km \\
   % \hline
   GW150914 ppE &  \cite{Shammi_2019} & $\lesssim 72.85$ km  \\
   % \hline
   EsGB constraints from LIGO/Virgo events and ParSpec formalism &  \cite{Carullo:2021dui} & $\lesssim 35$ km  \\
     % \hline
   EsGB constraints from EOB waveform model and ParSpec formalism &  \cite{Silva:2022srr} & $\lesssim 35$ km  \\
   % \hline
   X-ray binary orbital decay & \cite{Yagi_2012} & $\lesssim 14.1$ km \\
   % \hline
   LIGO SNR 30 detections (projection) & \cite{Stein_2014} & $\mathcal{O}(1-10)$ km \\
   % \hline
   Second order EdGB simulations for GW150914 & \cite{Okounkova:2020rqw} & $\lesssim 15.6$ km \\
   % \hline
   Compact star stability  & \cite{Pani_2011} & $\lesssim 7.6$ km \\
   % \hline
   First order EdGB scalar simulations for GW151226 & \cite{Witek_2019} & $\lesssim 3.8$ km \\
   % \hline
   EsGB constraints from BHBH GW events & \cite{Perkins:2021mhb} & $ \lesssim 1.7$ km \\
   % \hline
   EsGB constraints from BHNS GW events (GW200105, GW200115 and GW190814) & \cite{Lyu_2022} & $ \lesssim 1.18$ km \\
    \hline
    \multicolumn{3}{|c|}{\textbf{Coupling} $\bm{a_0}$} \\
    \hline
     Viking relativity experiment on signal retardation & \cite{1979ApJ...234L.219R, PhysRevLett.70.2220} & $\sqrt{10^{-3}}$ \\
    VBLI measurement of solar gravitational deflection  & \cite{Shapiro:2004zz} & $\sqrt{10^{-4}}$ \\
    Cassini Shapiro time delay & \cite{Bertotti_2022,Will:2014kxa} & $\sqrt{10^{-5}}$ \\
    \hline
     \multicolumn{3}{|c|}{\textbf{Theoretical constraints on} $\bm{\beta_0/m_1^2$} }\\
    \hline
   Shift-symmetric EsGB in spherical symmetry & \cite{Sotiriou:2014pfa} & $\lesssim 0.3$ \\
   % \hline
   Scalarized black holes in the full dynamical EdGB gravity  & \cite{Ripley:2019aqj} & $\lesssim 0.2$ \\
   \hline
   \end{tabular}
   \caption{Summary of current constraints on the EsGB coupling parameter 
      $\beta_{0}$, the dimensionless parameter $\beta_0 / m_1^2$
      and the conformal coupling $a_0$ (c.f.~\cite{Okounkova:2020rqw}). 
      Our conventions for $\beta_0$ and $a_0$ are given in Eq.~\eqref{eq:actionfinal}.
   }
\label{tab:constraints_gb}
\end{table*}

%==============================================================================
\section{\label{sec:numsetup}Numerical setup}

\subsection{Numerical relativity code and evolution equations}
\label{sec:num-evolution}

Here we describe our numerical set-
up to solve 
Eqs.~\eqref{eq:eomfinal1st} and \eqref{eq:eomfinal2nd}. 
For our simulations, 
we use a modified version of \texttt{GRChombo}~\cite{GRChombo2021, Clough:2015sqa, Radia_2022},
which is a publicly available finite difference numerical relativity code 
built on the \texttt{Chombo}~\cite{chombo} adaptive mesh refinement libraries.
The metric and the scalar field are evolved with the method of 
lines using fourth order finite difference stencils and Runge Kutta time integration. 
We use the covariant and
conformal Z4 (CCZ4) formulation~\cite{ccz4} with the moving puncture gauge~\cite{movingpuncture1, movingpuncture2} for evolving the Einstein equations
\eqref{eq:eomfinal1st}. 
The evolution equations for the Einstein field equations in the CCZ4 formulation can be found in Sec.~III F of~\cite{Z4eqs},
where we replace $\kappa_1 \to \kappa_1/\alpha$, 
in order to stably evolve BHs, 
and choose the constraint damping parameters to be 
$\kappa_1 = 0.1$, $\kappa_2 = 0$ and $\kappa_3 = 1$. 
The evolution equations for the scalar field~\eqref{eq:eomfinal2nd} read
\begin{eqnarray}
\label{eq:phit}
\partial_t \varphi 
& = &
\alpha \Pi + \beta^{i} \partial_{i} \varphi
,  \\
\label{eq:pit}
\partial_t \Pi 
& = &
  \alpha K \Pi + \beta^{i} \partial_{i} \Pi \nonumber \\
& {} &+ \tilde{\gamma}^{km}
\left(
 \chi \alpha \partial_{k} \partial_{m}\varphi
+ 
\chi \partial_{m}\alpha \partial_{k}\varphi 
-
\frac{1}{2} \alpha \partial_{m} \chi \partial_{k} \varphi
\right) \nonumber \\
& {} & -
\chi \alpha \tilde{\gamma}^{ij} \tilde{\Gamma}^{k}_{ij} \partial_{k} \varphi 
+
\alpha \left( \beta_{0} \mathcal{R}_{\rm{GB}}\right)
,
\end{eqnarray}
where $\Pi \equiv -1/\alpha (\partial_t \varphi
-\beta^m \partial_m \varphi)$, $\alpha$ is the lapse function, $\beta^{i}$ is the shift vector, $\tilde{\gamma}_{ij} \equiv \chi \gamma_{ij}$ is the conformally re-scaled metric constituent of the conformal factor $\chi = {\det(\gamma_{ij})}^{-1/3}$ and physical spatial metric $\gamma_{ij}$.
We rewrite the Gauss-Bonnet scalar in the gravito-electric, 
$E_{ij}$, and gravito-magnetic, $B_{ij}$, counterparts~\cite{Witek:2018dmd}
\begin{equation} \label{eq:decompRGB}
   \mathcal{R}^{GB} 
   = 
   8(E^{kl} E_{kl} -B_{kl}B^{kl})
   ,
\end{equation}
   where in CCZ4 variables they take the form of~\cite{Radia_2022}
\begin{equation} 
   \label{eq: egr}
\begin{aligned}
   E_{ij} 
   &= 
   \left[
      \mathcal{R}_{ij}  
      - 
      \tensor{K}{^s_i} K_{js} 
      + 
      K_{ij} (K - \Theta) 
      + 
      D_{(i} \Theta_{j)} 
   \right]^{TF}, \\
   B_{kl} 
   &=  
   \tensor{\epsilon}{_{ks(i}}D^s \tensor{K}{_{j)}^k} 
   .
\end{aligned}
\end{equation}
Here $\mathcal{R}_{ij}$ is the 3D Ricci tensor, $K_{ij}$ is the extrinsic curvature,
$\Theta = -n_{\mu}Z^{\mu}$ is the projection of the CCZ4 vector $Z^{\mu}$ 
onto the timelike unit normal $n^{\mu}$, $[..]^{\rm{TF}}$ 
denotes the tracefree part and $D_i$ 
is the covariant derivative compatible with the physical metric $\gamma_{ij}$. 
In the case of $\Theta = 0$, we recover the Baumgarte-Shapiro-Shibata-Nakamura-Oohara-Kojima formulation~\cite{Baumgarte_1998, Shibata:1995we, Nakamura:1987zz}.

%-----------------------------------------------------------------------------
\subsection{\label{sec:grid_and_ID}Grid setup and initial data}
For the results presented in this paper we 
set up a computational domain of size $1024M$.  
We use a grid spacing of $\Delta x= 2M$ in the outermost refinement level with 
eight additional refinement levels and a Courant factor of $1/4$. 
In this setup we roughly have 64 points covering the smallest BH in the 
most asymmetric run (mass ratio 1:2).
For boundary conditions, we use Sommerfeld boundary conditions and take 
advantage of the bitant symmetry to evolve only half of the grid. 
To validate our results, 
we estimated the discretization error for the amplitudes of $\Psi_4$ and 
$\varphi$ to be $\lesssim 2 \%$ and for phase of $\Psi_4$ to be 
$\lesssim 0.15$ radians. 
We further estimate the finite radius extraction error to be $ \lesssim 2 \%$ 
and thus a total error budget of $\lesssim 4 \%$.
See Figs.~\ref{Psi_convergence}--\ref{Psi_convergence_q05} of Appendix~\ref{sec:convergence} for 
a quantitative illustration of the convergence of the code.

For our initial data, we consider two nonspinning black holes with 
initial masses $m_1$ and $m_2$ (we assume $m_2>m_1$), 
initial linear momenta $\textbf{P}_{1}$ and $\textbf{P}_{2}$ 
and initial separation $D$. We prescribe quasicircular initial data for 
a black hole binary by using puncture initial data~\cite{puncture} of Bowen-York~\cite{bowenyork} type provided by the spectral initial data solver
\texttt{TwoPunctures}~\cite{mirenanomalies, spectralmarcus}.

We have investigated the effects of different initial separations on the 
eccentricity and found that $D=8M$ (resulting in four orbits) gave us an eccentricity 
estimator of no more than $0.012$ for all of the configurations considered here 
(we used Eq.~(20) of~\cite{Mroue:2010re} to calculate the eccentricity).

The initial data for the scalar field is chosen to be 
$\varphi = 0$ and $\partial_t\varphi=0$. 
The authors in \cite{Witek:2018dmd} found that using an initially vanishing scalar or superposing 
two hairy solutions gave identical results after sufficiently long evolution 
(see Fig.4 of~\cite{Witek:2018dmd}). Moreover, this choice of initial data
solves the exact constraint equations of EsGB gravity~\cite{East_2021,Ripley:2022cdh}.
We therefore do not investigate any other forms of initial data for the scalar field. 
We further pick three configurations for black holes of different mass ratios 
$q = m_1/m_2$, $(q \leq 1)$ summarized in Table~\ref{tab:simulations}.

\begin{table*}[hbt!]
\begin{tabular} {c  c  c  c  c  c  c  c  c}
 \hline
Run & $q$ & $m_1$ & $m_2$ & $\textbf{P}_{1}$ & $\textbf{P}_{2}$ & $M_{\rm{fin}}$ & $j_{\rm{fin}}$ & $j_{\rm{fin}}/M_{\rm{fin}}$\\ [1.5ex] 
 \toprule
\texttt{BBH-11} \quad & 1 & 0.5 & 0.5 & (-0.0013 -0.11 0.0) & (0.0013 0.11 0.0) & 0.9511 & 0.69 & 0.7255 \\ 
 % \hline
 \texttt{BBH-23} \quad & 2/3 & 0.4 & 0.6 & (-0.0012 -0.108 0.0)  & (0.0012 0.108 0.0)  & 0.9549  & 0.67 & 0.7016 \\ 
 % \hline
\texttt{BBH-12} \quad & 1/2 & 0.33 & 0.67 & (-0.0011 -0.1001 0.0) & (0.0011 0.1001 0.0) &  0.9611 & 0.62 & 0.6451  \\
 \hline
\end{tabular}
   \caption{Two puncture initial data used for three binary BH configurations 
   considered in this paper. 
   $M_{\rm fin}$ and $j_{\rm fin}$ denote the final mass and final spin of the 
   remnant BHs respectively, 
   which were calculated from balance arguments given in Eqs.~\eqref{eq:Mfin} and~\eqref{eq:jfin}.
   We estimate $0.01 \%$ error in the final mass and $0.05 \%$ error 
   in the final spin when compared to their Richardson-extrapolated values.
   In Appendix~\ref{sec:flux_convergence} 
   we present convergence plots for the radiated energy and angular momentum.
    \label{initconf}}
\label{tab:simulations}
\end{table*}

%==============================================================================
\subsection{\label{sec:extraction}Extraction of physical quantities}
To order $\mathcal{O}(\epsilon)$, the metric sector of our spacetime is 
determined by the Einstein equations of GR (see Eq.~\eqref{eq:eomfinal1st}),
so that the gravitational waveform will be no different from GR\@.
To calculate gravitational radiation we use the 
Newman-Penrose formalism~\cite{npformalism, Bishop:2016lgv}, 
and determine the outgoing gravitational radiation by
computing complex scalar $\Psi_4$. By interpolating the 
real and imaginary parts of $\Psi_4$ onto a sphere of fixed radius 
$R_{\rm{ext}} = 110M$, 
we decompose them in terms of spin-weighted spherical harmonics $Y^{s}_{lm}$
with $s=-2$~\cite{ulipunctures, spinweighted},
\begin{equation} 
   \label{eq:psi4}
   \Psi_{4, lm} (t, R) 
   = 
   \int_{S^2} \:
   \Psi_4(t, R, \theta, \phi) \:
   \overline{Y^{-2}_{lm}} (\theta, \phi) \text{d}\Omega,
\end{equation}
where $\rm{d} \Omega = \sin \theta \rm{d}\theta \rm{d}\phi$.
By similar decomposition into spherical harmonics $Y_{lm}$ 
and integrating over the sphere, we construct the scalar waveform
\begin{equation}
   \varphi_{lm} (t, R) 
   = 
   \int_{S^2} \:
   \varphi(t, R, \theta, \phi) \: \overline{Y_{lm}} (\theta, \phi) d\Omega
   .
\end{equation}

In what follows, we use the subscript
``$\rm{rad}$'' to denote quantities calculated from the start of
our simulation, which includes spurious ``junk'' radiation. 
Otherwise, we consider quantities from 
$t = R_{\rm{ext}} + 50M$ onwards. 
The mass and spin of the final BH are estimated from balance arguments~\cite{Witek_2010}
\begin{align}
\label{eq:Mfin}
M_{\rm{fin}} &= M_{\rm{ADM}} - E_{\rm{rad}}^{\rm{GW}} , \\
\label{eq:jfin}
j_{\rm{fin}} &= \frac{L - J^{z}_{\rm{rad}}}{M^2_{\rm{fin}}},
\end{align}
where $E_{\rm{rad}}^{\rm{GW}}$ denotes the total radiated energy of GWs, 
$J^{z}_{\rm{rad}}$ is the radiated angular momentum in the $z$ direction 
(by symmetry) and $L = P_{y} D$. 
In the above balance arguments the radiated energy and angular momentum are 
determined through~\cite{PhysRevD.59.124022, PhysRevD.76.041502}
\begin{align}
   E_{\rm{rad}}^{\rm{GW}} (t) 
   &= 
   \lim_{r \to \infty} \frac{r^2}{16 \pi} 
   \int_{t_0}^{t} dt' \oint_{S^2} d\Omega \:
   \mathbf{e}_r \left|\int_{-\infty}^{t'} dt'' \Psi_4 \right|^2
   , \label{eq:E_rad_GW}
   \\
   \mathbf{J}_{\rm{rad}}(t) 
   &= 
   - \lim_{r \to \infty}  \frac{r^2}{16 \pi} 
   \rm{Re} \int_{t_0}^{t'} 
   dt' \left\{\oint_{S^2} \left( \int_{-\infty}^{t'} 
   dt'' \Psi_4^{*} \right) \right. \nonumber \\
   &\left. \times \mathbf{J} 
   \left(\int_{-\infty}^{t'} 
      dt'' \int_{-\infty}^{t''} 
      dt''' 
      \Psi_4 
   \right) d\Omega \right \}, \label{eq:J_rad_GW}
\end{align}
where $\mathbf{e}_r$ is the unit radial vector of a sphere and 
$\mathbf{J}$ is the angular momentum operator for spin weight $s=-2$
\begin{align}
   \mathbf{J} 
   &=
   \left(\rm{Re} \mathbf{J}_{+}, \rm{Im} \mathbf{J}_{+}, \frac{\partial}{\partial \phi} \right)
   , \\
   \mathbf{J_{+}} 
   &= 
   \rm{e}^{ \rm{i} \phi} \left( 
      \rm{i} \frac{\partial}{\partial \theta} 
      - 
      \cot \theta \frac{\partial}{\partial \phi}
      + 
      2 \rm{i} \csc \theta
   \right)
   .
\end{align}

Apart from tensor energy, we expect some of the scalar field energy to be emitted too. 
Since the scalar field is of order $\varphi\sim\epsilon$, 
the radiated energy must be of order $E^{\varphi} \sim \epsilon^2$. 
We note that if we were to continue the order reduction scheme for our solution to 
$\mathcal{O}(\epsilon^2)$, 
there would be an additional radiated tensor energy sourced by 
$\mathcal{O}(\epsilon^2)$ corrections. 
At least in post-Newtonian (PN) theory this term is suppressed compared to
the scalar radiation, 
so we will ignore it as it is done in~\cite{Okounkova:2017yby,Witek:2018dmd}.
The energy flux of the scalar field through a two-sphere of radius $R$ is given via
\begin{equation} \label{eq:energy_flux}
   \frac{dE^{\varphi}}{dt} 
   = 
   \int_{S_R^2} T^{(\varphi)}_{ab} n^a dS^b
   ,
\end{equation}
where $n^a$ is the timelike unit normal and 
$T^{(\varphi)}_{ab} = \frac{1}{2} T_{ab} - \beta_{0} C^{\rm{GB}}_{ab} $ 
representing the right hand side of the modified Einstein's equations in Eq.~\eqref{eq:eoms_grav}. 
Assuming that a spherical surface is located in the radiation zone of 
an asymptotically flat spacetime and $\varphi$ behaves as outgoing radiation 
in the radiation zone, to leading order we find that 
$T^{(\varphi)}_{ab} = \frac{1}{2} T_{ab}$. 
Then,
to leading order Eq.~\eqref{eq:energy_flux} reads
\begin{equation} \label{eq:energy_flux_leading}
   \frac{dE_0^{\varphi}}{dt} 
   = 
   - 
   \int_{S^2} r^2 T^r_t d\Omega = \int_{S^2} [\dot{\varphi}(t-r)]^2 d\Omega
   ,
\end{equation}
where the dot denotes time differentiation. 
We therefore integrate Eq.~\eqref{eq:energy_flux_leading}
to find the total radiated scalar energy to leading order.

%==============================================================================
\subsection{\label{sec:numerical_results}Numerical results: Scalar waveforms}
\begin{figure*}[hbt!]
   \includegraphics[width=.5\linewidth]{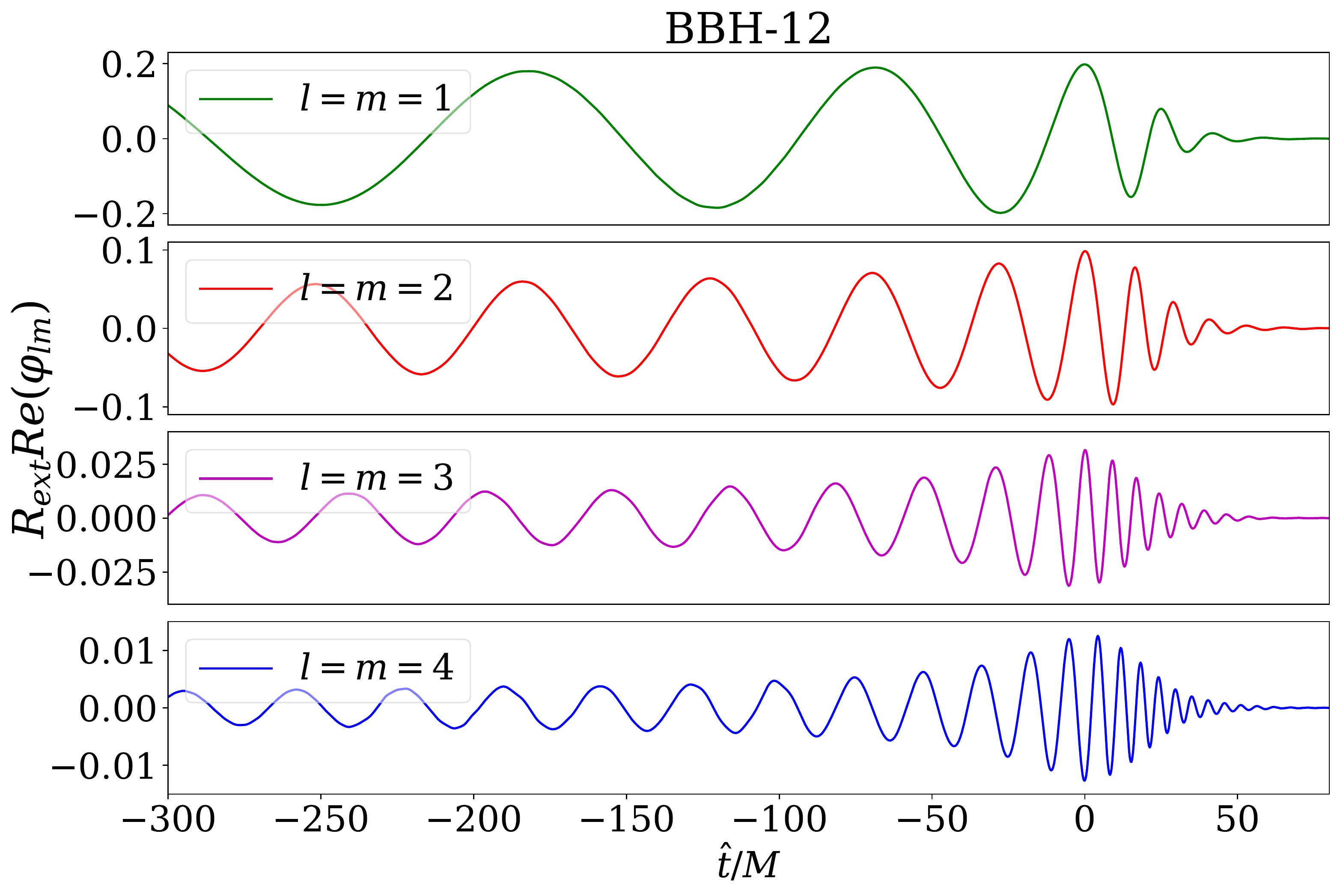}\hfill
   \includegraphics[width=.5\linewidth]{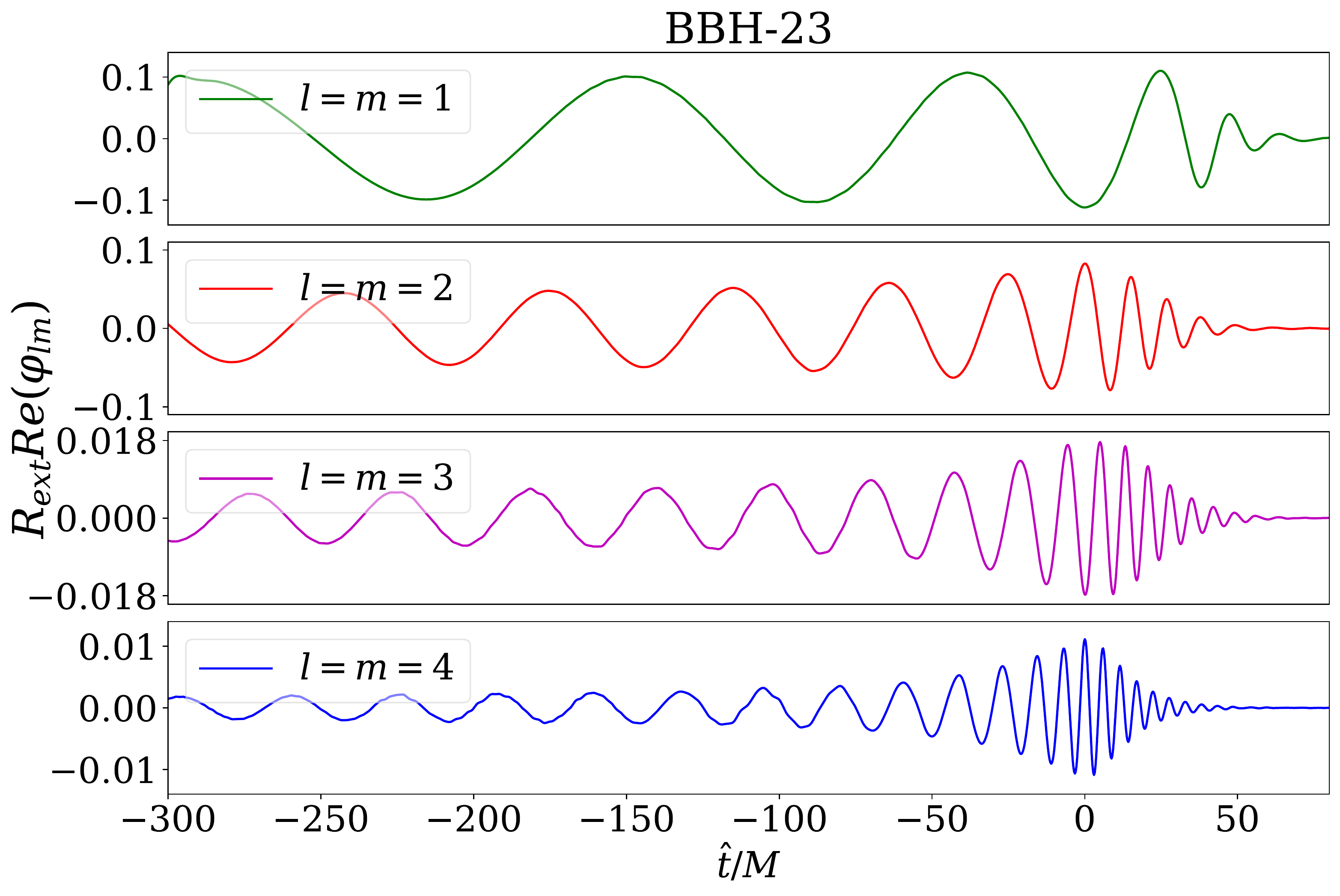}\par 
   \includegraphics[width=.5\linewidth]{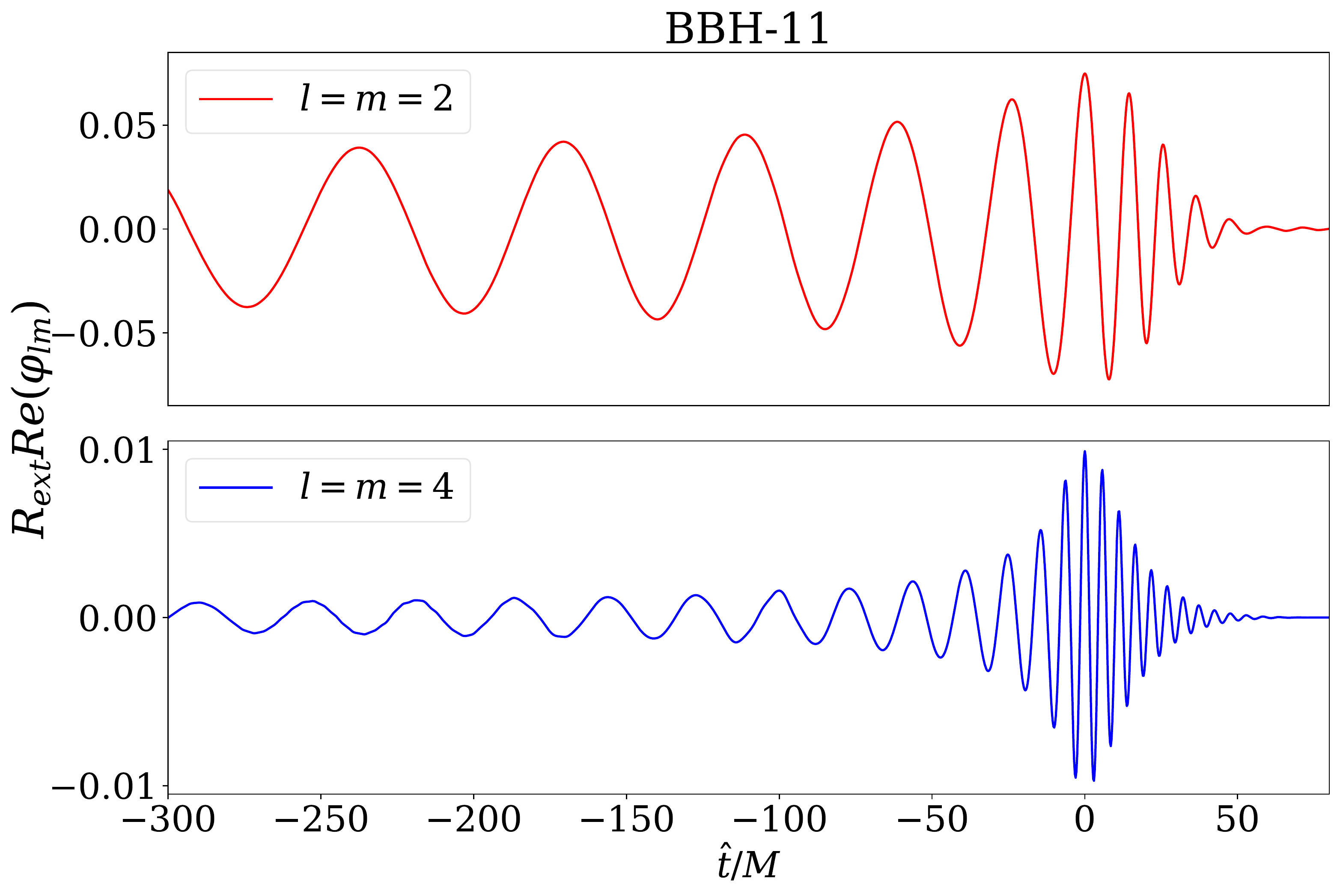}
   \caption{Scalar waveforms for $l \geq 1$ for mass ratios of 
   $q=1, 2/3, 1/2$ extracted at $R_{\rm{ext}} = 110$. 
   For unequal mass cases, the scalar field amplitude is more pronounced.  
   The waveforms have been aligned so that the amplitude peaks of the scalar field coincide at $\hat{t} = 0$,
    and we set $\beta_0/m_1^2 = 1$ for convenience
    (recall that since we are using an order-reduction scheme, there is no restriction on the
    value of $\beta_0/m_1^2$ in our evolution). 
    In our analysis, we choose a physically allowed value of $\beta_0/m^2$ 
    and rescale the scalar waveforms using Eq.~\eqref{eq:dimfulscalar}.}
\label{scalar_waveforms}
\end{figure*}

\begin{figure*}[hbt!]
   \includegraphics[width=0.5\linewidth]{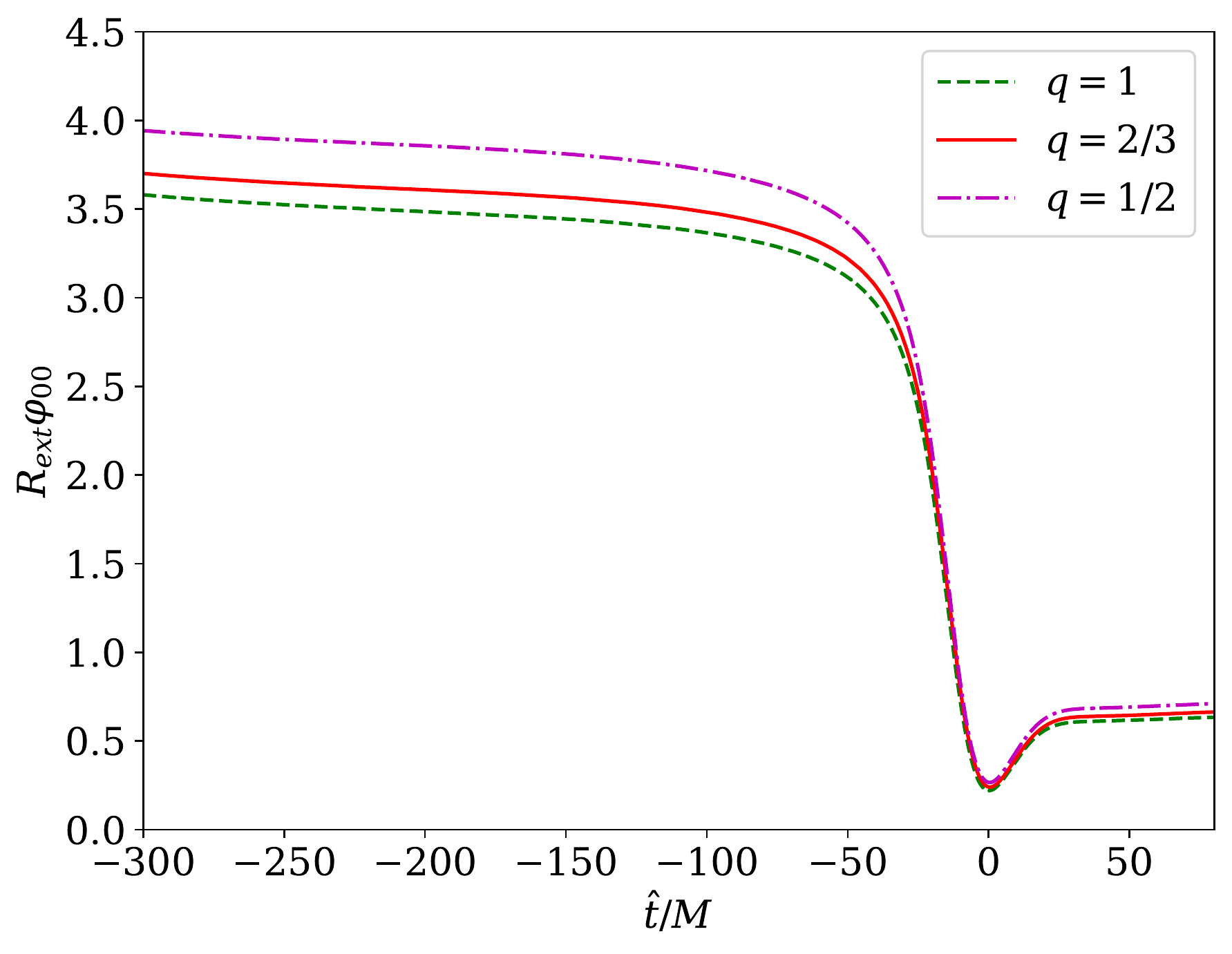}
   \caption{Scalar $l=0$ mode for binary mass ratios of 
   $q=1, 2/3, 1/2$, extracted at $R_{\rm{ext}} = 110$.}
\label{scalar_waveforms_00}
\end{figure*}

In this section we present our numerical results.
The initial data for the simulations are listed in Table~\ref{tab:simulations}.
We find good agreement with numerical results of~\cite{Witek_2019}, which
made use of the same decoupling approximation as we do. 
We show scalar waveforms for $l \geq 1$ modes 
in Fig.~\ref{scalar_waveforms} and for $l=0$ mode in Fig.~\ref{scalar_waveforms_00}, where we align the waveforms so that the amplitude peaks of the scalar field coincide at $\hat{t} = 0$ and choose $\beta_0/m_1^2$ for convenience.

\begin{figure*}[hbt!]
   \includegraphics[width=0.6\linewidth]{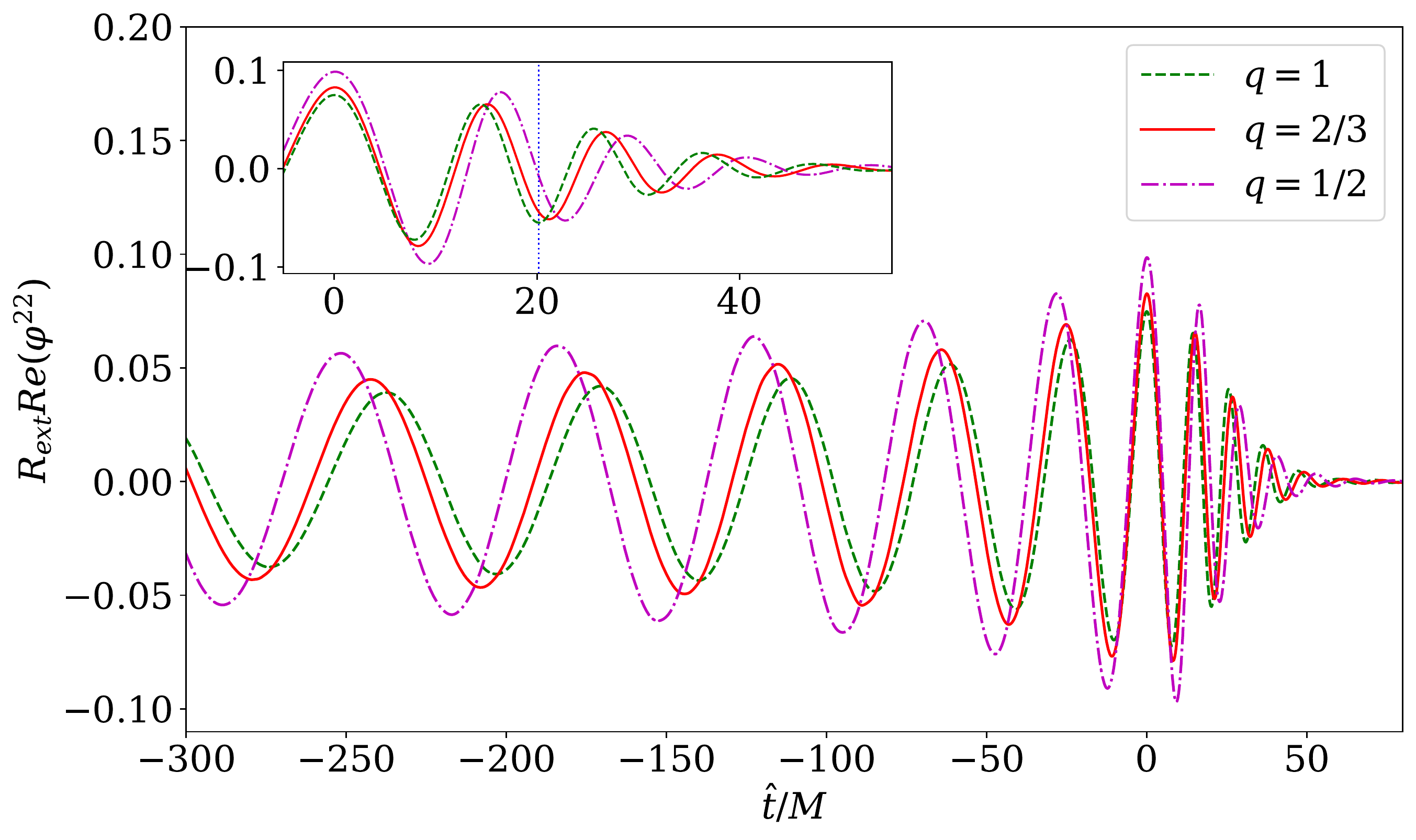}
    \caption{Scalar $l=2,m=2$ mode
    for mass ratios of $q=1, 2/3, 1/2$. 
   The inset shows the zoom-in on the ringdown portion of the signal. 
   Within the inset, the blue dotted line indicates the time where the amplitudes of 
   the $q=2/3$ and $q=1/2$ waveforms become smaller in amplitude 
   than the $q=1$ waveform.}
   \label{mode_22_comparison}
\end{figure*}

For equal-mass binaries we have only even $l$-modes present in the waveform, 
as equal-mass nonspinning binaries are symmetric under parity transformations 
($\vec{r}\to-\vec{r}$). 
Since $Y^m_{l=\rm{odd}}\left(\vartheta,\varphi\right)$ 
is antisymmetric and the EsGB coupling is invariant under parity, 
there must be no emission from odd $l$-modes for equal-mass binaries. 
On the other hand, in the unequal-mass case, odd $l$-modes are present and 
the dipole emission dominates all other higher modes of the scalar waveform for 
EsGB gravity (it enters at $-1$PN) order) 
\cite{Yagi:2011xp, Shiralilou:2020gah,Shiralilou:2021mfl}.  
In Fig.~\ref{scalar_waveforms_00}, we see that the $l=0$
mode looks unlike the other modes and is roughly constant before merger time
(this is consistent with leading order PN theory~\cite{Yagi:2011xp,Witek_2019}).
After merger it settles to almost the same value for all the mass ratios, 
since the remnant BHs formed have quite similar final masses and spins, and
thus the same amount of remnant scalar hair. 
Overall, from Figs.~\ref{scalar_waveforms} and~\ref{scalar_waveforms_00},
at fixed $\beta_0/m_1^2$, the scalar field radiation becomes stronger 
for more extreme unequal-mass binaries, especially in the inspiral 
and merger portions of the signal. 
End state BHs with smaller spin 
(\emph{i.e.}\ BHs formed from more extreme mass ratios)
have faster damping times than BHs with higher spin. 
As a result, the scalar waves in the ringdown
are more pronounced at late times for black holes produced from more symmetric-mass-ratio
($q \sim 1$) progenitors.
To illustrate this point, we align the $(22)$ modes in the amplitude peak of the scalar field for different mass ratios in 
Fig.~\ref{mode_22_comparison}, and zoom-in on the ringdown,
where the amplitude for more symmetric-mass-ratio binaries becomes larger than
for more asymmetric systems later in time.

\begin{figure*}
  \includegraphics[width=\linewidth]{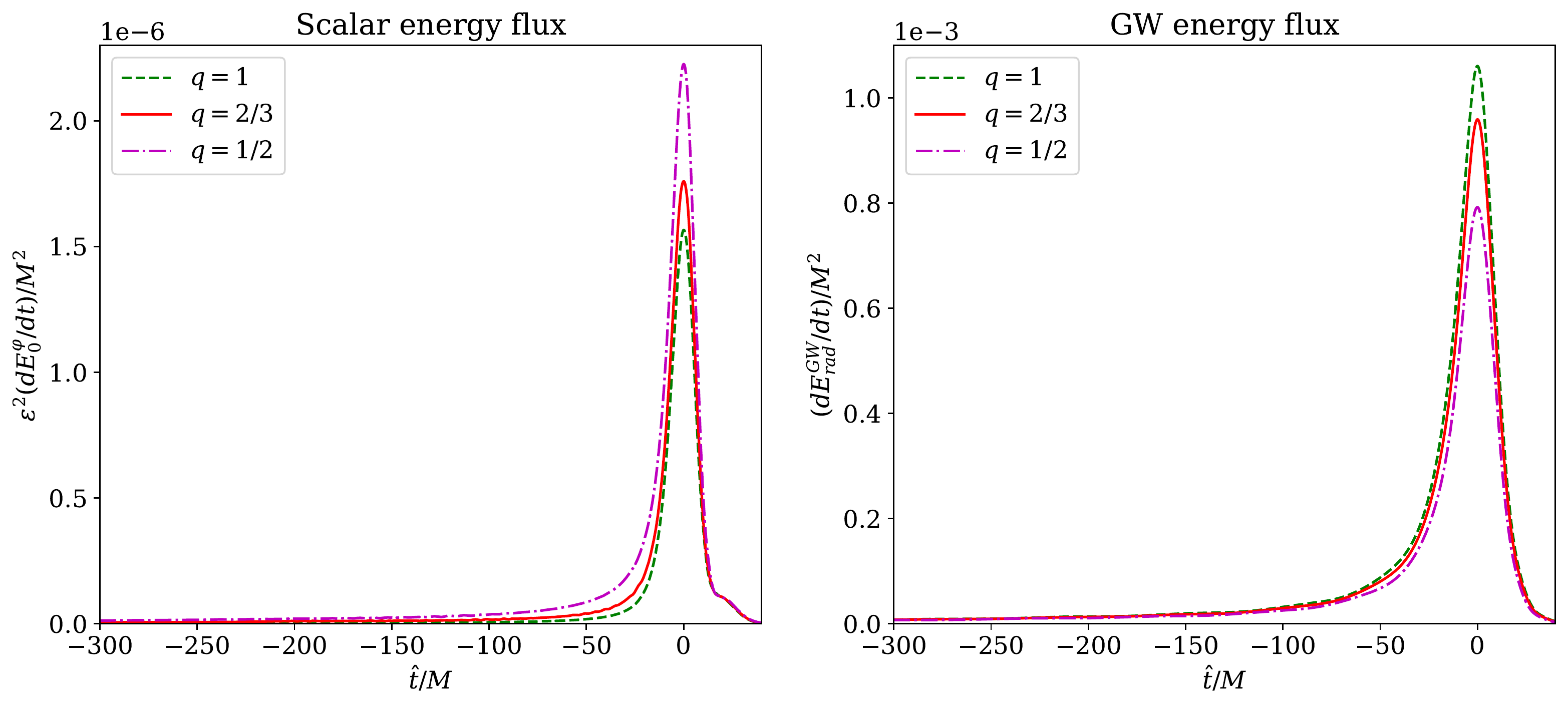}
 \caption{\textit{Left}: 
   scalar energy flux computed using Eq.~\eqref{eq:energy_flux_leading}.
    As in Fig.~\ref{scalar_waveforms}, we set $\beta_0/m_1^2 = 1$. 
   \textit{Right}: 
   gravitational wave energy flux computed using Eq.~\eqref{eq:E_rad_GW}.}
\label{fig:energy_fluxes}
\end{figure*}

Finally, the scalar energy flux, as computed from 
Eq.~\eqref{eq:energy_flux_leading}, is shown in Fig.~\ref{fig:energy_fluxes}.
We see that it is dominated by the contribution from $l=0$ mode,
which is larger for more unequal-mass-ratio binaries, as expected from
PN theory~\cite{Yagi:2011xp,Witek_2019}.
This is unlike the tensorial gravitational energy flux,
where the strongest flux comes from the equal-mass binary. 
This can be explained by the fact that to leading order the total radiated 
gravitational energy $E_{\rm{rad}}^{\rm{GW}} \sim \eta^2$, 
where $\eta = q/(1+q)^2$ is the symmetric mass ratio~\cite{Berti:2007fi}.
We note that the scalar energy flux measured here using Eq.~\eqref{eq:energy_flux_leading} is the second-order effect and so
is significantly suppressed---at merger,
it accounts for much less than $1 \%$ of the gravitational energy flux.

%==============================================================================
\section{Modeling the Scalar Ringdown Waveform}
\label{sec:ringdown}
When exploring the GW phenomenology of astrophysical systems in modified gravity,
the ultimate aim is to obtain an accurate waveform model for the signal 
(here the scalar ringdown) 
and to be in a position to observe the effects in question and 
infer information on the underlying theory from the observed GW data. 
In this section we model the scalar ringdown based on the NR data of
Sec.~\ref{sec:numerical_results}.
The gravitational waves emitted by the remnant BH formed by a merger,
known as the ringdown signal,
can be approximated as a superposition of damped sinusoids~\cite{1971ApJ...170L.105P}.
In GR, each mode of the tensorial ringdown signals can then be written as
\begin{equation}
   h^{lm} 
   = 
   \sum_{J=1}^{N} H_J e^{-i (\omega_J^{\rm{GR}} (t-t_0) + p_J)}
   , \label{eq:damped_sin_grav} 
   \\
\end{equation}
where $t_0$ is the start time of the ringdown, $J = (lmn)$ is the mode number,
$H_{J}$ and $p_{J}$ are real amplitude and phase of mode $J$ at $t=t_{0}$,
(which depend on the binary configuration and dynamics near merger), 
and $\omega_{J}$ are the complex frequencies of the $N$ most dominant QNMs. In a similar fashion, the scalar ringdown signal can be written as,
\begin{equation} 
\varphi_{lm} 
    = 
   \sum_{J=1}^N A_J e^{-i (\omega_J^{\rm{GR}} (t-t_0) + p_J)}
   \label{eq:damped_sin}
   ,
\end{equation}
where $A_J$ now denote scalar real amplitudes. For a mode $J$ corresponding to a given set of QNM indices $(lmn)$,
we write the real and imaginary parts as 
\begin{equation}
   \omega_{\rm{QNM}}  
   = 
   \omega_{lmn}- \frac{i}{\tau_{lmn}}
   .
\end{equation}
The integers $(lm)$ describe the angular properties of the emission,
while $n$ is the overtone number~\cite{Berti:2009kk}.
Modes with $n = 0$ are referred to as the \textit{fundamental modes},
while with $n \geq 1$ as \textit{overtones},
which (at least in GR)
have lower frequencies and greater damping times than their
corresponding fundamental modes.
We have denoted the total number of quasinormal modes with which we model the ringdown signal by $N$.

The no-hair theorem of GR states that
the damping times, $\tau^{lmn}$, and frequencies, $\omega^{lmn}$, for gravitational
wave perturbations around astrophysical (noncharged) BHs in GR, are uniquely determined by the mass and spin of
the final black hole.
The excitation coefficients also depend on the
initial conditions of the perturbation~\cite{Berti:2009kk}, in our case the progenitor parameters. For modified theories of gravity, the QNM spectrum of BH spacetimes generally depends on the theory in question,
and so observations of the ringdown signal can provide a way 
to discriminate between GR and possible modified theories of gravity. 
Unlike GR~\cite{teukolskyformalism}, for alternative theories of gravity there is typically no
set of separable master equations for linearized perturbations of black holes. 
Instead, the calculation of QNMs has relied on either the
assumption of spherical symmetry~\cite{Konoplya_2020,Bryant:2021xdh}, 
the slow spin approximation~\cite{Pierini:2021qnms, Pierini:2022eim},\footnote{However, there has been recent work on extending the Teukolsky formalism
to modified gravity theories to compute their quasinormal modes for fast-rotating
black hole solutions~\cite{Li:2022pcy}.}
or on numerical fits of time-domain waveforms computed from NR 
simulations~\cite{Witek:2018dmd, Okunkova:2019NRsecondorder}.
In this work we focus on the last method by performing numerical fits 
to the scalar and tensor waveforms.

In our analysis, we take a similar approach of modeling the scalar ringdown 
as a superposition of damped sinusoids given by Eq.~\eqref{eq:damped_sin}, 
except our QNM frequencies no longer correspond to just scalar QNMs of GR\@.
We note that the operator acting on $\varphi$ in its equation of motion
consists of the background GR metric in the decoupling approximation
we use (see Eq.~\eqref{eq:eomfinal2nd}).
Therefore, the homogeneous solution of the beyond-GR scalar perturbation equation 
around a BH spacetime should contain the same scalar QNMs as the corresponding Kerr BH in GR, $\omega_{lmn}^{\rm{Kerr, scalar}}$.
However, our equation is nonhomogeneous as we also have the Gauss-Bonnet scalar 
evaluated on the background GR metric, acting as a source term.
Particular solutions to the scalar equation introduce additional frequencies
into the ringdown QNM spectrum~\cite{Witek:2018dmd,Okunkova:2019NRsecondorder}.
The spectrum of these ``driven'' modes by the source term for $l\geq2$ has been shown to 
coincide with the quasinormal modes for tensor perturbations for Kerr,
$\omega_{lmn}^{\rm{Kerr, grav}}$;
while for $l=1$ the driven mode has been fitted empirically~\cite{Witek:2018dmd}.

Can our scalar ringdown waveforms be modeled by the gravitational and/or scalar QNMs of a Kerr black hole in GR as above findings suggest? To address this question we perform one-mode ($N=1$) and two-mode ($N=2$) fits, whilst keeping the QNMs fixed to their values in GR. These fixed values employed in our analysis can be found in Table~\ref{qnms_of_gr} of Appendix ~\ref{sec:gr_qnms}. For one-mode fits, we utilize Eq.~\eqref{eq:fit_summary_1}, where we check whether just one QNM fixed to either $\omega_{lmn}^{\rm{Kerr, scalar}}$ or $\omega_{lmn}^{\rm{Kerr, grav}}$ describes the data well. Next, we perform a two-mode fit using Eq.~\eqref{eq:fit_summary_2}, where we fix  $\omega_{lmn}^{\rm{Kerr, scalar}}$ and $\omega_{lmn}^{\rm{Kerr, grav}}$ simultaneously. We note that for the $l=1$ ringdown modes fitted with the $N=2$ mode fit we take a conservative approach and only use a fundamental scalar QNM and its first overtone:
   
\begin{widetext}
\begin{align} 
Re(\varphi^{lm}_{N=1}) &= 
   \begin{cases}
      A_{1}^{lm0} e^{-(t-t_0) / \tau_{lm0}} \text{cos}(\omega_{lm0}(t-t_0)+p_1^{lm0}), 
      \quad 
      \omega_{\rm{QNM}} \in 
      \left\{\omega_{\rm{QNM}}^{\rm{Kerr, scalar}} , \omega_{\rm{QNM}}^{\rm{Kerr, grav}}\right\} 
      &
      l,m>1
      \\
      A_{1}^{lm0} e^{-(t-t_0) / \tau^{\text{Kerr, scalar}}_{lm0}}  \text{cos}(\omega_{lm0}^{\text{Kerr, scalar}} (t-t_0)+p_1^{lm0})
      &
      l=m=1
      ,
   \end{cases}  \label{eq:fit_summary_1}
   \\
   \vspace{2cm}
   Re(\varphi^{lm}_{N=2}) &= 
   \begin{cases}
   \begin{aligned}
      A_{1}^{lm0} e^{-(t-t_0) / \tau^{\rm{Kerr, scalar}}_{lm0}}  \text{cos}(& \omega_{lm0}^{\rm{Kerr, scalar}} (t-t_0) + p_1^{lm0}) \\ 
      & + 
      A_{2}^{lm0} e^{-(t-t_0) / \tau^{\rm{Kerr, grav}}_{lm0}}  \text{cos}(\omega_{lm0}^{\rm{Kerr, grav}} (t-t_0) + p_2^{lm0})
      &
      l,m>1
   \end{aligned}
      \\
   \begin{aligned}
      A_{1}^{lm0} e^{-(t-t_0) /\tau^{\rm{Kerr, scalar}}_{lm0}} \text{cos}(& \omega_{lm0}^{\rm{Kerr, scalar}} (t-t_0) + p_1^{lm0}) +  \\
     &  A_{2}^{lm0} e^{-(t-t_0) / \tau^{\rm{Kerr, scalar}}_{lm1}} \text{cos}(\omega_{lm1}^{\rm{Kerr, scalar}} (t-t_0)+p_2^{lm1})
      &
      l=m=1
      .
   \end{aligned}
   \end{cases} \label{eq:fit_summary_2}
\end{align}
\end{widetext}

Equations~\eqref{eq:fit_summary_1} and \eqref{eq:fit_summary_2} fit to the real parts of the scalar multipoles, as their imaginary parts will have the same amplitudes and phases. We make use of all scalar mode waveforms presented in Sec.~\ref{sec:numerical_results},
 and therefore perform the fitting on the 
 $(lm) = (11), (22), (33), (44)$ modes.

We sample the parameter space using \texttt{emcee}~\cite{emcee}.
Dropping the $(lmn)$ indices in Eqs.~\eqref{eq:fit_summary_1} and \eqref{eq:fit_summary_2}
and keeping just the mode number $N \in [1,2]$, 
we use uniform priors in our inference, namely $\log{A_i} \in [-3, 1]$ and 
$p_i \in [0, 2\pi]$ for $i \in [1,N]$. We compute the frequencies $\omega_{lmn}$
with the \texttt{qnm} package~\cite{Stein:2019mop}
using our final spin 
calculations from Table~\ref{initconf}.

There is an ongoing debate around when the optimal
time is to begin fitting quasinormal ringdown (see \emph{e.g.}~\cite{Bhagwat:2017tkm}), as different
start times can give different answers for the mode fits, and have also caused some controversy regarding the confidence of a subdominant QNM observation in GW data~\cite{Capano:2021etf}.
While overtones are typically quick to decay, they can play an important role as we approach the time of merger~\cite{London:2014cma,Giesler_2019,JimenezForteza:2020cve,MaganaZertuche:2021syq,Isi:2021iql,Cotesta:2022pci}.
To avoid the intricacies of having to account for overtones, as well as nonlinear effects in the early ringdown~\cite{Cheung:2022rbm,Mitman:2022qdl}, we take a conservative approach and start the ringdown analysis $t_0 = 16M$
after the time of the peak amplitude 
of the scalar mode\footnote{We have investigated the choice of other 
start times ranging from $10M$ to $16M$ and found similar results 
on the estimated parameters.}. 
By starting too early we risk including nonlinear effects from 
the merger and by starting too late the SNR of the signal may significantly be lower.
In our analysis, the damped sinusoid model is most accurate as
an approximation in the late ringdown and so we sacrifice on
the strength of the signal.
The end time of our analysis is determined by the point in the 
late ringdown when the signal becomes very faint and the 
numerical noise becomes more significant.
The point when this happens varies on the mode-by-mode basis, 
with higher modes being prone to numerical noise earlier than others. 
However, the typical length of our ringdown varies in the range of $\sim 15-30M$.

\begin{figure*}
\begin{minipage}{.5\textwidth}
  \includegraphics[width=1\linewidth]{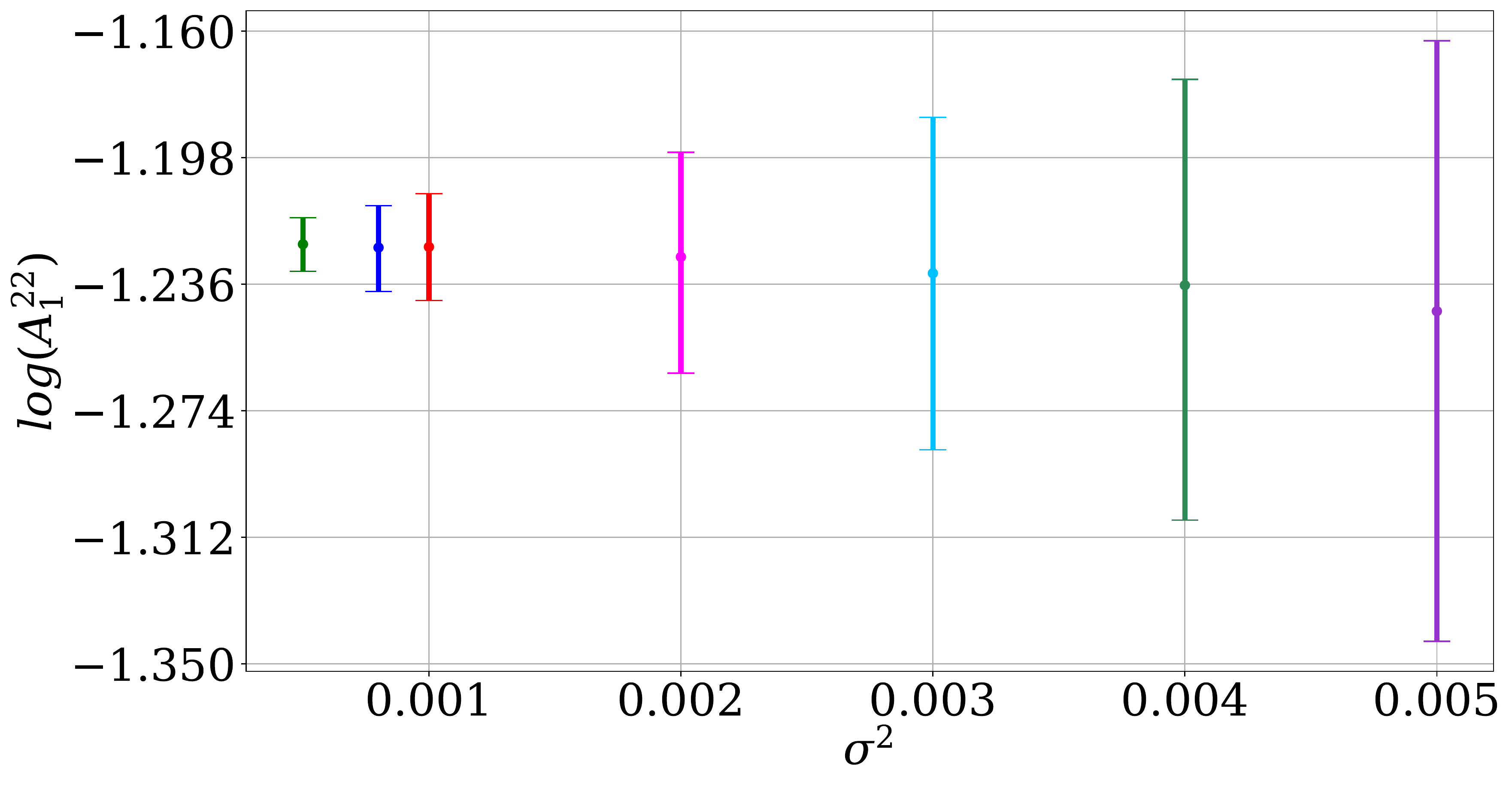}
\end{minipage}%
\begin{minipage}{.5\textwidth}
  \includegraphics[width=1\linewidth]{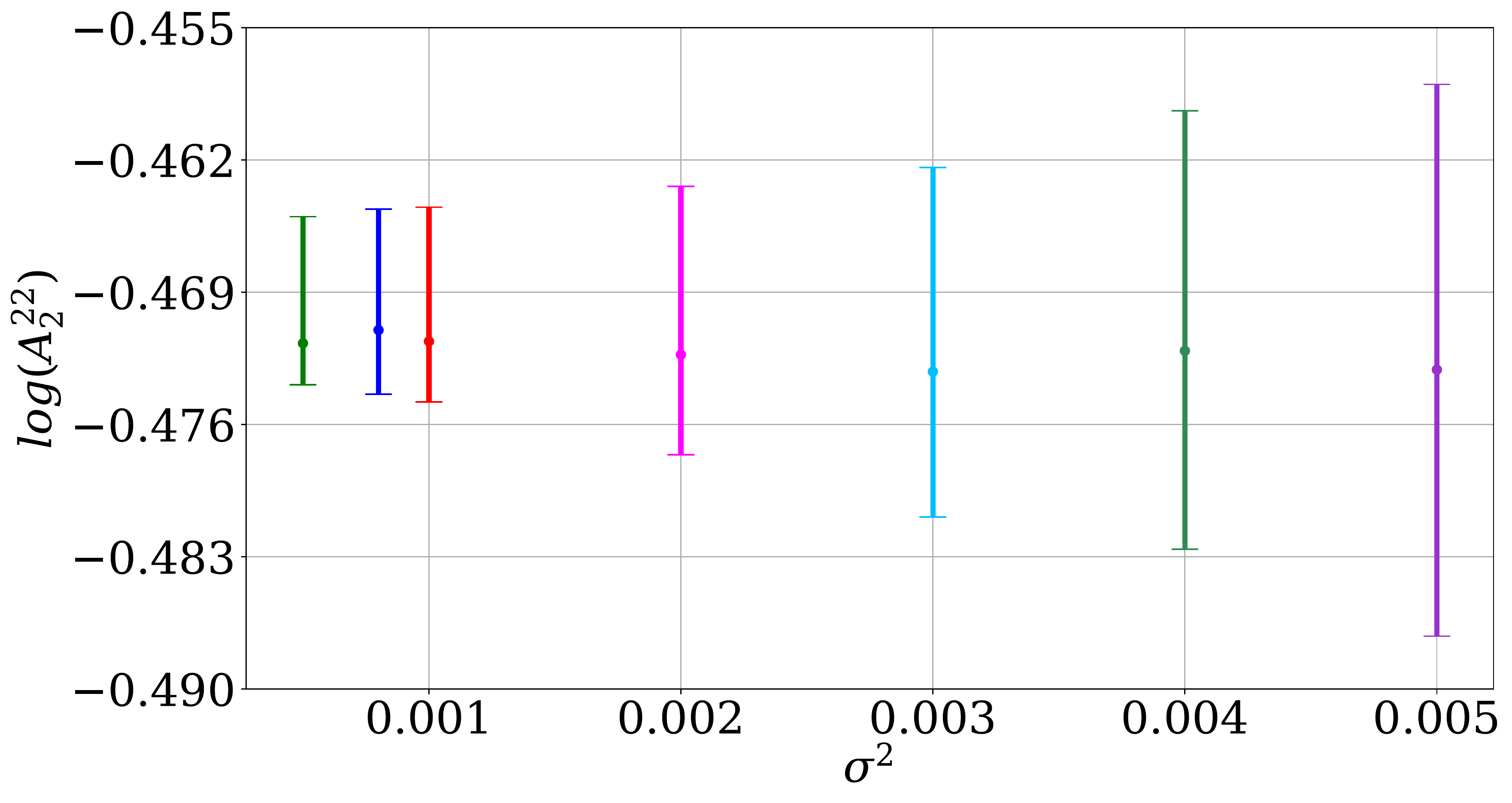}
\end{minipage}
 \caption{Convergence of the estimated amplitudes with decreasing variance 
   $\sigma^2 = \{0.005, 0.004, 0.003, 0.002, 0.001, 0.0008, 0.0005 \}$ 
   of Eq.~\eqref{eq:likelihood} for the $(22)$ mode of \texttt{BBH-12} run. 
   The variance value estimated as $4 \%$ from the peak amplitude takes 
   the value of $\sigma^2=0.003$ in our fit for this configuration. 
   We have verified that the same convergence holds qualitatively for the 
   $N=1$ mode fit too.}
\label{fig:sigma_conv}
\end{figure*}

In the optimization process, 
we use a least-squares likelihood in the time domain, 
with a flat noise spectral density
\begin{equation}\label{eq:likelihood}
   L(d;  \bm{\theta}) 
   \propto  
   \sum_{i} {\exp\left(-\frac{(\varphi_{\vec\theta}(t_i)-d(t_i))^2}{2 \sigma^2} \right)}
   ,
\end{equation}
where $d(t_i)$ is the time sequence of numerical data, 
$\sigma^2$ is the variance and $\bm{\theta} = \{A_i, p_i \}$. 
We estimate the variance by taking our total error budget percentage (\emph{i.e.}\ $4 \%$)
from the peak of the amplitude for each mode. 
As such, the error we account for through the variance parameter varies on 
a mode by mode basis.
This choice for error estimation in our analysis assumes that the total 
NR error is not correlated in time.
In reality, this is not likely to be an accurate assumption to make as 
in certain regions of the waveform we may in fact 
be overestimating or underestimating the allowed deviation of the data from the true values. 
Ideally, our likelihood function should contain some sort of correction 
to account for such correlation of the data with time. 
However, this is out of the scope of this work and we leave this for future study. 
We have verified the robustness of our choice of variance in the likelihood 
function of Eq.~\eqref{eq:likelihood} by gradually decreasing its value and
assessing its effect on the estimation of the modes' amplitudes. 
As expected, with decreasing variance, the amplitudes converge to some 
fixed values with smaller error bars, representing an ideal scenario where our 
NR error is very small. 
These idealized values provide the relative error we make in our 
fitting procedure when estimating amplitudes using our actual total NR error 
budget percentage to calculate fixed variance. Figure~\ref{fig:sigma_conv}
shows this convergence of $A_i$ for $i \in [1,2]$ of the $(22) $ mode of the 
\texttt{BBH-12} configuration, where the fixed variance of 
$\sigma^2 = 0.003$ used in the fitting procedure results in 
median amplitudes lying within the error bars of the smallest variance.

In Table~\ref{tab:fits_results} we summarize the results of our fits.
We note that we fixed the dimensionless coupling $\beta_0/m_1^2$ 
to unity for convenience there. 
Therefore, when choosing a different value for the coupling, appropriate rescaling to the amplitudes has to be done 
according to Eq.~\eqref{eq:dimfulscalar}. 
We find that the ratio of amplitudes of scalar Kerr QNMs to 
gravitational Kerr QNMs varies on a mode-by-mode basis for $l>1$ modes.
In the case of the $l=1$ mode fit, the amplitude of the overtone $n=1$
dominates the amplitude of its fundamental mode as also found in the analysis 
of gravitational ringdown of Ref.~\cite{Giesler_2019}. Unless indicated by dots in Table ~\ref{tab:fits_results}, we find that for certain modes a $N=1$ mode fit is able to describe 
the data accurately enough.
This happens in the cases where the amplitude of one of the modes in the $N=2$ mode fit is dominant, 
or when the frequencies of the $N=2$ mode fit lie close to each other. 

We additionally fit the amplitudes of our gravitational waveforms in Table 
\ref{tab:fits_results_grav}, 
where we use $(lm) = (11), (22), (33), (44), (55), (66)$ 
modes in the fitting procedure. 
We find that the $N=1$ mode fit of Eq.~\eqref{eq:damped_sin_grav}
is sufficient to describe the data. 
Since the Einstein equations to leading order remain unchanged 
(see Eq.~\eqref{eq:eomfinal1st}), 
we have fixed the frequency of the $N=1$ mode fit to the
fundamental gravitational Kerr QNM of GR\@.

In summary, the leading order QNMs of the scalar and tensor 
components of the ringdown waveform in EsGB gravity
are accurately described by QNMs of GR\@.
The next correction in $\epsilon$
to the scalar QNM spectrum would depend on the 
nature of the coupling to the Gauss-Bonnet term, 
while the correction to gravitational QNM spectrum would happen at 
$\mathcal{O}(\epsilon^2)$. 
In the case of subleading corrections to the scalar QNMs, 
for EdGB gravity with exponential coupling, $\beta(\varphi) = e^{\varphi}$, 
the next correction will come from second order in $\epsilon$, 
while for shift-symmetric Gauss-Bonnet gravity with coupling 
$\beta(\varphi) = \varphi$, at the third order in $\epsilon$
\cite{Corman:2022xqg}.
Given the smallness of $\epsilon$,
we expect any kind of deviation from GR 
in the scalar QNM spectrum in Gauss-Bonnet gravity to be very small.

\begin{table*}[hbt!]
\setlength{\extrarowheight}{2pt}
\begin{tabular}{ |c | c| c | c| c | c|  }
\hline
   Run & $(lm)$ mode 
   &  \multicolumn{2}{|c|}{$N=2$} & \multicolumn{1}{|c|}{$N=1$, 
   $\omega_{lm}^{\rm{Kerr,grav}}$ fixed} & \multicolumn{1}{|c|}{$N=1$, 
   $\omega_{lm}^{\rm{Kerr,scalar}}$ fixed}\\
\cline{3-6}                                            
 & & $\log_{10}A_1$   &    $\log_{10}A_2$   &  $\log_{10}A_1$    & $\log_{10}A_1$    \\
 \hline
\texttt{BBH-11} & 22 & $-0.81_{-0.02}^{+0.26}$ & $-0.67_{-0.09}^{+0.01}$ & \ldots &  \ldots \\
						 & 44 & $-1.62_{-0.48}^{+0.04}$& $-2.41_{-0.39}^{+0.77}$ & \ldots &  $-1.58_{-0.34}^{+0.01}$\\
\hline   
\texttt{BBH-23} \quad & 11 & $-0.50_{-0.02}^{+0.01}$ & $0.07_{-0.04}^{+0.08}$ & \ldots & \ldots \\ 
  			 \quad & 22 & $-0.89_{-0.03}^{+0.03}$ & $-0.61_{-0.02}^{+0.01}$ &  \ldots & \ldots \\
  			 \quad & 33 & $-1.34_{-0.09}^{+0.03}$ & $-1.39_{-0.07}^{+0.03}$ & \ldots & \ldots \\
  			 \quad & 44 & $-1.76_{-0.16}^{+0.04}$ & $-2.59_{-0.28}^{+0.46}$ & \ldots & $-1.72_{-0.58}^{+0.01}$ \\        
\hline
\texttt{BBH-12} \quad & 11 & $-0.10_{-0.01}^{+0.01}$ & $0.69_{-0.12}^{+0.18}$ & \ldots & \ldots \\ 
  			 \quad & 22 & $-1.23_{-0.05}^{+0.05}$ & $-0.47_{-0.01}^{+0.01}$ & $-0.48_{-0.01}^{+0.01}$ & \ldots\\
  			 \quad & 33 & $-1.30_{-0.05}^{+0.17}$ & $-1.29_{-0.07}^{+0.04}$ & $-1.09_{-0.13}^{+0.01}$ & $-1.03_{-0.01}^{+0.01}$ \\
			\quad & 44 & $-1.90_{-0.12}^{+0.46}$ & $-1.40_{-0.11}^{+0.03}$ & $-1.40_{-0.32}^{+0.01}$ & \ldots\\ 
\hline
\end{tabular}\\
\caption{Estimated amplitudes from the $N=1$ mode and $N=2$ mode fits summarized by Eqs.~\eqref{eq:fit_summary_1} and \eqref{eq:fit_summary_2}. As in Fig.~\ref{scalar_waveforms}, we set $\beta_0/m_1^2=1$ for convenience. 
    The dots indicate that a specific fit could not describe the data accurately. In all cases with an $N=2$ mode fit, $A_1$ represents the amplitude of the scalar Kerr QNM. For $(11)$ mode, $A_2$ is the amplitude of the scalar Kerr overtone, while for all other modes, $A_2$ represents the amplitude of gravitational Kerr QNM. The upper and lower limits on the estimated parameters lie within 90\% confidence interval and the central value is the median.}
\label{tab:fits_results}
\end{table*}

\begin{table}[hbt!]
\begin{tabular} {|c | c | c|} 
 \hline
Run & $(lm)$ mode & $\log_{10}A_1$\\ [1.5ex]
 \hline
\texttt{BBH-11} \quad & 22 & $-0.81_{-0.01}^{+0.01}$\\ 
              \quad & 44 & $-1.62_{-0.01}^{+0.01}$\\ 
              \quad & 66 & $-2.36_{-0.01}^{+0.01}$\\ 
\hline
\texttt{BBH-23} \quad & 22 & $-1.04_{-0.01}^{+0.01}$ \\
  			 \quad & 33 & $-1.58_{-0.003}^{+0.003}$ \\
  			 \quad & 44 & $-1.75_{-0.01}^{+0.01}$ \\      
  			 \quad & 55 &  $-1.93_{-0.25}^{+0.01}$\\  
  			 \quad & 66 & $-2.51_{-0.02}^{+0.01}$ \\
\hline
\texttt{BBH-12} \quad & 22 & $-1.03_{-0.01}^{+0.01}$ \\
  			 \quad & 33 & $-1.34_{-0.01}^{+0.01}$ \\
  			 \quad & 44 &  $-1.74_{-0.001}^{+0.002}$\\      
  			 \quad & 55 &  $-1.91_{-0.01}^{+0.01}$\\  
  			 \quad & 66 & $-2.29_{-0.03}^{+0.02}$ \\
\hline
\end{tabular}
\caption{Estimated gravitational amplitudes of the $N=1$ mode fit of Eq.~\eqref{eq:damped_sin_grav}, where the fundamental gravitational QNMs are kept fixed. The upper and lower limits on the estimated parameters lie within 90\% confidence interval.}
\label{tab:fits_results_grav}
\end{table}

%==============================================================================

\section{Observational Prospects}
\label{sec:obs}

\subsection{Measurability of the scalar ringdown}
\label{sec:measurability}

Having formulated a model for the scalar polarization component of the EsGB
ringdown waveform, we now make a first attempt to assess the measurability
of its presence in compact binary coalescence observations by the current 
and future networks of GW detectors. 
Returning to Eq.~\eqref{eq:wave_zone_expansion}, 
we recall that the second term predicts an additional scalar contribution, 
whose strength is controlled by the magnitude of the conformal coupling 
$a_0$ defined in Sec.~\ref{sec:polarization}
and the amplitude of the scalar field $\varphi$, 
which may be rescaled using the dimensionless parameter 
$\beta_0/m_1^2$ according to Eq.~\eqref{eq:dimfulscalar}. Therefore, the choice of the rescaling applied to the amplitude of the scalar field will affect the strength of the scalar polarization, and our choices of the re-scaling will be detailed in the following sections. 
However, before making predictions on the strength of the scalar signal, it is necessary to make the conversion to physical units.
The scalar polarization now takes the form of  
\begin{equation} 
   \label{eq:scalar_strain}
   h_{\rm{S}} 
   = 
   \frac{M_{\rm{fin}}}{D_L} a_0 \varphi
   ,
\end{equation}
where $M_{\rm{fin}}$ is the final BH mass and $D_L$ is its luminosity distance from 
the Earth. 
We note that $\varphi$ is the sum of all $(lm)$ modes and is dependent 
on the inclination angle $\iota$, 
which is defined as the angle between orbital angular momentum 
and the line-of-sight
\begin{equation}
   \varphi 
   = 
   \sum_{lm} \varphi^{lm} Y_{lm}(\iota)
   .
\end{equation}
In our analysis we set $\iota = 60^{\circ}$.
As we discussed in Sec.~\ref{sec:polarization},
each polarization has a distinct geometrical imprint on a GW detector. 
We therefore begin with projecting the scalar strain, 
$h_{\rm{S}}$, onto the detector $D$ located at $\mathbf{x}_D$
\begin{equation}
   h_{\rm{D}}(t) 
   = 
   h_{\rm{S}} (t, \mathbf{x}_{\rm{D}}) F^{\rm{S}}
   .
\end{equation}
Here $F^{\rm{S}}$ is the response, or antenna pattern, 
of a detector $D$ to scalar polarization. 
It can be given in terms of the polar angle $\theta$ and the azimuthal angle $\phi$ in the detector frame
(\emph{i.e.}\ $\theta$ and $\phi$  measured with respect to detector arms
along the $x$ and $y$ axes) in the following form
\begin{equation}
   F^{\rm{S}} (\theta, \phi)
   = 
   - \frac{1}{2} \sin^2\theta \cos 2\phi
   .
\end{equation}
\begin{figure}[ht!]
\includegraphics[width=\linewidth]{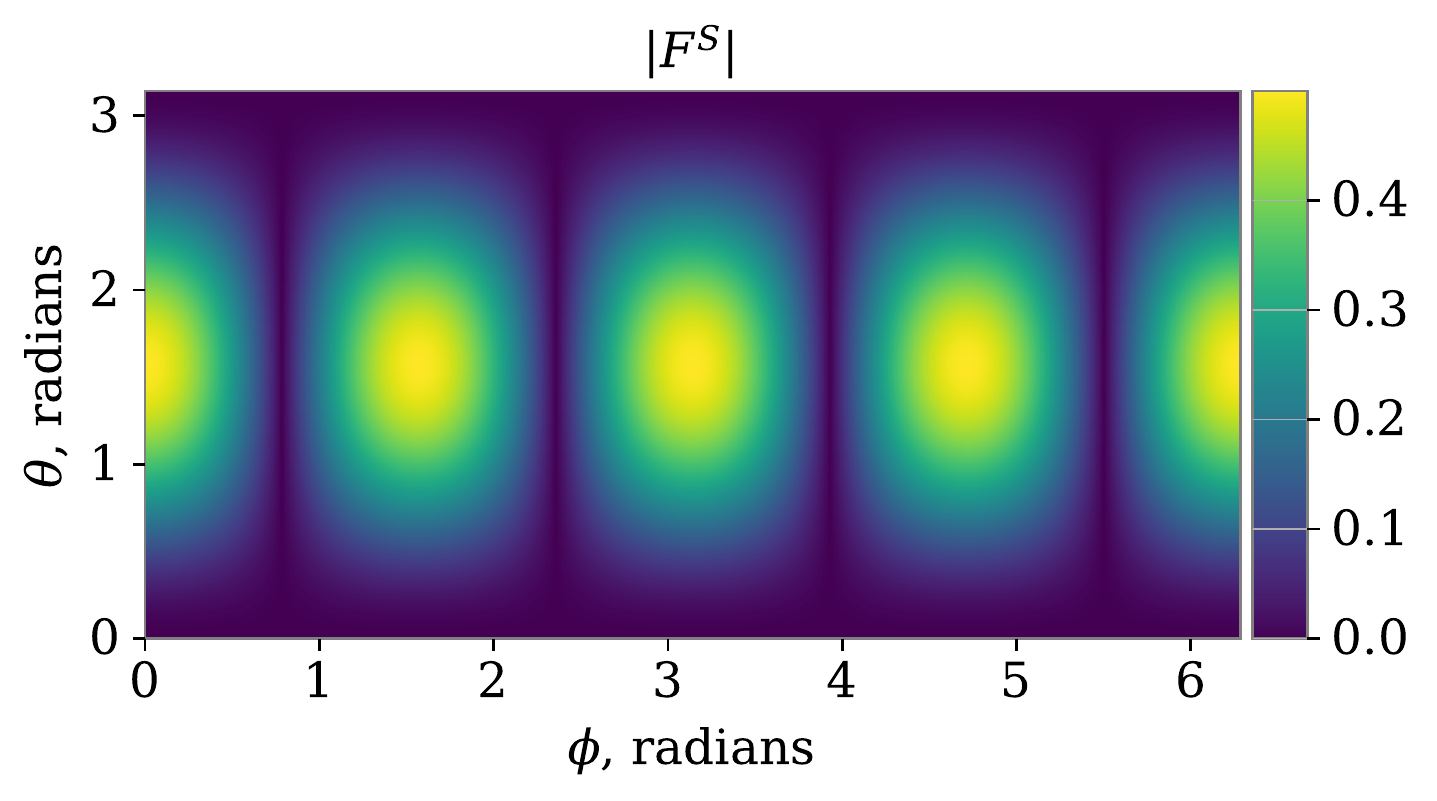}\hfill
\caption{Scalar antenna pattern for various choices of 
   polar and azimuthal angles. 
   Yellow regions indicate the maximum response from the detector 
   from the incoming scalar-polarized GW.}
\label{scalar_response}
\end{figure}

Note the absence of dependence on the polarization angle $\psi$ that defines the orientation of the source projected on the celestial sphere; this is a direct consequence of the spin-0 nature of the scalar field, in contrast with the tensorial GWs that behave as a spin-2 field.
The response to scalar polarization depends on the polar and azimuthal angles at which the GW comes in with respect to the detector frame.
Figure \ref{scalar_response} illustrates the change of the antenna pattern for various angles.
We position the source at an optimal sky location for one of the detectors (LIGO Hanford in particular), which is straight down one of its arms, \emph{i.e.}\
$\theta=\frac{\pi}{2}$, $\phi = 0$.
We then compute the squared SNR of the scalar signal defined via
\begin{equation} \label{eq:snr_def}
   \rho^2 
   = 
   \int_{0}^{\infty} {\frac{4 |\tilde{h}_{\rm{D}}(f)|^2}{S_n(f)} df }
   ,
\end{equation}
where $\tilde{h}_{\rm{D}}$ is the signal in the Fourier domain and $S_n(f)$ 
is the noise power spectral density (PSD) of the detector,
which we set to the estimated noise curve of i.\ the current network of LIGO (Hanford and Livingston) and Virgo detectors at design sensitivity~\cite{LIGOScientific:2014pky,KAGRA:2013rdx} and ii.\ the Einstein Telescope (ET)~\cite{Punturo_2010} in the ET-D configuration~\cite{Hild:2010id}.
We list luminosity distance and the range of masses chosen
for each of the detectors in Table~\ref{tab:SNR_param}.

\begin{table}[h]
\begin{tabular} {c c c }
 \hline
Detector & $D_L / \rm{Gpc}$ & $M_{\rm{fin}}/M_{\astrosun}$\\ [1.5ex] 
 \toprule
LIGO/Virgo \quad & 1 & [100, 1000] \\ 
ET \quad & 1 & [10, 1000] \\ 
 \hline
\end{tabular}
   \caption{Distance and final BH mass range targeted with each of
   the detectors considered in this study.}
\label{tab:SNR_param}
\end{table}

In the computation of SNR from Eq.~\eqref{eq:snr_def},
the signal is weighted by the noise PSD curve in the frequency domain. 
Having a frequency of the scalar waveform close to the minima of the PSD 
would therefore increase the integrand of SNR for a fixed amplitude 
of the waveform. 
However, for varying mass, the amplitude is not fixed but scales
according to Eq.~\eqref{eq:scalar_strain}.
Further, the amplitude of the scalar signal is affected by the mass ratio: as suggested by our NR simulations from Sec.~\ref{sec:numerical_results}, amplitude increases for more
asymmetric binaries.

We now move on to estimate the actual signal strength for two different scenarios with respect to the values of coupling parameters in EsGB\@.

\subsubsection{Maximally allowed couplings\label{sec:maximally_allowed_couplings}}
Our prior for the range allowed for the couplings $a_0$ and $\beta_0/m_1^2$, 
is controlled by the theoretical and observational constraints described 
in Sec.~\ref{sec:constraints}.
In this section we make the conservative choice of 
$\max(\beta_0/m_1^2) = \epsilon_{\rm{thr}} \approx 0.2$ and
$a_0^{\max} = \sqrt{10^{-3}} \approx 0.0316$ 
to be maximally allowed values for our couplings 
(\emph{i.e.}\ the blue region of Fig.~\ref{fig:validity}
is where the theory can be said to be no longer predictive) and investigate in more detail the dependence of the SNR of the scalar ringdown signal on the total mass $M_{\rm{fin}}$ and mass ratio $q$ of the progenitors. Following common practice~\cite{Abbott_2017, Iacovelli:2022bbs},
we assume SNR $\geq 8$ is required for a detection.

\begin{figure*}[hbt!]
\includegraphics[width=.5\linewidth]{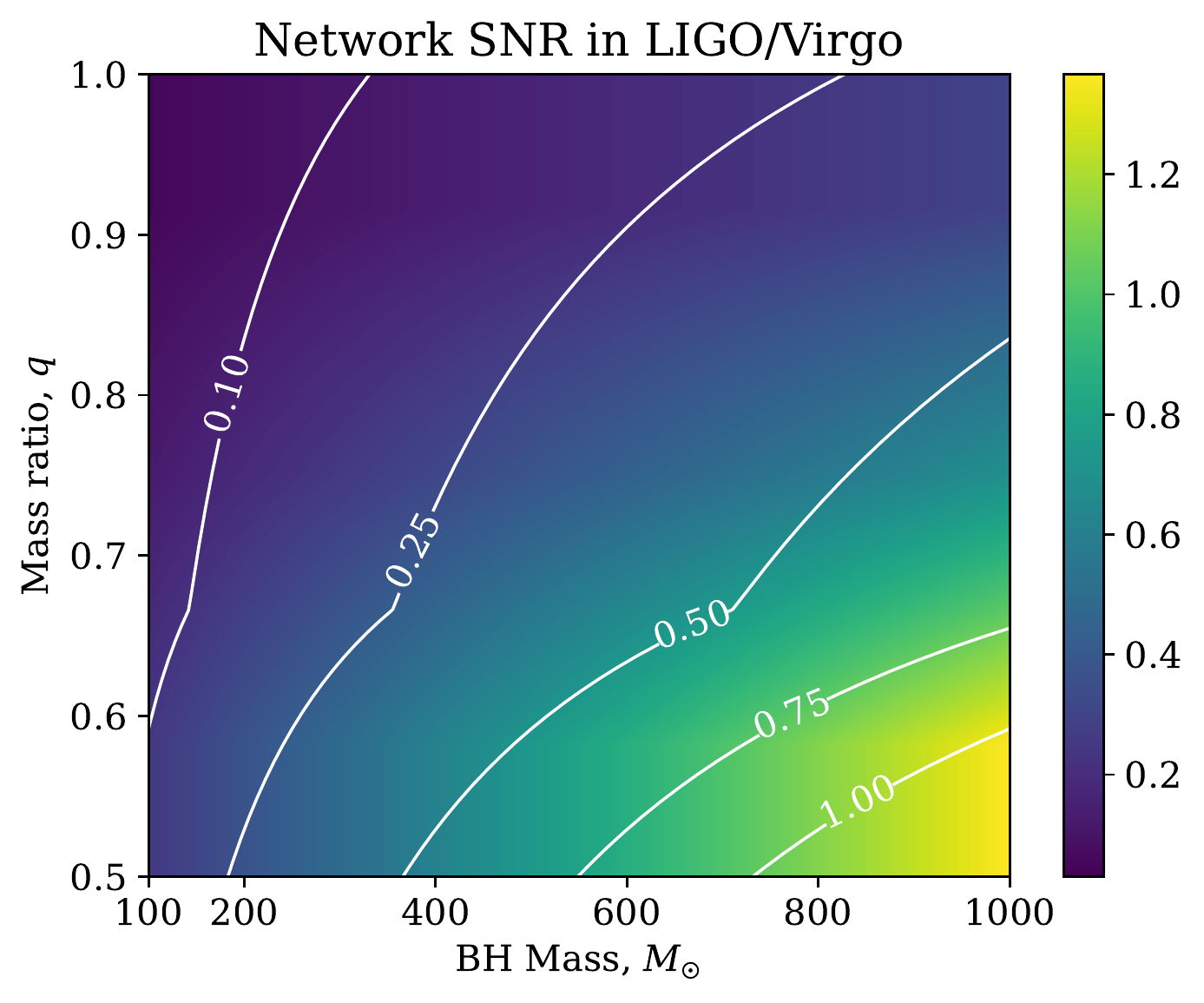}\hfill
\caption{Contour plot of network SNR in the
   $M_{\rm{fin}}\times q$ plane for the scalar ringdown of a BBH at 1 Gpc as observed by the Virgo, Livingston and Hanford network of detectors at design sensitivity.
   Here the couplings $a_0, \beta_0 /m_1^2$ are fixed to their
    maximally allowed values as described in Sec.~\ref{sec:maximally_allowed_couplings}. 
    NR data are available along horizontal lines of $q = \{1/2,2/3,1\}$, 
    based on which SNR values are estimated by interpolation on the plane.}
\label{snr_network}
\end{figure*}

Figure~\ref{snr_network} shows the network SNR for the current LIGO  (Livingston and Hanford) and Virgo detector network, calculated via quadrature,
% \begin{equation}
$\rho_{\rm{network}} = \sqrt{\sum_j \rho_j^2 }$,
% \end{equation}
where $\rho_j$ represents the SNR in a single detector. Here the values of numerically calculated SNR span three horizontal lines of constant $q=\{1/2,2/3,1\}$ and are interpolated
on the parameter space of $M_{\rm{fin}} \times q$.
The typical scalar signal is weaker than its tensorial counterpart by at least 3 orders of magnitude, and the choice of maximally allowed couplings results in increased SNR for larger masses.
The largest network SNR of $1.8$ is produced from the smallest mass ratio of $q=1/2$ and the final BH mass of $M_{\rm{fin}} = 1000 M_{\astrosun}$. We note, however, that our choice of 
$t_0 = 16M$ to be the ringdown start time reduces the 
strength of the signal. As seen in Sec.~\ref{sec:ringdown},
the amplitudes for highly asymmetric mass ratios become significantly damped at later times. Therefore, one may expect larger SNR, if an earlier ringdown start time was chosen.

\begin{figure*}[hbt!]
\includegraphics[width=.5\linewidth]{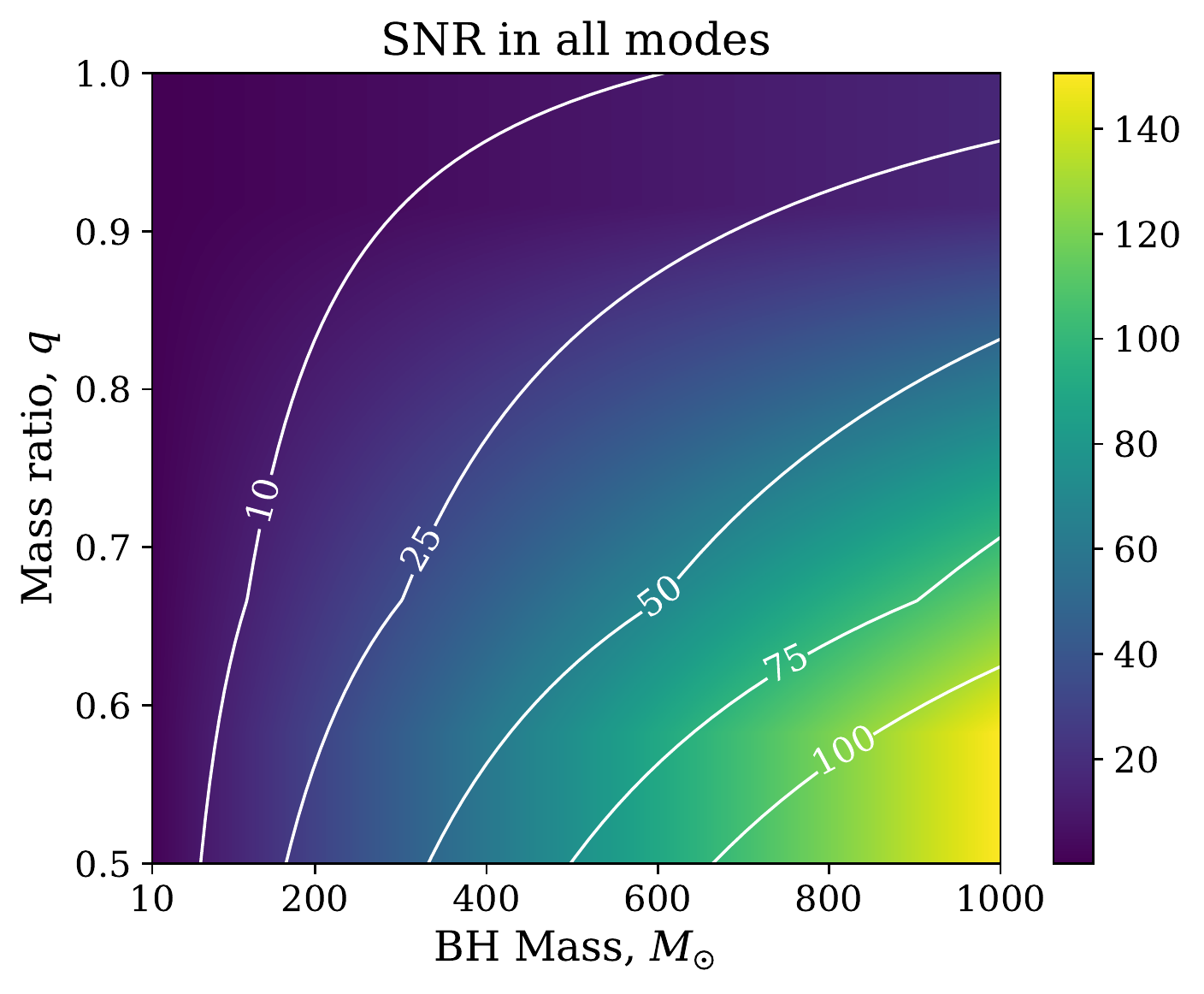}\hfill
\includegraphics[width=.5\linewidth]{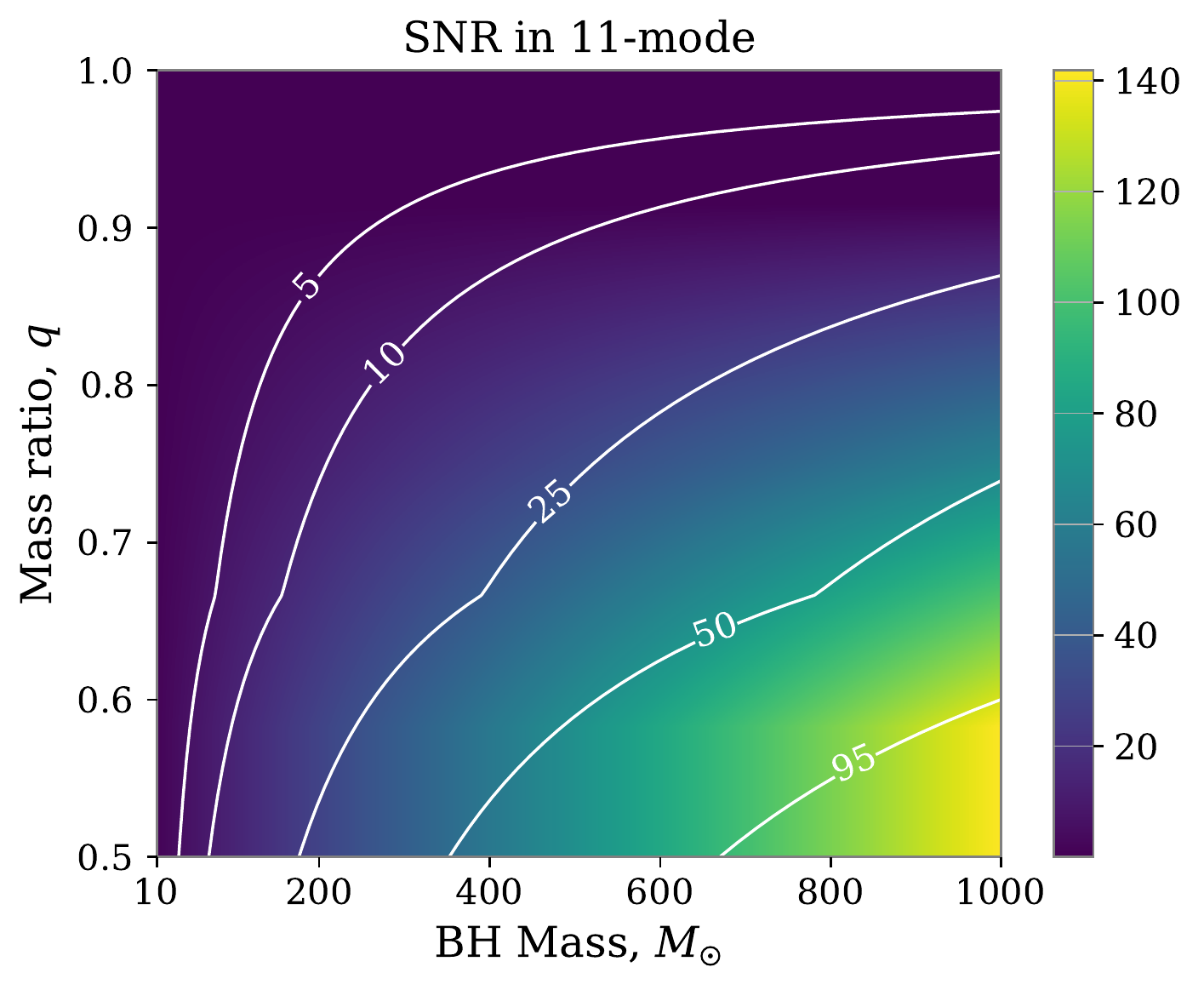}\par 
\includegraphics[width=.5\linewidth]{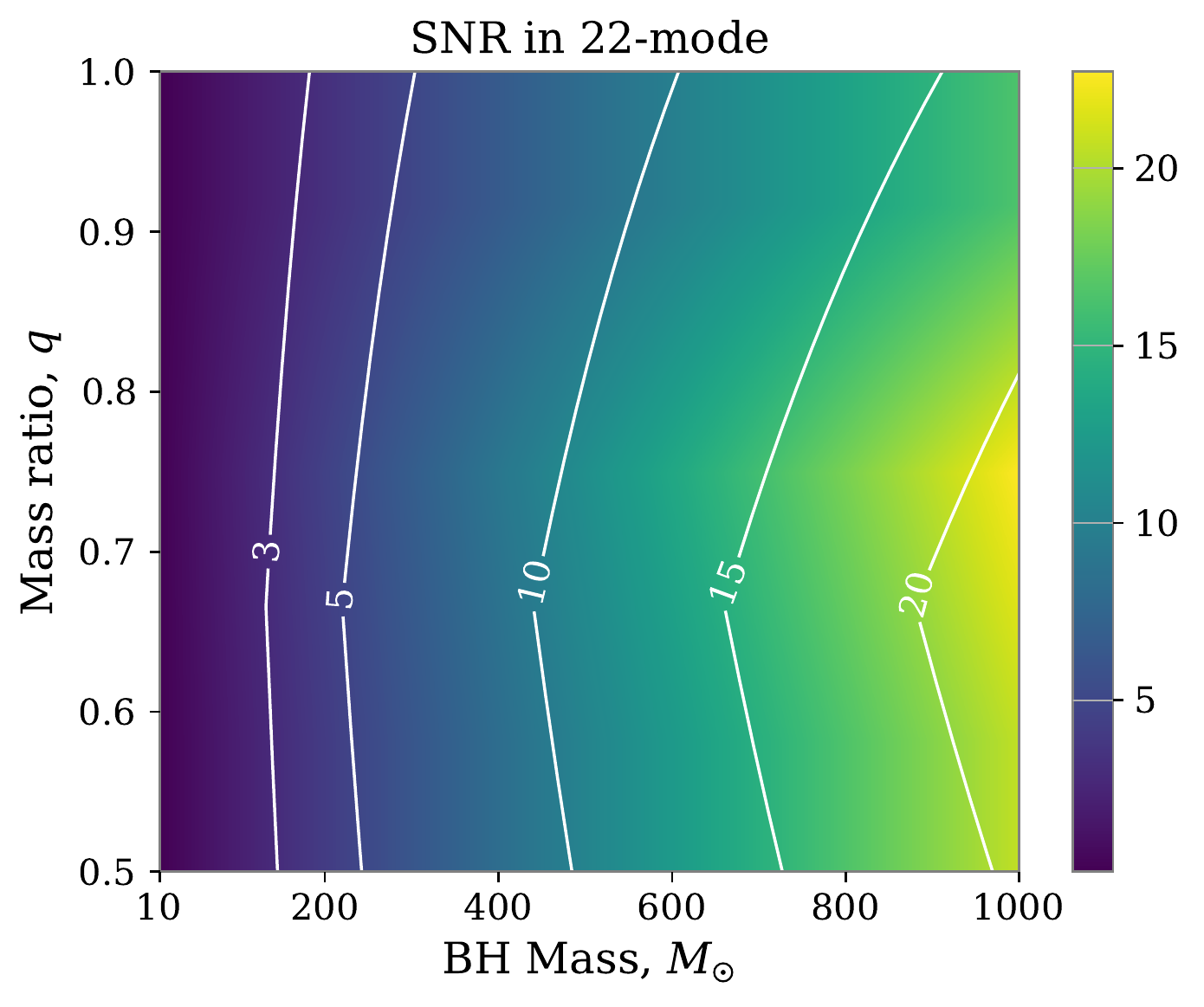}\hfill
\includegraphics[width=.5\linewidth]{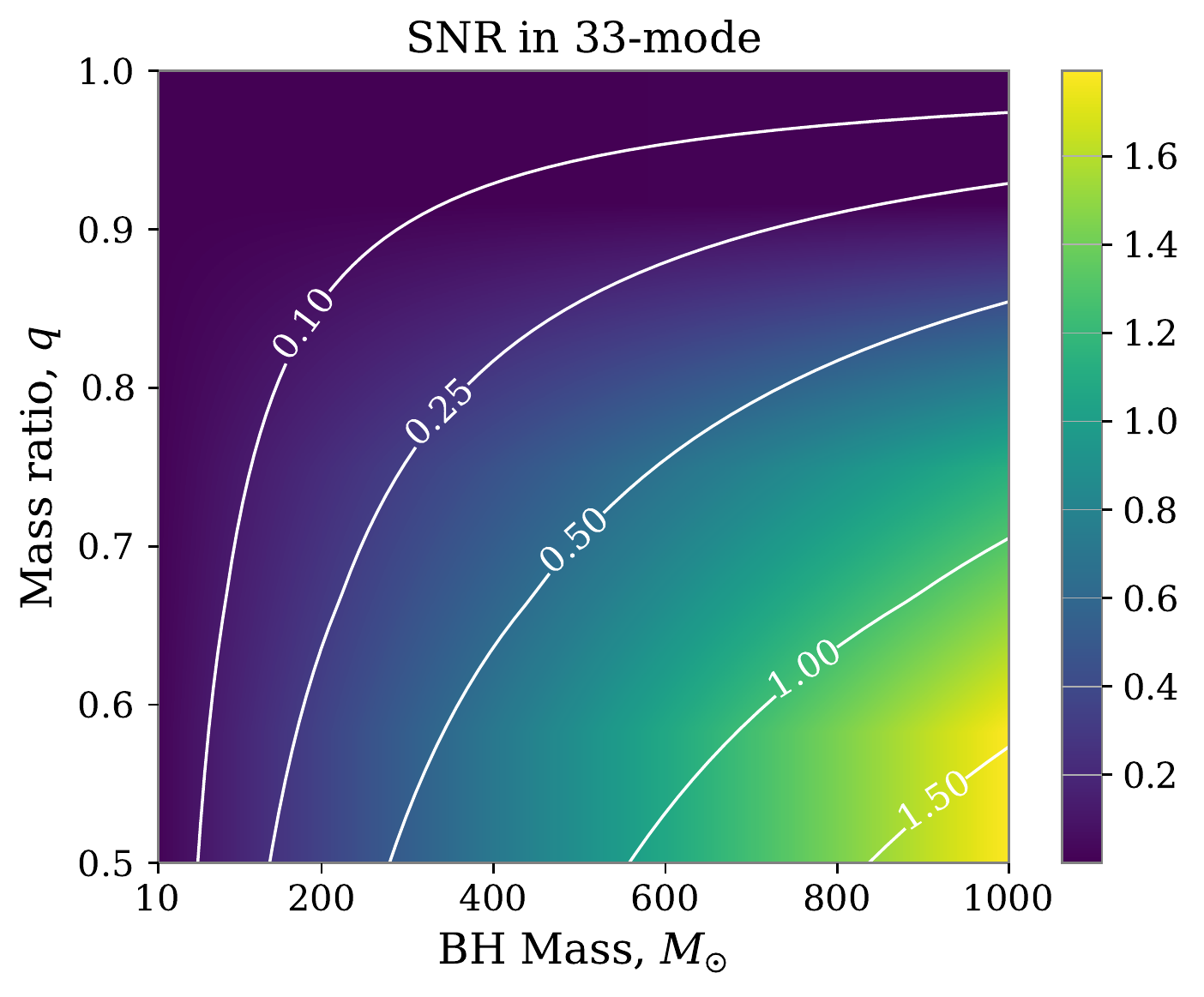}
\caption{Contour plot of network SNR in the
   $M_{\rm{fin}}\times q$ plane for the scalar ringdown of a BBH at 1 Gpc as observed by the Einstein Telescope.
    The couplings $a_0, \beta_0 /m_1^2$ are fixed to their maximally allowed values as described in 
    Sec.~\ref{sec:maximally_allowed_couplings}. 
    The top left panel shows the total SNR, while the individual scalar mode SNRs are 
    shown in the remaining three panels. 
    NR data are available along horizontal lines of $q = \{1/2,2/3,1\}$, 
    based on which SNR values are estimated by interpolation on the plane.}
\label{snr_q_vs_mass}
\end{figure*}

The strength of the signal significantly improves in the ET, which we illustrate in Fig.~\ref{snr_q_vs_mass}. Here we further disentangle contributions of each mode to the total SNR by plotting SNR for
the strongest individual modes, \emph{i.e.}\ $(lm) = (11), (22), (33)$.
We observe that some cross terms between modes may contribute destructively in the integrand of SNR (see Eq.~\eqref{eq:snr_def}) and as such, the total SNR is not simply a sum in quadrature of the individual modes' contributions. We find that the strongest total SNR calculated from all modes is 
recovered from the most extreme mass ratio considered in our simulations, 
$q=1/2$. 
This is as expected since its modes' amplitudes are significantly 
larger than for milder mass ratios. 
Assuming luminosity distance of $D_{L} = 1 \rm{Gpc}$ for \texttt{BBH-12} configuration, the total SNR of the scalar ringdown is weak for BHs with masses $M_{\rm{fin}} \lesssim 50M_{\astrosun}$ and falls beyond the detectability threshold. However, taking parameters of GW151226 event with 
$q \sim 1/2$, $D_{L} = 440 \, \rm{Mpc}$ and $M_{\rm{fin}} \sim 20.8 M_{\astrosun}$
\cite{PhysRevLett.116.241103}, 
we find the SNR in ET to be around $7.15$. Interestingly, the SNR of individual modes has a more complicated structure and there are several competing factors contributing to their strength.
First, the final BH mass determines where the QNM frequencies lie in relation to the high-sensitivity region of the detector's noise curve.
Second, the hierarchy and relative amplitudes of the scalar modes are determined by the mass ratio of the progenitor (\emph{e.g.}\ with odd-$m$ modes vanishing for symmetric binaries).
Third, higher modes are less long-lived for more extreme mass ratios, meaning that at later ringdown start times these modes become significantly damped.
Finally, the inclination angle contributes an additional $l$-dependent geometric factor to the relative mode amplitudes as seen by the observer.
As such, in our simulated analysis, the SNR of the $(22)$ mode is the strongest for $q=2/3$ and $(11)$ and $(33)$ modes are the strongest for the $q=1/2$ one.

\subsubsection{Imposing the most pessimistic constraint, $\sqrt{\beta_0} = 1.18\, \rm{km}$\label{sec:pessimistic_constraint_beta}}
So far we explored the strength of the signal in the optimistic scenario, where the strength of the EsGB dimensionless parameter is set at the limit of the theoretical bound and appears to be strong enough for certain binary configurations.
Constraints on the EsGB coupling $\beta_0$ coming from data analyses of astrophysical observations (see Table~\ref{tab:constraints_gb}) further limit the range of the coupling and therefore are expected to significantly lower the strength of the scalar signal.
We now follow the analysis of~\cite{Lyu_2022} on GW signals from the inspiral of BBH and BH-NS binaries and set the value of the EsGB coupling to their quoted 90\% combined upper bound, \emph{i.e.}\ $\sqrt{\beta_0} = 1.18 \, \rm{km}$.
This result is based on an analytically approximated inspiral model, where the leading-order correction to the waveform at $-1$PN (\emph{i.e.}\ $\delta\Psi \propto (\beta_0^2/M^4) v^{-7} $) is inferred and mapped to a bound on $\beta_0$.
Higher-order corrections to an incomplete 2PN level can vary the resulting bounds by more than 10\%, depending on the binary parameters, therefore this value should not be considered as a robust upper bound at 90\% confidence.
Nevertheless, we expect that the upcoming fourth observing run (O4) of the LVK collaboration~\cite{KAGRA:2013rdx} will probe this value range with much higher confidence.

\begin{figure*}[hbt!]
\includegraphics[width=.5\linewidth]{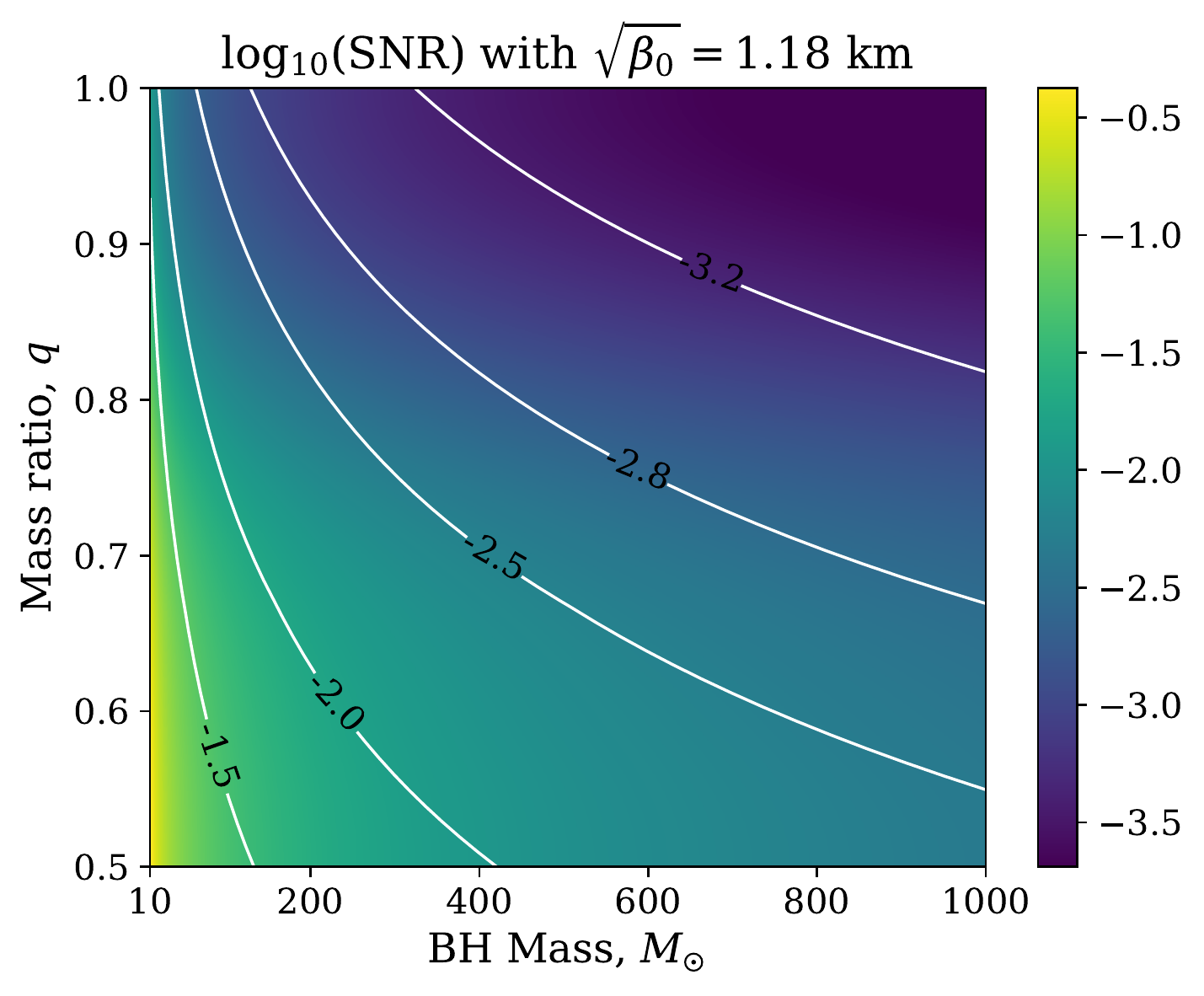}\hfill
\includegraphics[width=.5\linewidth]{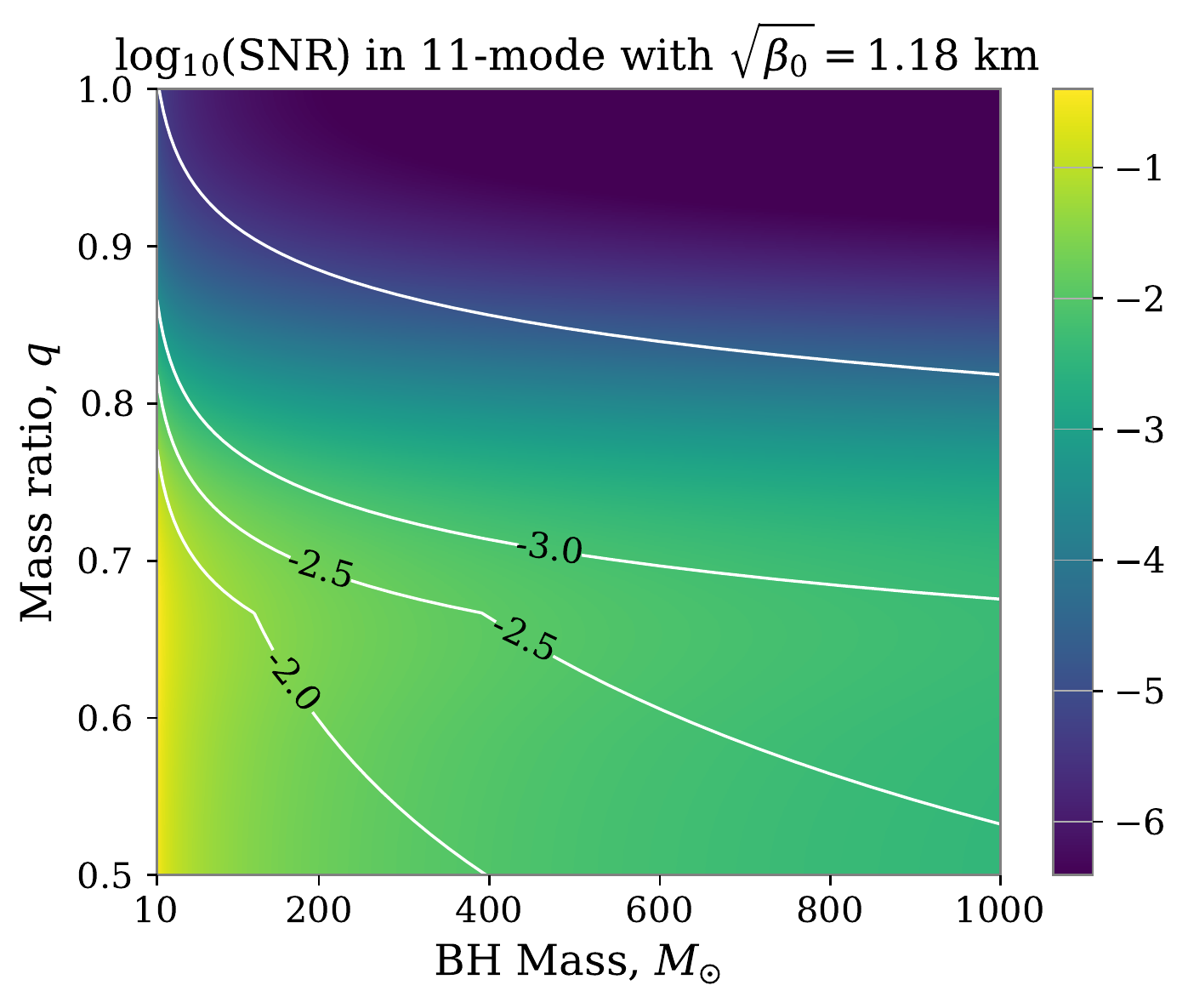}\par
\includegraphics[width=.5\linewidth]{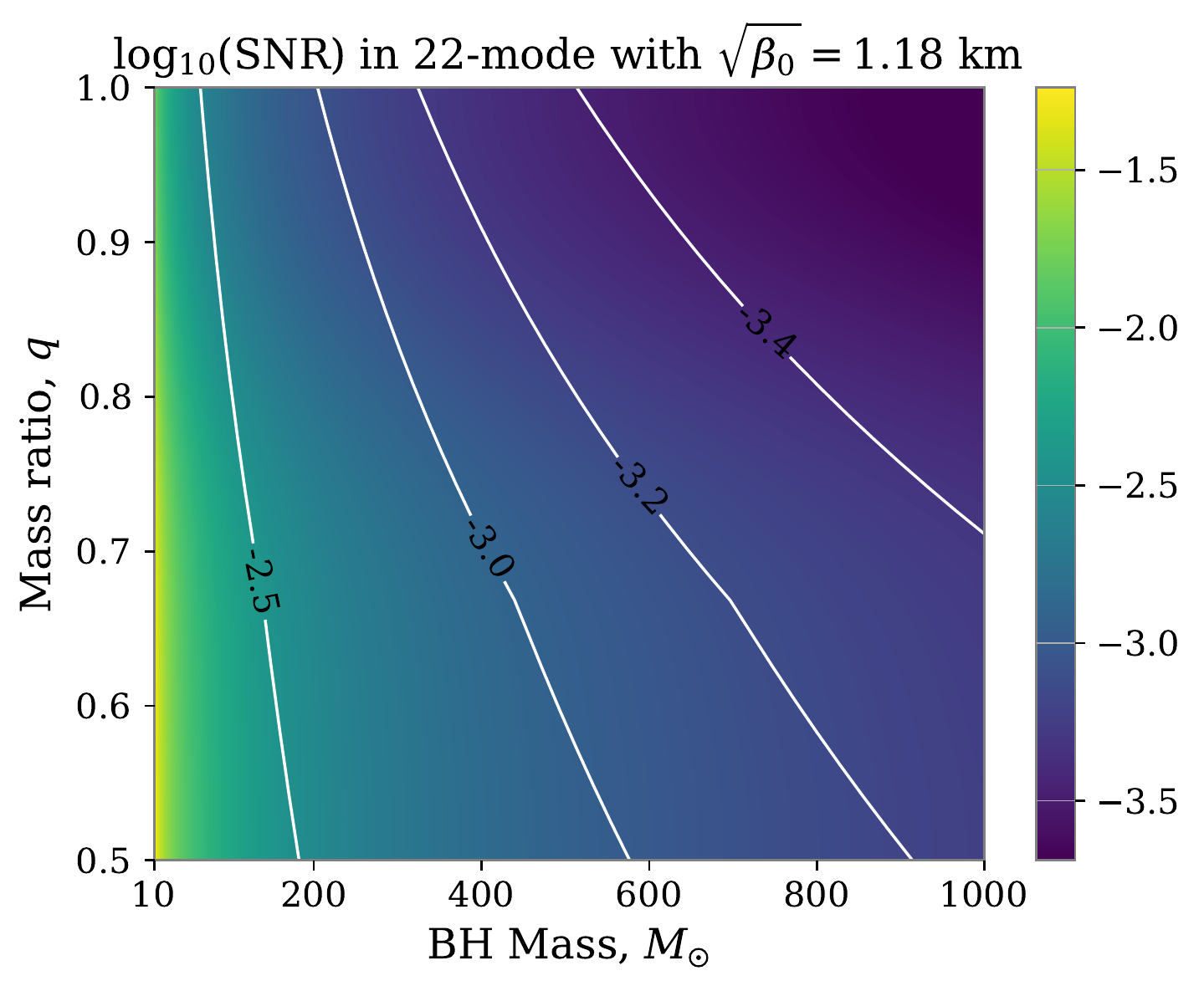}\hfill
\includegraphics[width=.5\linewidth]{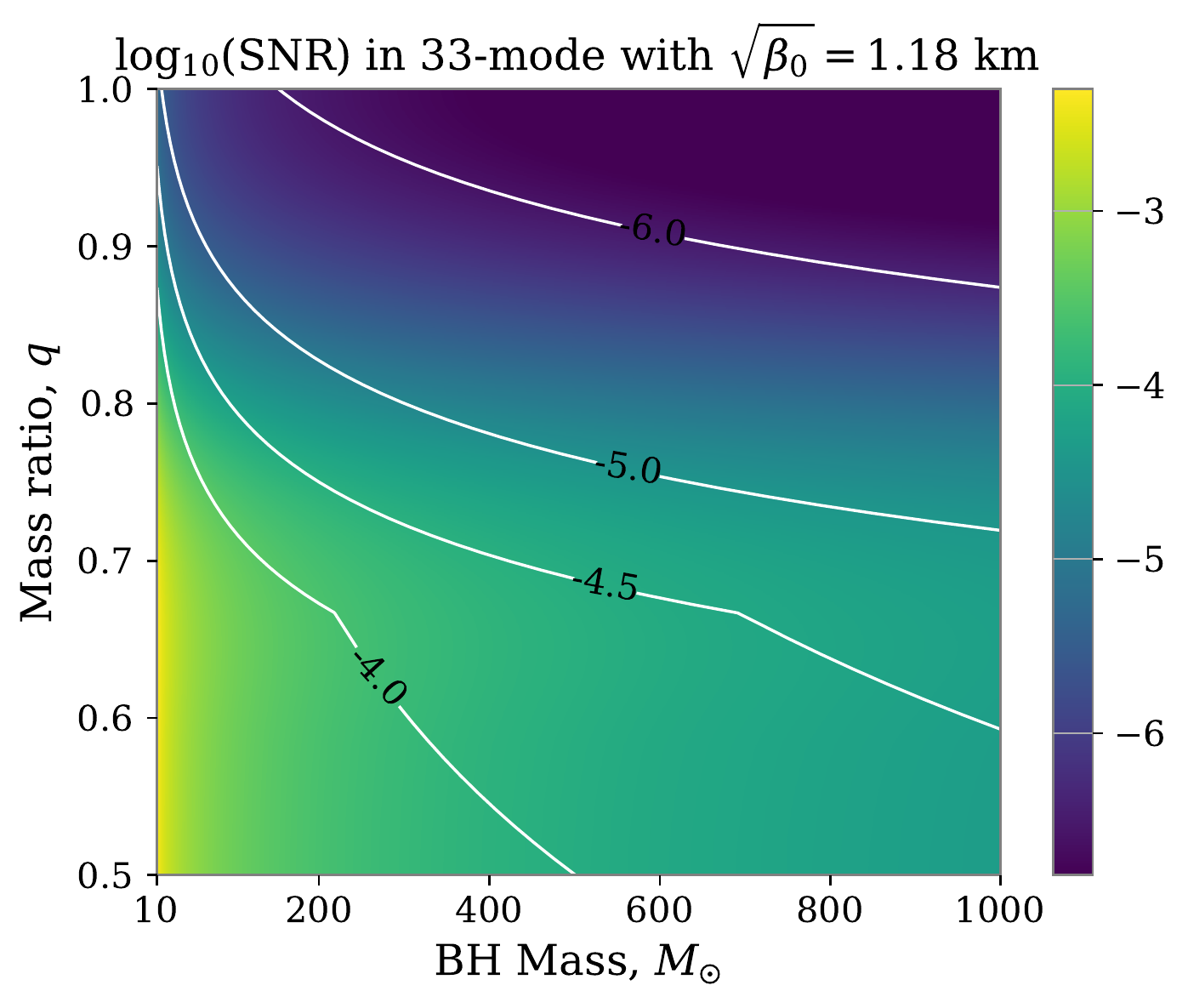}
\caption{Same as Fig.~\ref{snr_q_vs_mass}, except here the Gauss-Bonnet coupling $\sqrt{\beta_0} = 1.18 \, \rm{km}$ is fixed and the plot is given on a logarithmic scale. For more discussion see Sec.~\ref{sec:pessimistic_constraint_beta}.}
\label{snr_beta}
\end{figure*}

In this scenario the SNR of the scalar ringdown becomes significantly suppressed, which we demonstrate in Fig.~\ref{snr_beta} on a logarithmic scale.
Unlike the previous case, where we bounded $a_0$ and $\beta_0/m_1^2$, here, smaller values of the final BH mass and more asymmetric mass ratios yield a stronger scalar polarization signal, as expected.
In particular, the largest SNR of $0.42$ is observed for a binary of mass ratio $q=1/2$ with the smallest final mass of $M_{\rm{fin}} = 10 M_{\astrosun}$.
Unless the source is observed at very small luminosity distance of 
$D_{\rm{L}} \sim 52.5 \, \rm{Mpc}$,
the scalar ringdown signal of this binary configuration will not be detectable even with third generation detectors.
Our predictions are less pessimistic for lower-mass binaries at even higher mass ratios, however, in this work we do not consider BH-NS binaries or BHs with masses lower than $5 M_{\odot}$.

%==============================================================================
\section{Bayesian Inference on the Scalar Ringdown Signal}
\label{sec:injection}
We now perform full Bayesian inference on a hypothetical scalar signal 
present in the bands of the LVK detector network and ET\@.
We model the data stream in each detector as
\begin{equation} \label{eq:stream_data}
d(t) = h(t) + n(t),
\end{equation}
where $h(t)$ is the signal as described in Sec.~\ref{sec:ringdown} and $n(t)$ is a realization of detector noise.
Although we use continuum-appropriate notation, in practice the above are realized as discrete time series with a sampling rate of $4096$ Hz, which is sufficient for our purposes in this work.

\subsection{Analysis setup}
\label{sec:analysis-setup}

Before delving into the details of our analysis, let us first describe the simplifications performed here.
First, we assume that the GR part of the signal is well modeled and has been successfully reconstructed and removed from the data, without being particularly interested in the details of that part of the analysis.
In practice this means that we have reduced the problem of performing an apples-to-apples comparison between GR and EsGB to the problem of searching for the presence of our EsGB scalar-ringdown signal in the residual data.
This, in turn, makes use of the decoupling-limit assumption, that is, GW generation is driven by GR dynamics and therefore the tensorial ringdown signals in GR and EsGB are identical.
Furthermore, we factorize the analysis of multiple QNMs into individual analyses for each $(lm)$ pair, here performed using $N=1$ or $N=2$ modes for each pair that can capture the presence of a ``gravity-led'' mode (or even an unusually strong overtone).
This simplification assumes that the frequency content of \emph{e.g.}\ the $l=m=1$ and $l=m=2$ modes that are studied here can be separated into disjoint regions in the frequency domain, as was done in recent searches for subdominant modes~\cite{Capano:2021etf}.
Although this latter assumption may not always hold, prior information on the final BH parameters, and hence on the expected scalar spectrum, will always be available from the analysis of the much louder tensorial signal (``$+$'' and ``$\times$'' polarizations), thus allowing for the scalar analysis setup to be adapted accordingly.
Details related to the above considerations will not be discussed here, but are left to be explored in future work.

We thus define our two competing hypotheses as follows:
\begin{enumerate}
  \item $\mathcal{H}_0$ is our null hypothesis, corresponding to the data consisting of pure Gaussian noise, according to the PSD of each detector in our network.
  \item $\mathcal{H}_1$ is our scalar-ringdown signal hypothesis, corresponding to the presence of a scalar signal in addition to noise in our data, where the scalar model is described in Sec.~\ref{sec:ringdown} and is parametrized by the real QNM amplitudes and phases $(A_k^{lm}, p_k^{lm})$, while the complex mode frequencies are fixed to their expected values (see details in Eqs.~\eqref{eq:fit_summary_1} and \eqref{eq:fit_summary_2}).
\end{enumerate}

\begin{figure*}[hbt!]
\includegraphics[width=.5\linewidth]{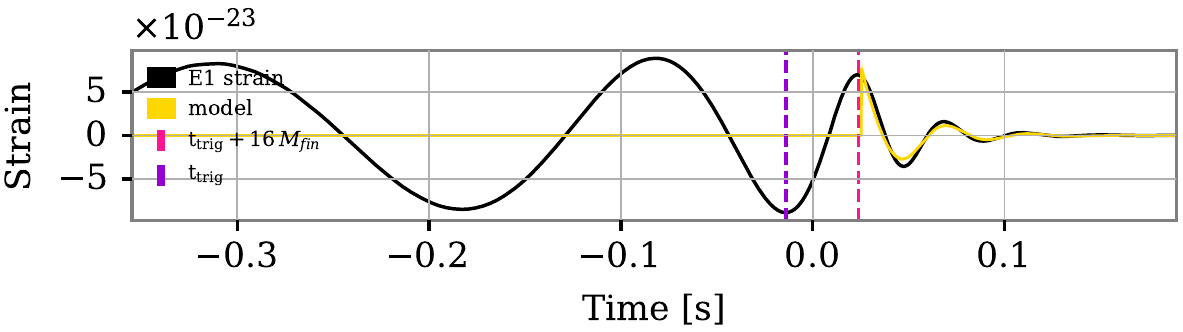}\hfill
\includegraphics[width=.5\linewidth]{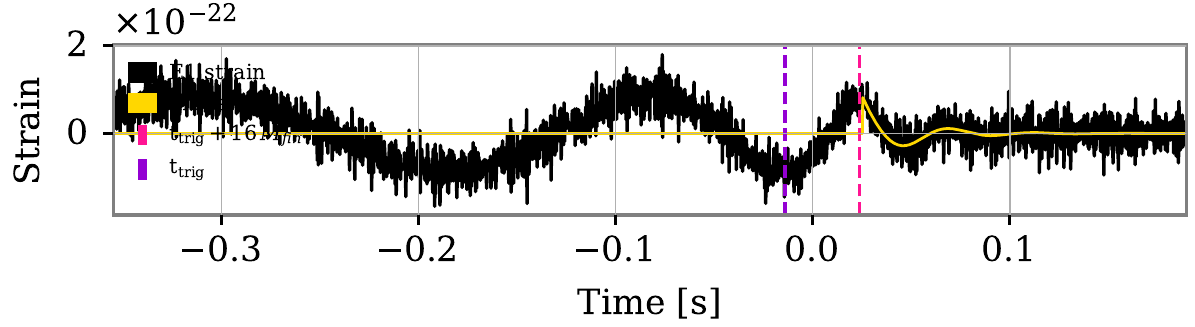}\hfill
\includegraphics[width=.5\linewidth]{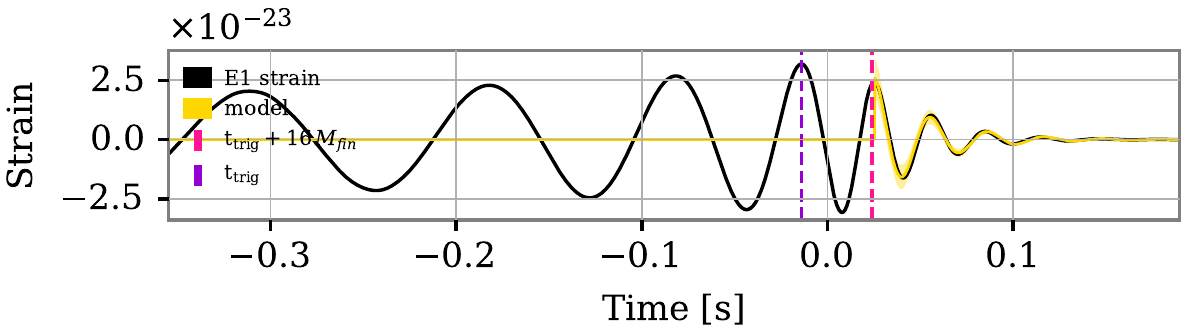}\hfill
\includegraphics[width=.5\linewidth]{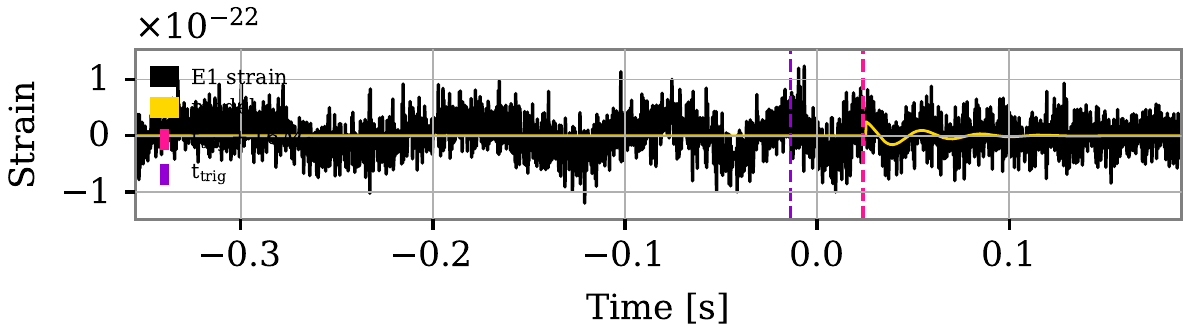}\hfill
\caption{Reconstructed waveforms for $(11)$ (top) and $(22)$ (bottom) modes for 
    the scalar field for an injection into the Einstein Telescope. 
    The left panel illustrates zero-noise injections and the right panel Gaussian-noise injections. The yellow line shows the recovered damped sinusoid model for each of the modes (see the main text for the details), the dashed purple and pink lines show the trigger time (defined as the reference time in the reference detector) and ringdown start time respectively. For more discussion see Sec.~\ref{sec:results}.}
\label{fig:recovery}
\end{figure*}

Our first task is then to infer whether our hypothetical scalar signal could be identified as being present in the data---according to~\eqref{eq:stream_data}, where $h(t) = h_S(t; \vec{\theta})$ is consistent with our scalar signal model---or the data is consistent with pure noise \emph{i.e.}\ $h(t)=0$.
For scalar signals that are detected with confidence against the noise hypothesis, our analysis will also estimate the parameters $\vec{\theta}$ of our scalar ringdown model as posterior probability distributions on the model parameter space $p(\vec{\theta} | d, \mathcal{H}_1)$.
To accomplish this, we inject our NR scalar waveforms $h_S(t)$ into data streams of simulated Gaussian noise and perform time-domain ringdown analysis using the \texttt{pyRing} 
pipeline~\cite{LIGOScientific:2020ibl,Carullo_2019,Isi_2019}, 
which relies on a Python implementation of the nested sampling 
algorithm~\cite{2004AIPC..735..395S} called \texttt{CPNest}~\cite{cpnest}.
The built-in calculation of the evidences $P(d|\mathcal{H}_{i})$ in nested sampling, 
by statistically integrating each likelihood function $p(d|\vec\theta, \mathcal{H}_{i})$ over the respective parameter space $\Sigma_{i}$, enables model selection between the signal and noise hypotheses, by means of the Bayes factor,
\begin{equation}
   \mathcal{B}_0^1 
   = 
   \frac{P \left(d| \mathcal{H}_1 \right)}{P \left(d|\mathcal{H}_0 \right)} 
   = 
   \frac{ 
      \int_{\Sigma_{1}} p(d | \vec{\theta_1}, \mathcal{H}_1) p(\vec{\theta_1} | \mathcal{H}_1) \text{d} \vec{\theta_1}
   }{
      \int_{\Sigma_{0}} p(d | \vec{\theta_0}, \mathcal{H}_0) p(\vec{\theta_0} | \mathcal{H}_0) \text{d} \vec{\theta_0}
   }
   .
\end{equation}
Since $\mathcal{H}_0$ here is the noise hypothesis, its parameter space is the empty set and the integral in the denominator reduces to a product of noise likelihoods $p(d|\mathcal{H}_0) \propto \exp\{-\frac{1}{2} \sum\limits_{i,j} d_i \, C^{-1}_{ij} d_j\}$, one for each detector, where $C_{ij}$ is the two-point autocovariance matrix, which characterizes the detector noise in the time domain as a discrete stochastic process that is assumed to be Gaussian and stationary~\cite{Carullo_2019}.
Similarly, the pointwise likelihood for $\mathcal{H}_1$ is computed using the same expression, where we replace $d \rightarrow d - h_S(t;\vec{\theta})$.
That is, the signal hypothesis states that the residual is pure noise.

We adopt two approaches in the injection of the scalar signal. 
First, we perform a zero-noise injection by setting $n(t)=0$ in~\eqref{eq:stream_data}, in order to isolate possible biases due to realization-specific effects of the noise
in our analysis and disentangle them from potential systematic biases inherent to our model.
Note, however, that since the likelihood is weighted by the noise PSDs of our GW detectors,
the overall effect of the noise on the uncertainty of our measurements is still incorporated in our method.
Second, we inject the scalar waveform into simulated noise and assess 
its effects on the parameter estimation. 
We have followed the standards of the LIGO Scientific Collaboration Algorithm Library (LAL)~\cite{https://doi.org/10.48550/arxiv.1703.01076} to perform NR scalar waveform injections, for which we have used a modified version of \texttt{LALsuite}~\cite{lalsuite} to accommodate the construction of the scalar waveform.

We present results for the injection of the \texttt{BBH-12} binary configuration, which gives us the strongest scalar wave signal.
We place the source along one of the arms of the detector, as described in Sec.~\ref{sec:measurability},
at a luminosity distance of $D_L = 100 \, \rm{Mpc}$,
and choose a final mass of $M_{\rm{fin}} = 500 M_{\astrosun}$.
We choose to inject the two loudest modes $(lm) = (11)$ and $(lm) = (22)$ and further take maximally allowed values for the theory couplings $\beta_0 / m_1^2$ and $a_0$ as described in Sec.~\ref{sec:measurability}, \emph{i.e.}\ $\max(\beta_0/m_1^2) = 0.2$ and $a_0^{\max} = \sqrt{10^{-3}} \approx 0.0316$.
The SNR value for the configuration involving the $(11)$ mode is roughly $700$ and for the $(22)$ is around $100$.
We use Table~\ref{tab:fits_results} of Sec.~\ref{sec:ringdown} to determine which recovery model of Eqs.~\eqref{eq:fit_summary_1} and \eqref{eq:fit_summary_2} to use for each of our injections.
As such, for the $(11)$ mode, we use the $N=2$ fit with frequencies fixed to the fundamental and first overtone scalar Kerr QNMs, while for the $(22)$ mode we use the $N=1$ mode fit with the frequency fixed to the fundamental gravitational Kerr QNM\@.

\subsection{Results}
\label{sec:results}

\begin{figure*}[hbt!]
\includegraphics[width=.55\linewidth, valign=b]{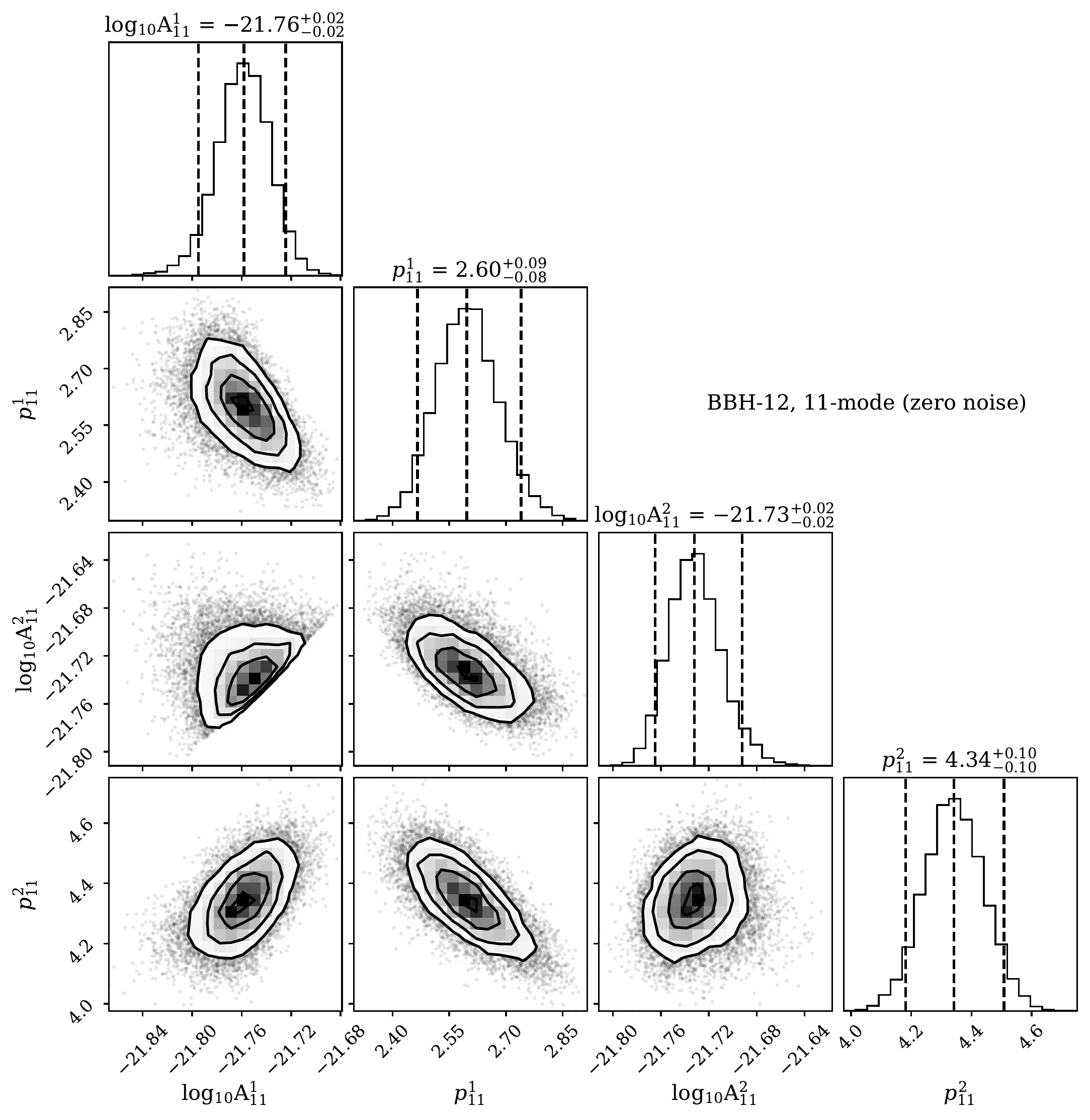}\hfill
\includegraphics[width=.45\linewidth, valign=b]{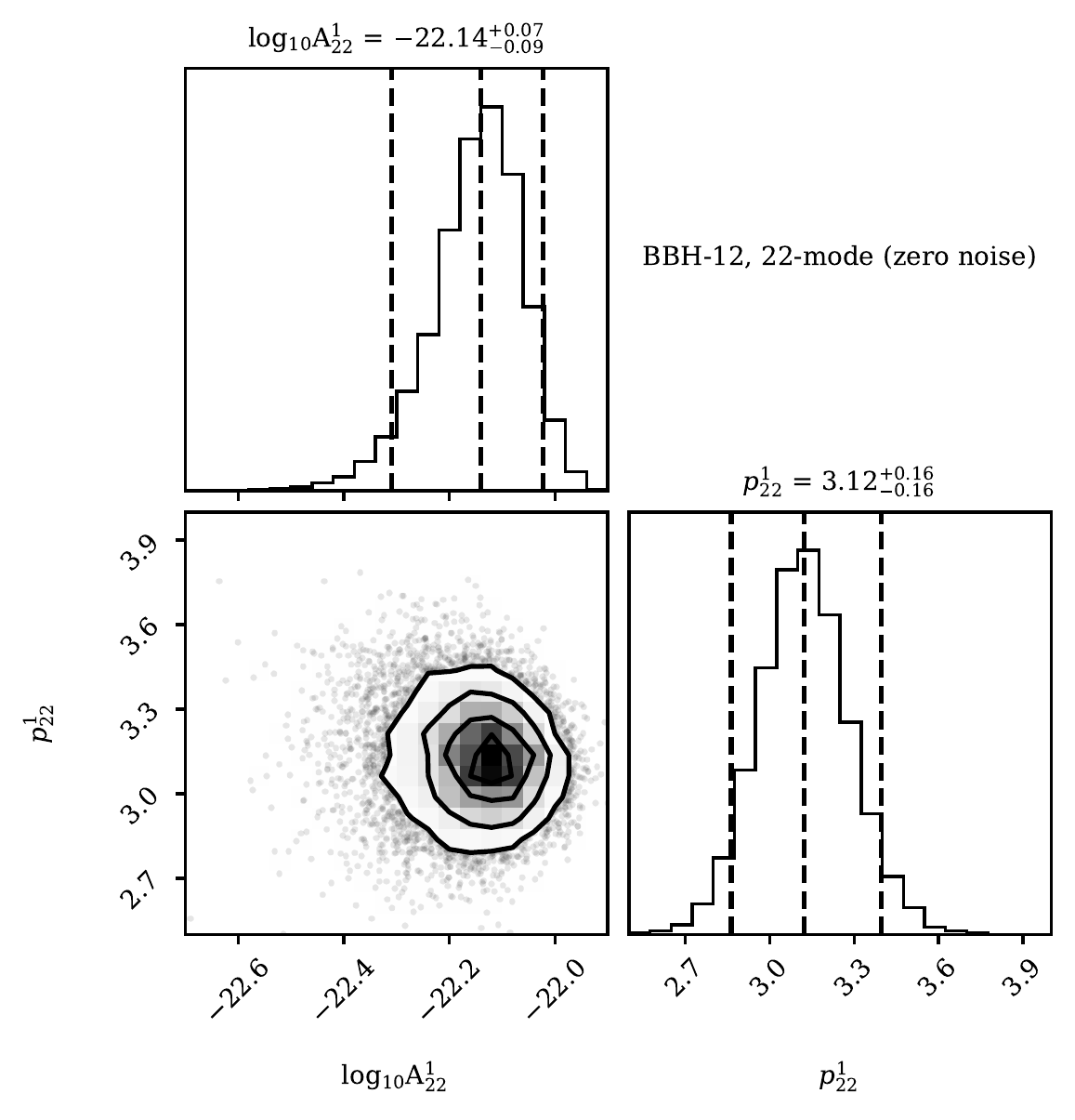} \\
\includegraphics[width=.55\linewidth, valign=b]{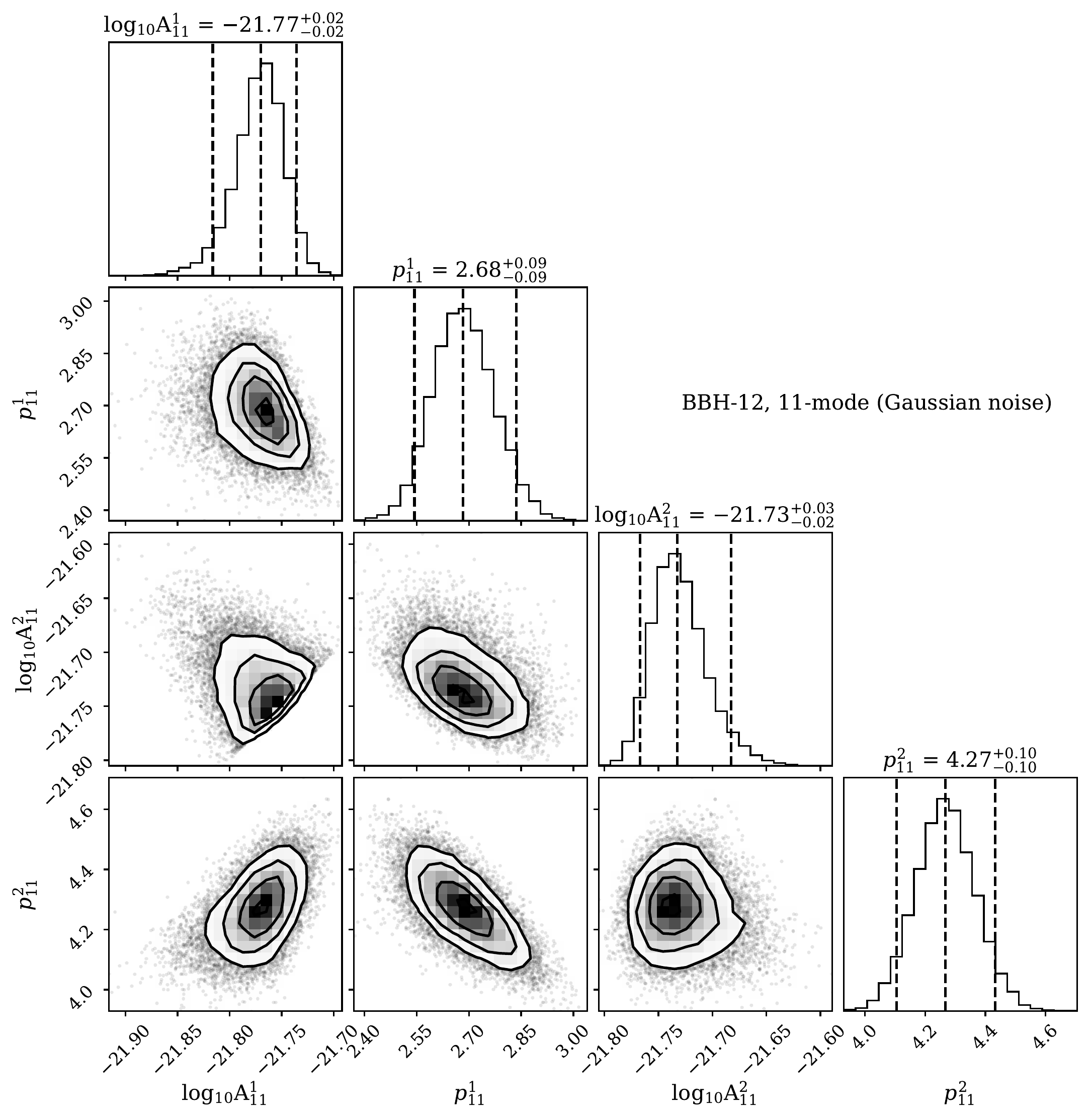}\hfill
\includegraphics[width=.45\linewidth, valign=b]{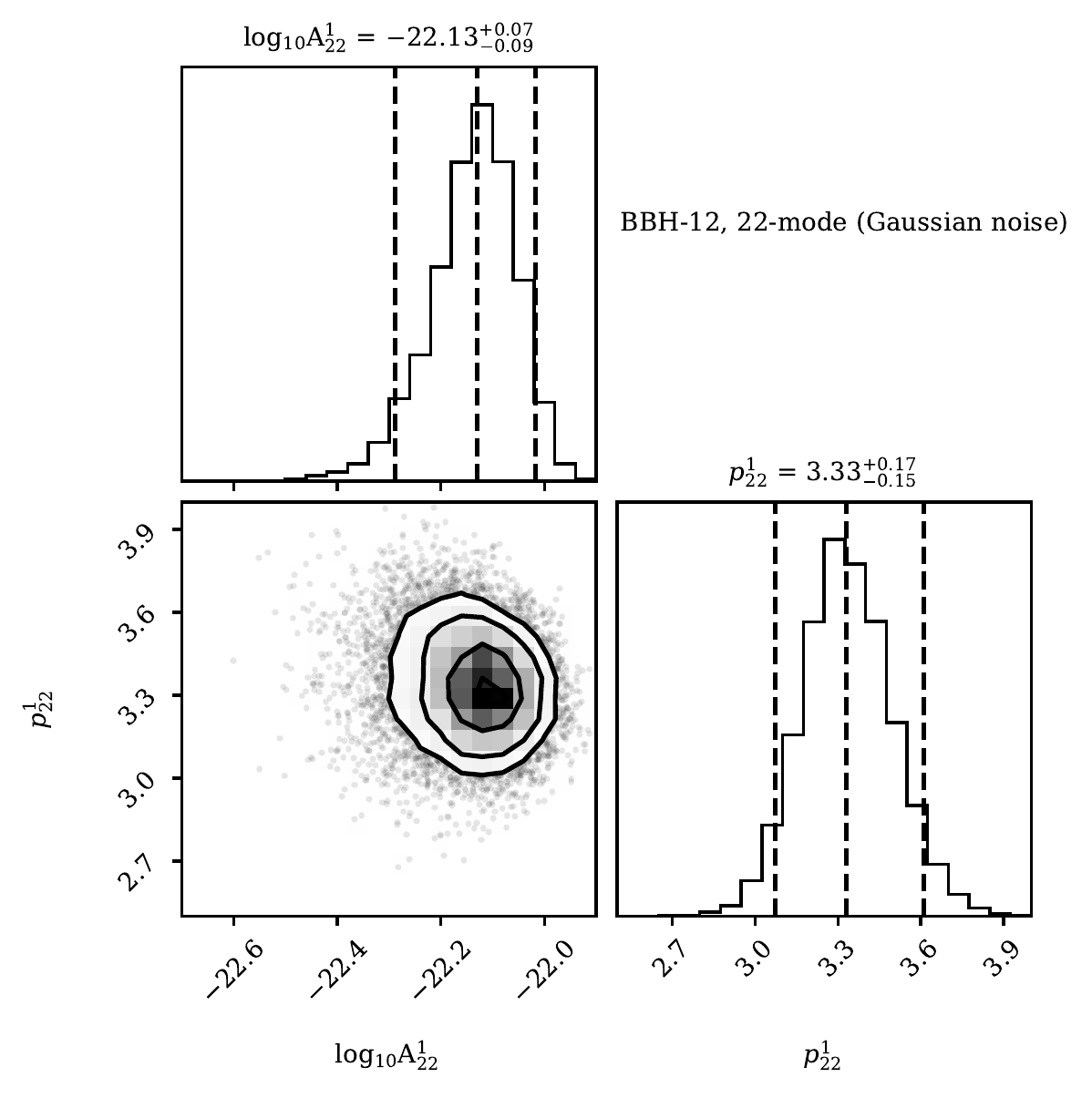}
\caption{Joint posterior distributions for the amplitudes and phases recovered for $(11)$ and $(22)$ modes. The upper panel shows the parameters for the zero-noise injections and the lower panel for the Gaussian noise injections. The upper and lower limits on the estimated parameters lie within 90\% confidence interval. Parameter estimation from both injections clearly demonstrates that noise does not bias the recovered amplitudes and phases significantly.
    For more discussion see Sec.~\ref{sec:results}.}
\label{fig:param_estimation}
\end{figure*}

Using the two hypotheses, $\mathcal{H}_1$ and $\mathcal{H}_2$, as defined in Sec.~\ref{sec:analysis-setup}, in Fig.~\ref{fig:recovery}, we present the results of the recovery of the $(11)$ and $(22)$ modes with zero-noise and colored Gaussian noise injections in ET\@.
The results suggest that the scalar-ringdown signal hypothesis $\mathcal{H}_1$ is favored for both injections, \emph{i.e.}\ the data contains a scalar-polarized GW present at the expected frequencies within the detector's sensitivity band.
The Bayes factors for the zero-noise and Gaussian noise injections are $\ln\mathcal{B}^1_0 \sim 16771$ and  $\ln\mathcal{B}^1_0 \sim 16127$, and
$\ln\mathcal{B}^1_0 \sim 47$ and  $\ln\mathcal{B}^1_0 \sim 15$  for the $(11)$ and $(22)$ modes, respectively.
Finally, in Fig.~\ref{fig:param_estimation},
we demonstrate the effect of Gaussian noise 
on the estimation of the modes' amplitudes and phases. 
Overall, with the source parameters of our choice we conclude
that Gaussian noise does not compromise our ability to measure scalar polarization, neither does it introduce significant biases in recovering the modes.

As we have seen in Sec.~\ref{sec:measurability}, for a fixed set of coupling parameters $a_0$ and $\beta_0 / m_1^2$, larger total masses or more extreme mass ratios are needed to achieve higher SNRs at a given distance.
From the observational point of view, the astrophysical population statistics for BBH systems allow for a rather faint probability of either placing stringent bounds on EsGB or making a direct detection, based on the scalar ringdown signal of a single BBH source, even with a 3G/XG detector.
Having multiple detectors in the network and combining information from multiple BBH events across a wide mass range will improve our overall sensitivity to the consistent presence of such signals.
As the rate of BBH events with SNR>100 in the next-generation network is expected to reach $\mathcal{O}(10^2 - 10^3)$ per year~\cite{Borhanian:2022czq}, joint information from multiple events, even if they may be weak,
will statistically improve the confidence of detecting a scalar signal as well as the precision of the inference by roughly a factor of $\sim \sqrt{n}$, $n$ being the number of detected scalar signals~\cite{PhysRevD.99.124044}.
While the presence of scalar polarization in the ringdown spectrum will be a smoking gun (unlike the inspiral dephasing), indicating the existence of a fundamental scalar field, more care will be needed in disentangling this weak effect from noise artifacts and other systematics.

%==============================================================================
\section{\label{sec:conclusion}Conclusion}

Through the presence of a conformal coupling to matter, black hole
binaries in EsGB gravity emit gravitational waves that have a scalar polarization.
The family of theories we consider in this work is controlled by two coupling parameters: 
the Gauss-Bonnet coupling $\beta_0$ and the conformal coupling $a_0$, 
which both impact the strength of the scalar-polarized gravitational wave.
We have studied the extent to which these couplings could be constrained
through gravitational wave observations of the scalar polarization during
binary black hole ringdown.
By exploring the parameter space in the range of validity of our couplings,
and calculating respective signal-to-noise ratios,
we conclude that the scalar polarization has a low chance of being detected.
There is a stronger possibility of detection with next generation detectors 
like the Einstein Telescope,
targeting larger masses or more asymmetric binary progenitors,
given the observed amplification in the scalar polarization amplitude 
for those systems from our NR simulations.
Furthermore, future observations with 
LISA~\cite{https://doi.org/10.48550/arxiv.1702.00786} may possibly allow
for more precise measurements of ringdown from extreme mass-ratio inspirals,
although the much higher mass range that LISA will probe may not be favorable 
for sourcing sufficiently strong scalar modes in theories like the ones studied here. 
The SNR is greatly dependent on the value of the dimensionless EsGB parameter,
$\beta_0/m_1^2$. Even at the largest allowed theoretical value for the dimensionless coupling 
($\beta_0/m_1^2\sim 0.2$), 
one would need to have total mass of roughly $\gtrsim 50M_{\astrosun}$ and a fairly asymmetric
mass ratio ($q\sim 1/2$) in order for there to be a detection with a future generation GW detector.
For smaller values of $\beta_0/m_1^2$, the strength of the signal drops beyond the detectability threshold
for any inspiral scenario we considered, even for the Einstein Telescope. 
Given the present observational constraint of $\sqrt{\beta_0}\lesssim1.18$km, and that $M_{\astrosun}\sim1.5$km in geometrized units, detecting a scalar polarization in a $50M_{\astrosun}$ event requires the smaller
mass black hole to have a mass of $\sim 2.5M_{\astrosun}$ in order for a competitive constraint
on $\beta_0$ to be possible from a ringdown measurement with the Einstein Telescope.

The most stringent constraint on most modified theories of gravity comes 
from the inspiral, through measurements of the phase of the gravitational
wave signal~\cite{Perkins:2021mhb,Lyu_2022}.
This being said, a measurement of the inspiral phase does not completely determine
the nature of a potential deviation from GR, as the dephasing could come
from the emission of energy in other energy channels beyond 
(for example) that of a gravitational scalar field.
Measuring the ringdown signal provides a complementary test to the measurement
of the inspiral phase. 

Apart from performing a targeted test on the GW data for signs of our EsGB model, the method that we propose here can be applied to a more general setting.
EsGB with a conformal scalar coupling
in its decoupling limit \cite{Benkel:2016kcq, Silva:2020omi} 
is a prime example of a well-motivated theory,
in which the spin-0 QNMs of the Kerr metric are excited and 
observed via radiation of scalar GWs.
Since backreaction can be neglected in the limit of a small coupling $\beta_0/m_1^2$, 
part of the scalar quasinormal mode spectrum is identical to that of the spectrum of
a (massless) scalar about a Kerr black hole~\cite{Berti:2009kk,Witek:2018dmd}.
More generally, any alternative theory featuring an additional 
scalar field that couples to the matter-field metric
will also radiate spin-0 QNM frequencies, provided that the theory respects the
``no-backreaction'' property (\emph{i.e.}\ to leading order, the GR sector determines the background
metric and is driving the dynamics).
In short, the general strategy amounts to using the information from the observed tensorial QNM frequencies to measure the remnant BH parameters, fixing the spin-0 QNM frequencies, and subsequently searching for the presence of any residual scalar GW signal on those frequencies, while being agnostic about the amplitude of each mode.
The mode amplitudes are sensitive to the details of a given theory, since the field equations will determine how strongly each mode is sourced from the initial perturbation.
Therefore, following the detection of a scalar signal, measurements of the relative mode amplitudes, combined with measurements on progenitor parameters, will narrow down the specifics of the theory and its couplings.
A similar approach could also cover theories featuring a gravitational vector field.
A thorough analysis of this strategy is beyond the scope of this article.

There are several more directions for future work.
We have only considered the impact of the scalar polarization on the
ringdown; it would be interesting to see if stronger constraints could
be placed on $\beta_0$ and $a_0$ from polarization measurements of the
inspiral waveform.
We have solved for the equations of motion of EsGB gravity in the
decoupling limit. While this approximation has been shown to be
accurate for modeling the scalar waveform in EsGB gravity,
at least during the inspiral phase of evolution (and ignoring
the effects of dephasing)~\cite{Witek:2018dmd},
it would be interesting to compare it to full solutions to EsGB 
gravity~\cite{East_2021,AresteSalo:2022hua,Corman:2022xqg}.
It will also be important to extend our study to spinning progenitors and see whether the scalar mode amplitudes' dependence on spins follows a functional form similar to that of the tensorial modes \cite{Kamaretsos:2012bs, London:2018gaq}.
Finally, while we have argued that EsGB gravity with a conformal scalar
coupling to the metric may give the largest scalar polarization
signal for black hole binaries among scalar-tensor gravity theories,
it would be interesting to numerically model other modified theories of gravity
featuring scalar or vector-polarized gravitational waves and devise similar targeted tests.
%==============================================================================
\acknowledgements

We thank Ulrich Sperhake and Daniela Doneva for useful conversations on aspects of this project, Miren Radia for the help with \texttt{GRChombo}, Danny Laghi and Gregorio Carullo for getting started with \texttt{pyRing} and Maxence Corman for reviewing our preprint.
We also thank the entire \texttt{GRChombo} collaboration for their support and code development work.

T.E. is supported by the Centre for Doctoral Training
(CDT) at the University of Cambridge funded through the STFC\@.
M.A. is supported by the Kavli Foundation.
J.L.R. acknowledges support
from STFC Research Grant No. ST/V005669/1, 
and from the Simons Foundation through Award No. 896696.

This research project was conducted using
computational resources at the Cambridge Service for Data Driven Discovery (CSD3) system at the University of Cambridge, provided by Dell EMC and Intel using Tier-2 funding from the Engineering and Physical Sciences Research Council (Capital Grant No. EP/T022159/1), and DiRAC funding from the Science and Technology Facilities Council (www.dirac.ac.uk), and Cosma7 of Durham University through Grants No. ST/P002293/1, No. ST/R002371/1 and No. ST/S002502/1, Durham University and STFC operations Grant No. ST/R000832/1. Some of the simulations presented in this article were also performed  on computational resources managed and supported by Princeton Research Computing,  a consortium of groups including the Princeton Institute for Computational Science and Engineering (PICSciE) and the  Office of Information Technology's High Performance Computing Center and 
Visualization Laboratory at Princeton University.
%==============================================================================
\appendix 
%==============================================================================

\appendix

%==============================================================================
\section{Convergence testing and error estimation}
%==============================================================================
\subsection{\label{sec:convergence}Gravitational and scalar waveforms}
\begin{figure*}[!htb]
  \includegraphics[width=0.8\linewidth]{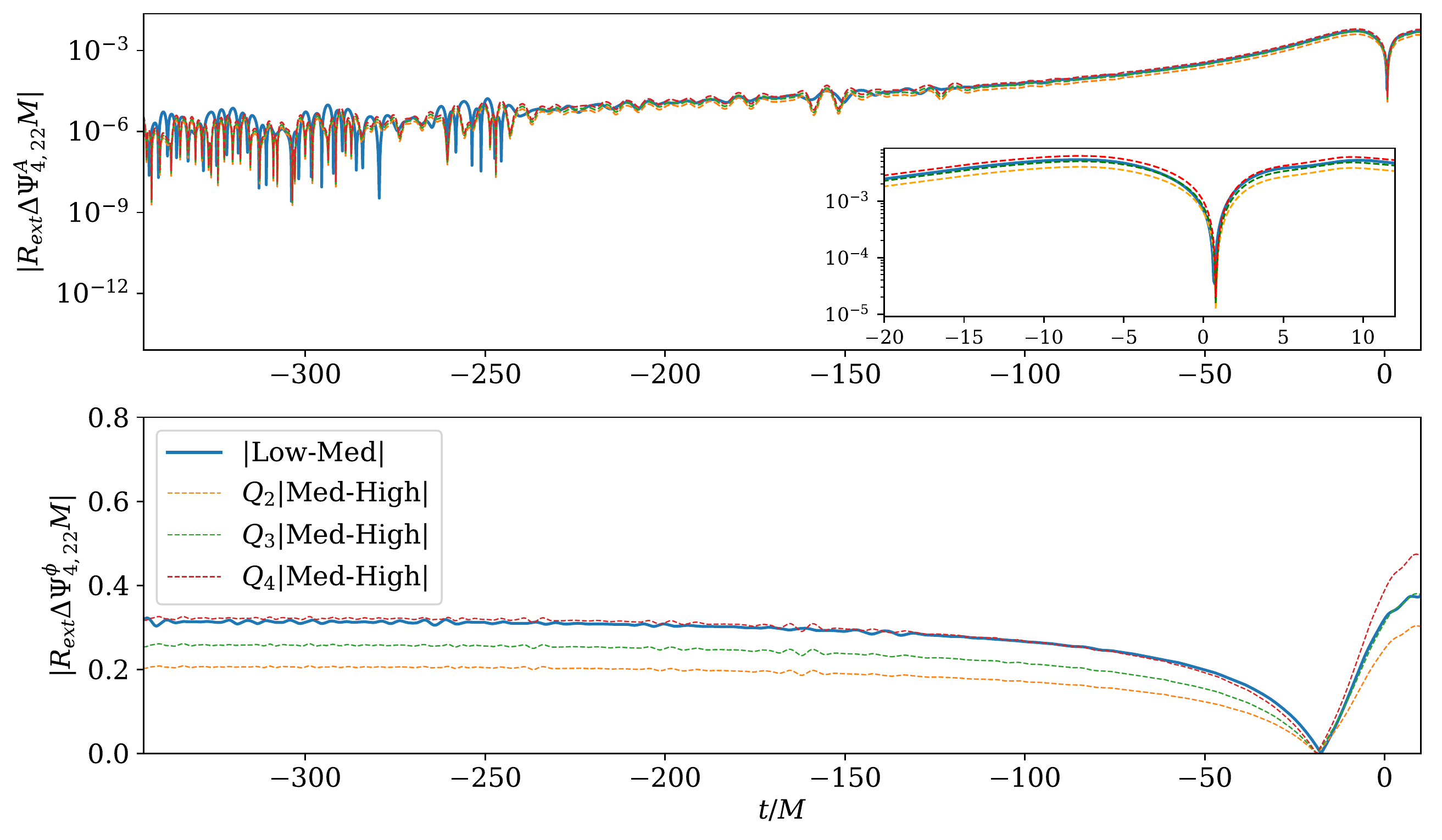}
\caption{Convergence of $\Psi_{4, 22}$ extracted at $R_{\rm{ex}} = 110M$ 
   for \texttt{BBH-11} configuration. 
   \textit{Top}: convergence of the amplitude, $\Psi_{4,lm}^{A}$. 
   The difference for medium/high resolutions is rescaled according to second 
   and third orders of convergence. The inset shows an interval around the merger. 
   \textit{Bottom}: convergence of the phase, $\Psi_{4,lm}^{\phi}$. 
   The difference for medium/high resolutions is rescaled according 
   to third and fourth orders of convergence.}
\label{Psi_convergence}
\end{figure*}

In this section we follow the methodology of~\cite{Radia_2022}.
We start with our results on the convergence of the code and the 
discretization error due to finite resolution for simulation configuration 
\texttt{BBH-11}. 
We use $\Delta x_{\rm{high}} =2M$, $\Delta x_{\rm{med}} = 2.67M$ and 
$\Delta x_{\rm{low}} = 3.2M$ on the coarsest level of 
high, medium and low resolution runs, respectively. 
We note that the results presented in the main body were obtained 
from evolving the higher resolution configuration. 
The order of convergence is estimated for the  amplitude and phase of $\Psi_4$:
\begin{equation}
   \Psi_{4, lm} 
   = 
   \Psi_{4,lm}^A e^{i \Psi_{4,lm}^{\phi}}
   .
\end{equation}
In Fig.~\ref{Psi_convergence}, where we have already multiplied the
difference between medium and high resolutions by an appropriate factor, 
we plot the convergence order for the amplitude and phase of $\Psi_4$.  
The convergence factor of order $n$ for a given quantity $\tau$ 
is calculated by its difference at low/medium and medium/high resolutions:
\begin{equation} 
   Q_n 
   = 
   \frac{\tau_{\rm{low}} - \tau_{\rm{med}}}{\tau_{\rm{med}} - \tau_{\rm{high}}}
   .
\end{equation}
In the continuum limit the truncation error typically approaches zero as a 
power of the discretization parameter $\Delta x$, 
that is for a given approximation error of order $n$,
\begin{equation}
   \lim_{\Delta x \to 0} \tau_{\Delta x} = \Delta x^n \times \rm{const}.
\end{equation}
Therefore, in the continuum limit the convergence factor reduces to
\begin{equation}
Q_n = \frac{\Delta x_{\rm{low}}^n - \Delta x_{\rm{med}}^n}{\Delta x_{\rm{med}}^n - \Delta x_{\rm{high}}^n}.
\end{equation}

\begin{figure*}[!htb]
\includegraphics[width=0.5\linewidth]{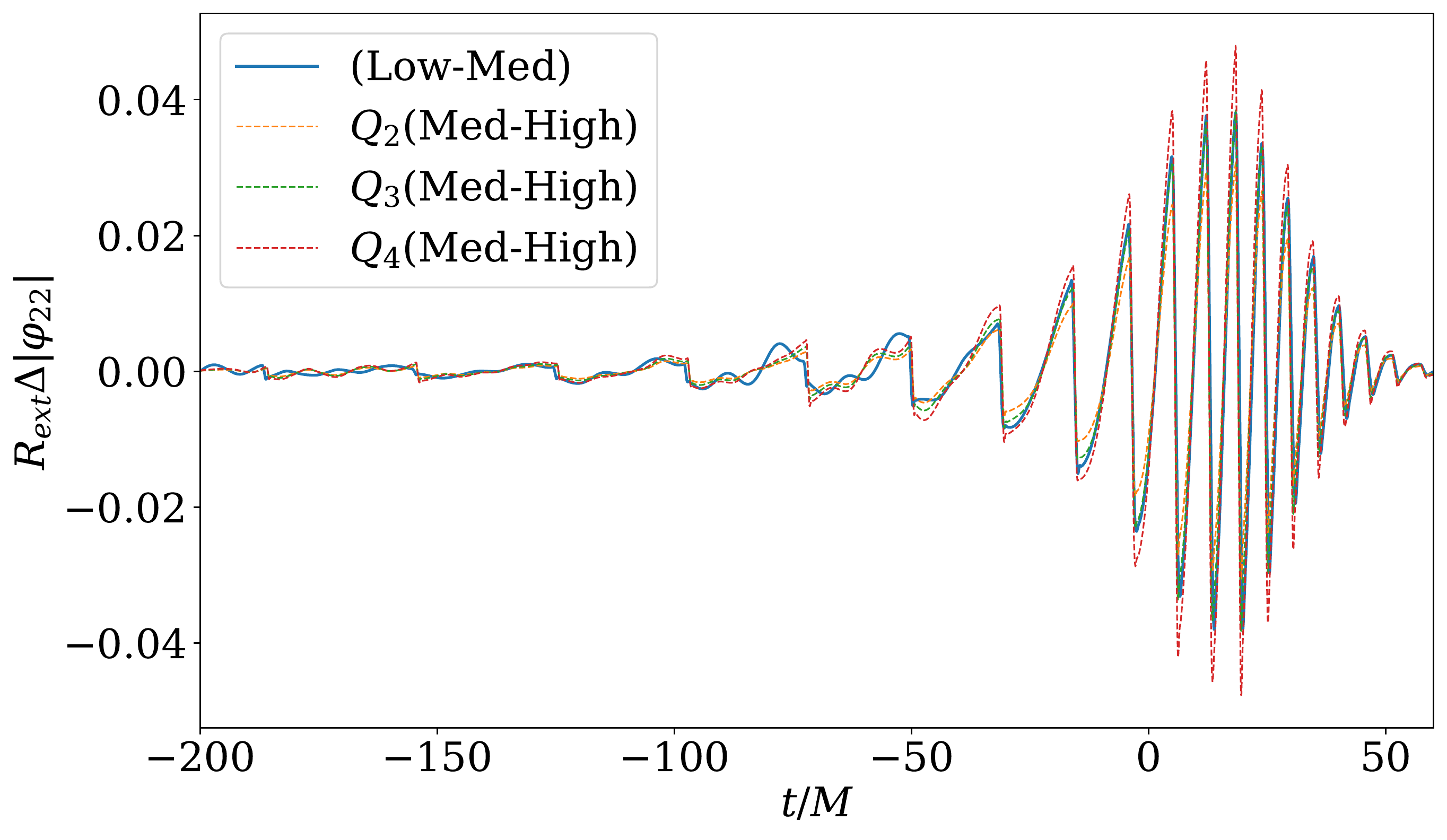}\hfill
\includegraphics[width=0.5\linewidth]{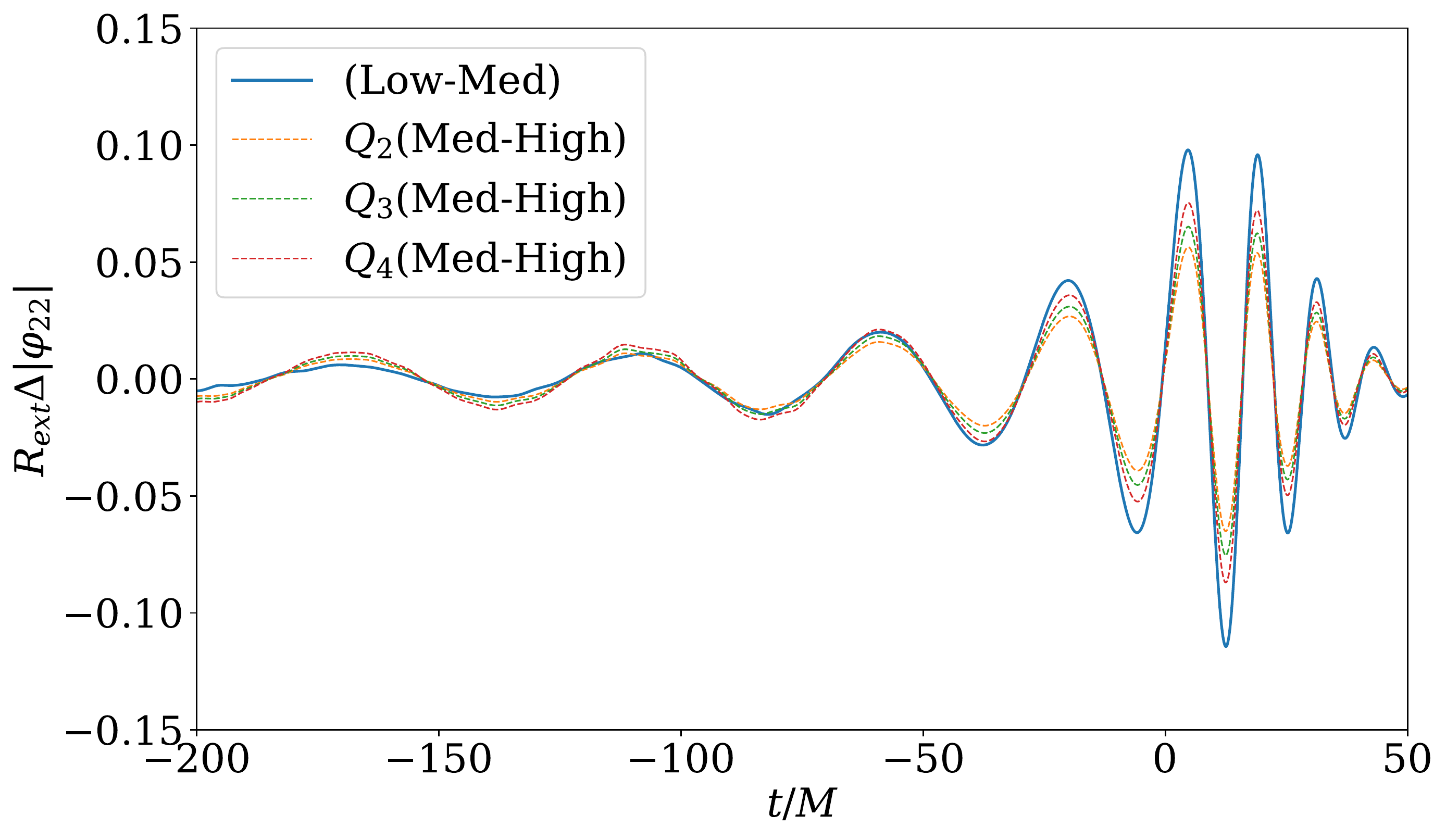}
\caption{\textit{Left:} convergence of the amplitude of $\varphi_{22}$ 
   extracted at $R_{\rm{ex}} = 110M$ for \texttt{BBH-11} configuration. 
   The difference for medium/high resolutions is rescaled from second to fourth 
   orders of convergence. 
   \textit{Right:} same as left but for \texttt{BBH-12} configuration.}
   \label{amp_scalar_convergence}
\end{figure*}

For the amplitude of $\Psi_4$ we find the convergence order to be between second 
and third order in the inspiral, 
which then increases to between third and fourth in the late inspiral and merger. 
For the phase of $\Psi_4$ we find convergence consistent with fourth order dropping 
to third around merger. We perform a similar analysis for the amplitude of the 
scalar field amplitude and find the order of convergence fluctuating between second 
and third as indicated in the left panel of Fig.~\ref{amp_scalar_convergence}.

To estimate the discretization error for the amplitude $e_A$, 
we measure the relative percentage error between the finest resolution result 
and Richardson extrapolation, 
while for the phase error $e_{\phi}$ we simply give it as the difference,
\begin{align} 
   \label{eq:errorsamp}
   e_A 
   &= 
   \frac{
      |\Psi_{4,lm}^{A} - \Psi_{4,lm}^{A, \rm{Richardson}}|
   }{
      \Psi_{4,lm}^{A, \rm{Richardson}}
   } 
   \times 100\%,  
   \\ 
   \label{eq:errorsphase}
   e_{\phi} 
   &= 
   |\Psi_{4,lm}^{\phi} - \Psi_{4,lm}^{\phi, \rm{Richardson}}|
   .
\end{align}
Finally, to estimate the error due to finite radius extraction we compute 
a given radiated quantity at several finite extraction radii and 
extrapolate to infinity by fitting a polynomial in $1/r$. 
We use first order extrapolation of polynomial order $n=1$:
\begin{equation} \label{eq:extrinf}
   \Psi_{4,lm}^{i, n=1} 
   = 
   \frac{\Psi_{4,lm}^{i}}{r}
   .
\end{equation}
We then compute the corresponding errors for the amplitude and phase 
as given in Eqs.~\eqref{eq:errorsamp} and~\eqref{eq:errorsphase},
where Richardson extrapolated quantities are now replaced by the 
extrapolated quantities prescribed by Eq.~\eqref{eq:extrinf}.
\begin{figure*}[!htb]
  \includegraphics[width=0.8\linewidth]{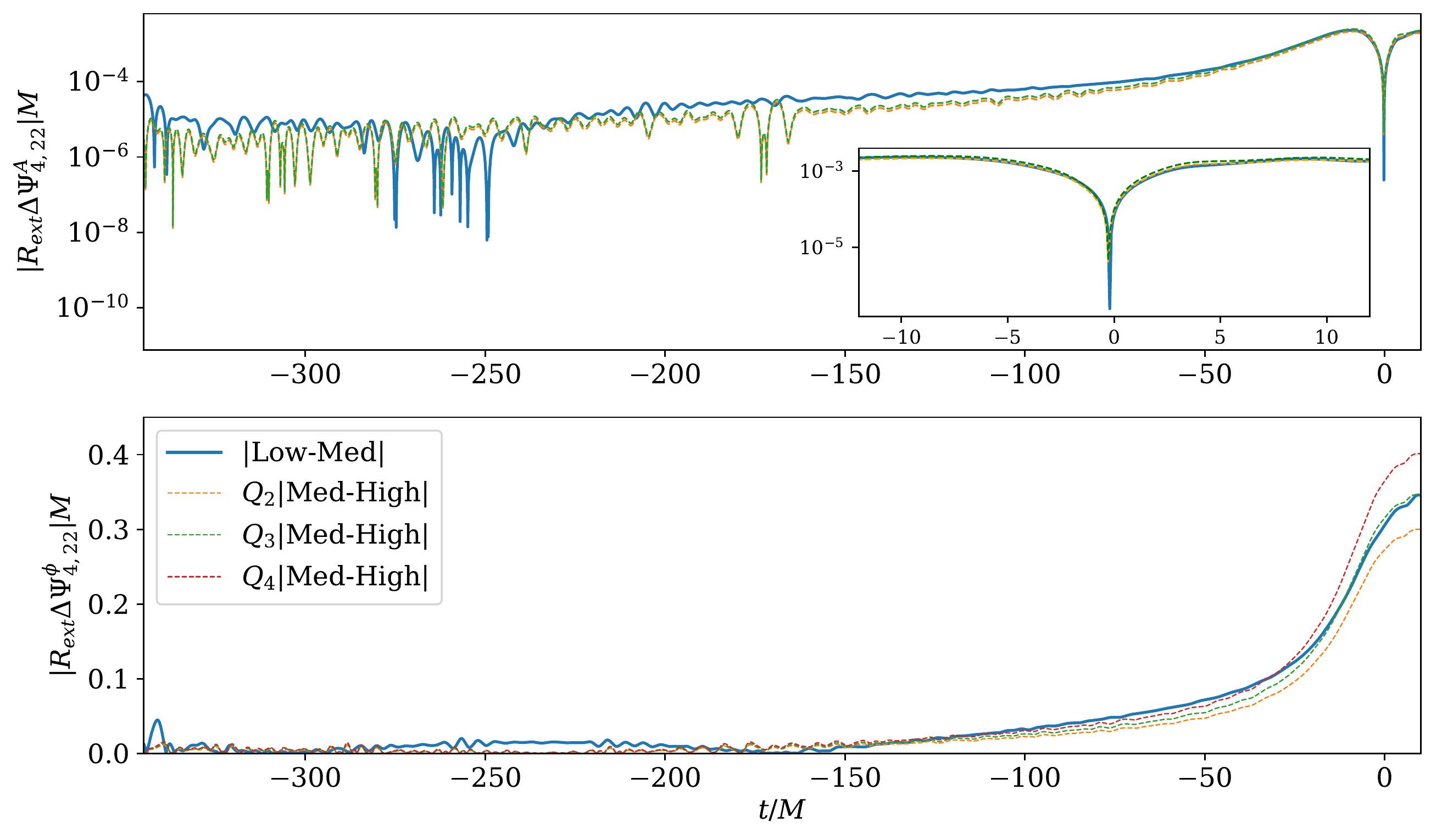}
\caption{convergence of $\Psi_{4, 22}$ extracted at $R_{\rm{ex}} = 110M$ 
   for \texttt{BBH-12} configuration. 
   \textit{Top}: convergence of the amplitude, $\Psi_{4,lm}^{A}$. 
   \textit{Bottom}: convergence of the phase, $\Psi_{4,lm}^{\phi}$.}
\label{Psi_convergence_q05}
\end{figure*}

Additionally, we present convergence results for \texttt{BBH-12} 
configuration, where we now use 
$\Delta x_{\rm{high}} =2M$, $\Delta x_{\rm{med}} = 2.29M$ 
and $\Delta x_{\rm{low}} = 2.67M$ on the coarsest level of 
high, medium and low resolution runs respectively. 
We present our results in Fig.~\ref{Psi_convergence_q05} and the right panel of Fig.~\ref{amp_scalar_convergence} for the gravitational $(22)$ mode of 
$\Psi_4$ and the scalar field amplitude respectively. We find similar results to \texttt{BBH-11} configuration. 
We observe between second and third orders of convergence in the 
amplitude and between third and fourth orders of convergence for the phase. 
The convergence order for the scalar field amplitude is 
second order in the inspiral and fourth order in the merger and ringdown. 

%==============================================================================
\subsection{\label{sec:flux_convergence}
   GW energy and angular momentum flux convergence}
Finally, we present convergence plots for \texttt{BBH-11} of GW energy flux 
and GW angular momentum flux computed with the use of Eqs.~\eqref{eq:E_rad_GW} and~\eqref{eq:J_rad_GW}, respectively.
We employ the same set of resolutions as in Sec.~\ref{sec:convergence}
and compute the total radiated energy and the total angular momentum flux, 
as described in Sec.~\ref{sec:extraction} of the main text.

\begin{figure*}
\includegraphics[width=0.5\linewidth]{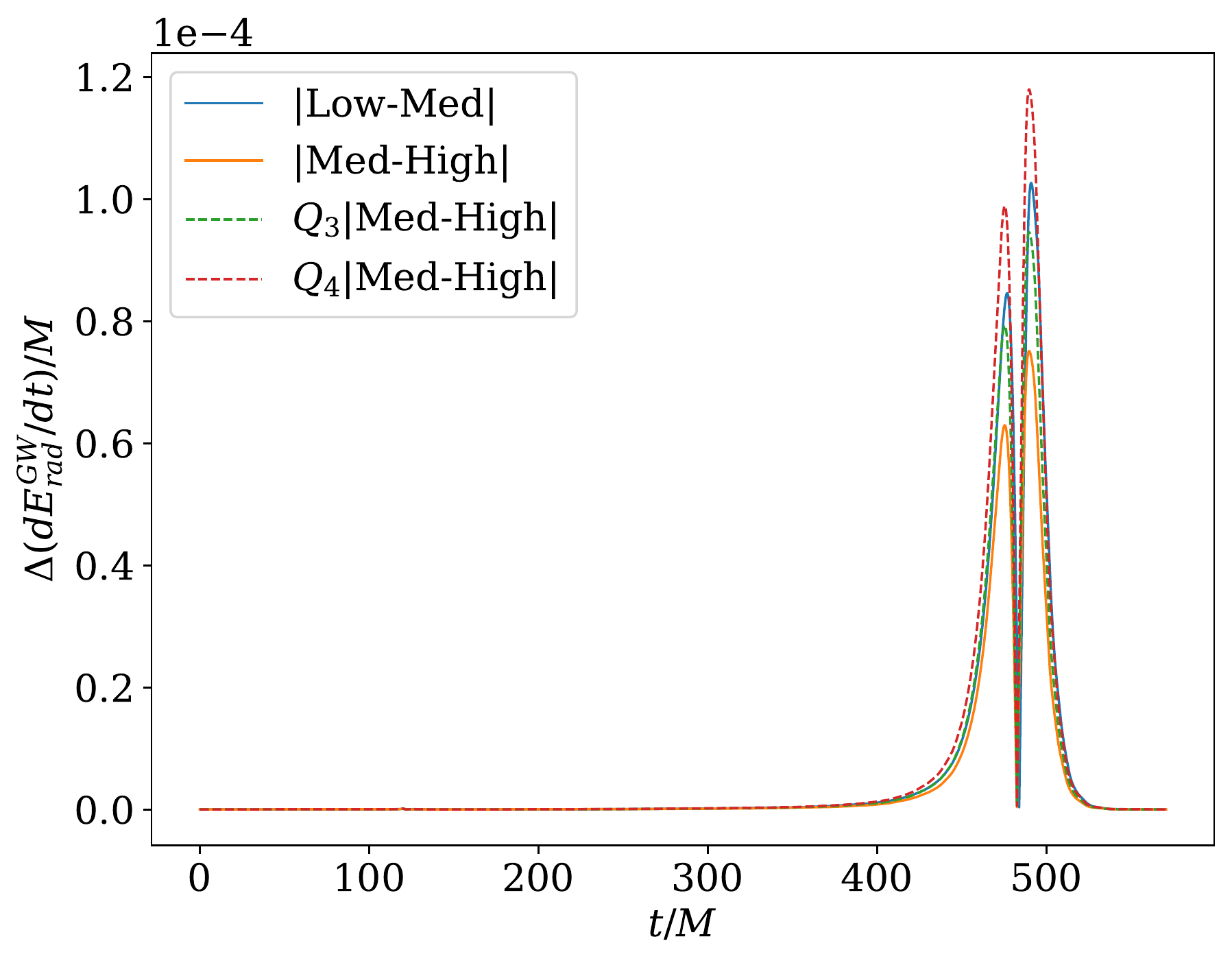}\hfill
\includegraphics[width=0.5\linewidth]{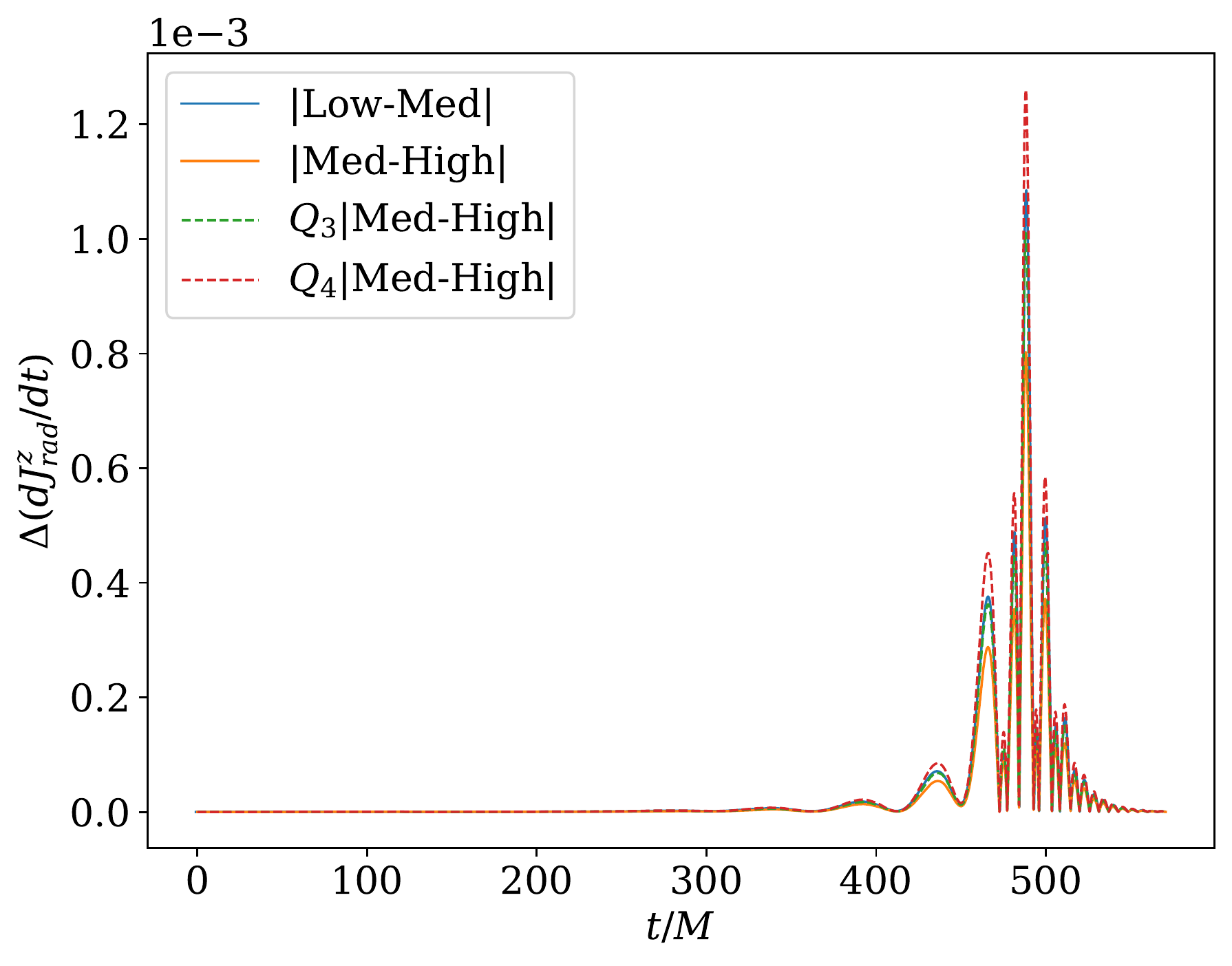}
\caption{\textit{Left}: convergence of the energy flux extracted at 
   $R_{\rm{ext}} = 110M$ for \texttt{BBH-11}.
   The difference for medium/high resolutions is rescaled according to third and 
   forth orders of convergence. \textit{Right}: 
   same as left but the diagnostic quantity is the angular momentum flux in 
   the $z$ direction extracted at $R_{\rm{ext}} = 110M$ for \texttt{BBH-11}.}
   \label{conv_ang_flux}
\end{figure*}

From Fig.~\ref{conv_ang_flux},
we conclude that both the energy flux and angular momentum flux have an order 
of convergence between third and fourth. 
By using third order Richardson extrapolation, 
we estimate the error for final mass and spin to be of 
$0.01 \%$ and $0.05 \%$, respectively.
%==============================================================================
\section{\label{sec:codevalidation}Code Validation} 
\begin{figure*}[!htb]
   \minipage{0.8\textwidth}
   \includegraphics[width=\linewidth]{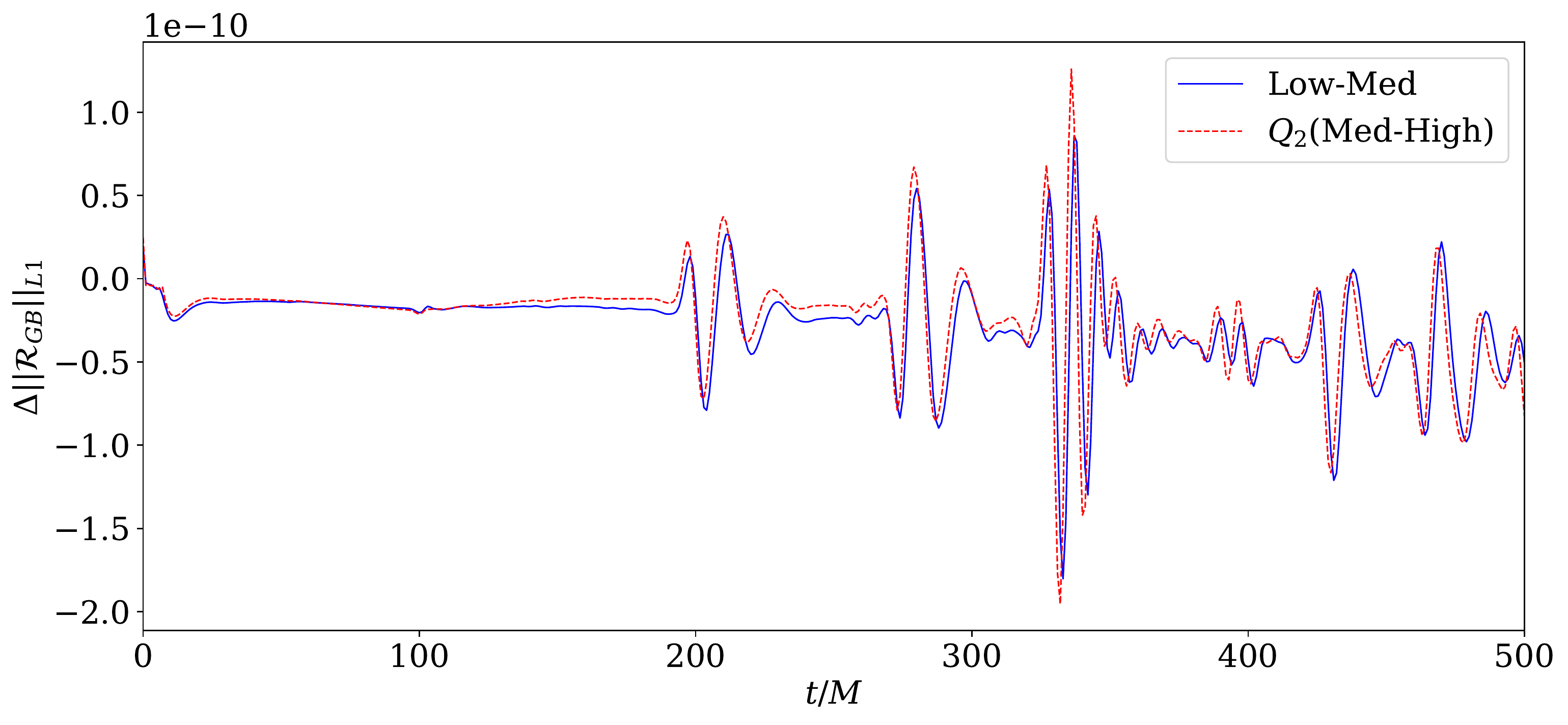}
   \endminipage\hfill
   \caption{Self-convergence of the $L1$-norm normalized over the total volume 
      for the Gauss bonnet term. 
      The difference for medium/high resolutions is rescaled according to the 
      second order convergence.
   }\label{self_convergence_GB}
\end{figure*}

In this section we address the tests on the implemented Gauss-Bonnet term in 
\texttt{GRChombo}. 
First of all, we performed a self-convergence test by evolving a scalar field on a 
Schwarzschild background. 
We use $\Delta x_{\rm{high}} =1.347M$, $\Delta x_{\rm{med}} = 1.6M$ and 
$\Delta x_{\rm{low}} = 2M$ on the coarsest level of 
high, medium and low resolution runs respectively. 
Fig.~\ref{self_convergence_GB}
demonstrates our results and overall we find second order of convergence.

\begin{figure*}
   \minipage{\textwidth}
   \includegraphics[width=0.8\linewidth]{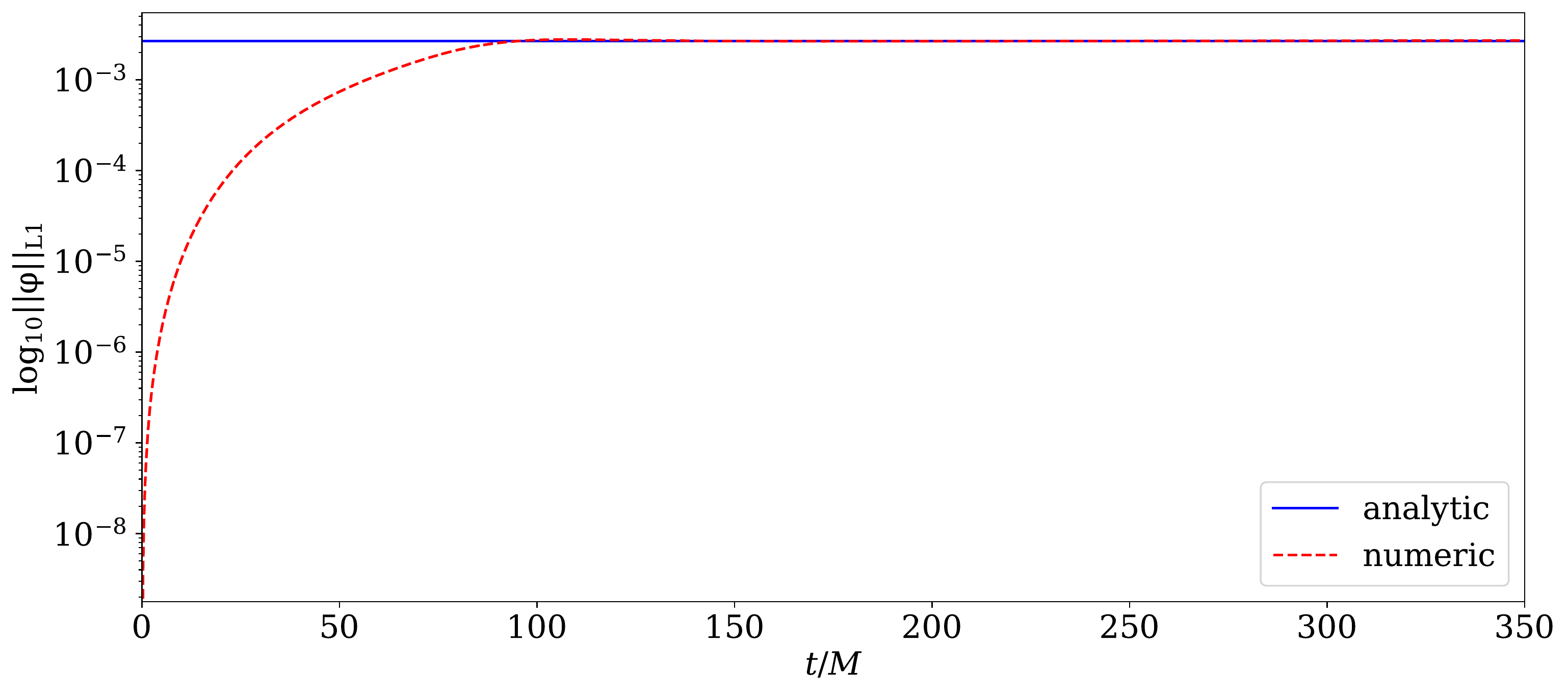}
   \endminipage\hfill
   \caption{Comparison of the $L1$-norm of the scalar field profile 
   normalized over total volume for numeric and analytic solutions.
   }\label{analytic_convergence}
\end{figure*}

Finally, we compare our numeric result for a scalar field evolved on 
a Schwarzschild background to the analytic solution in the 
Painlev$\grave{\text{e}}$-Gullstrand (PG) coordinates. 
By starting with the scalar field equation
\begin{equation}
   \Box \varphi 
   + 
   \beta_0 \mathcal{R}_{GB} 
   = 
   0
   ,
\end{equation}
 we impose a regularity condition of $\partial_r \phi$ at the horizon $r = 2M$ 
 and the falloff of the scalar field at infinity, 
 $\lim_{r\to \infty} \phi = 0$. 
 We find an analytic solution given in terms of PG coordinates by
\begin{equation} \label{eq:phianalytic}
   \varphi(x) 
   = 
   \frac{2 \beta_0}{M} \left(
      \frac{1}{x} 
      + 
      \frac{1}{x^2} 
      + 
      \frac{4}{3x^3} 
   \right), 
   \qquad 
   \text{where } x 
   = 
   \frac{r}{M}
   .
\end{equation}
We then calculate the $L1$-norm for the analytic and numeric solutions and 
normalize it over the total volume. 
Figure~\ref{analytic_convergence} shows good agreement between solutions
after $t \sim 100M$, when the scalar has grown and settled to a constant value.

%==============================================================================
\section{\label{sec:gr_qnms}Quasinormal modes of a Kerr black hole in GR} 
In Sec.~\ref{sec:ringdown} and in particular Eqs.~\eqref{eq:fit_summary_1} and \eqref{eq:fit_summary_2}, we utilize QNMs of a Kerr black hole in GR to perform numerical fits to our data. In Table \ref{qnms_of_gr} we present the values of the frequencies ($\omega_{lmn}$) and damping times ($\tau_{lmn}$), of the gravitational quasinormal modes of the remnant black hole. These modes were computed using the \texttt{qnm} package~\cite{Stein:2019mop}.

\begin{table*}
\begin{tabular} {|c|c|c|c|c|c|} 
 \hline
Run & $(lmn)$ mode & $\omega^{\rm{Kerr, scalar}}_{lmn}$ & $\tau^{\rm{Kerr, scalar}}_{lmn}$ & $\omega^{\rm{Kerr, grav}}_{lmn}$ & $\tau^{\rm{Kerr, grav}}_{lmn}$ \\ [1.5ex]
 \hline
\texttt{BBH-11} \quad & (220) & $0.6526$ & $11.3627$ & $0.5291$ & $12.3320$\\ 
              \quad & (440) & $1.2083$ & $11.4250$ & $1.1356$ & $11.8200$\\ 
\hline
\texttt{BBH-23} \quad & (110) & $0.3730$ & $1.1043$ & \ldots & \ldots \\
  			 \quad & (111) & $0.3570$ & $3.6328$ & \ldots & \ldots \\
  			 \quad & (220) & $0.6434$ & $11.2444$ & $0.5197$ & $12.2165$ \\      
  			 \quad & (330) & $0.9159$ & $11.2855$ & $0.8246$ & $11.8901$\\  
  			 \quad & (440) & $1.1892$ & $11.3026$ & $1.11690$ & $11.6904$ \\
\hline
\texttt{BBH-12} \quad & (110) & $ 0.3654$ & $10.9405$ & \ldots & \ldots \\
  			 \quad & (111) & $0.3485$ & $3.5715$ & \ldots & \ldots \\
  			 \quad & (220) & $0.6279$ & $11.0667$ & $0.5043$ & $12.0431$ \\      
  			 \quad & (330) & $0.8923$ & $11.10375$ & $0.8016$ & $11.6951$\\  
  			 \quad & (440) & $1.1574$ & $11.1192$ & $1.0858$ & $11.4945$ \\
\hline
\end{tabular}
\caption{QNMs of the remnant Kerr BH in GR, used in the fitting procedure described in Sec~\ref{sec:ringdown}.} \label{qnms_of_gr}
\end{table*}

\newpage
\clearpage

%\input{main.bbl}
%==============================================================================
\bibliography{thebib_T}

%apsrev4-2.bst 2019-01-14 (MD) hand-edited version of apsrev4-1.bst
%Control: key (0)
%Control: author (8) initials jnrlst
%Control: editor formatted (1) identically to author
%Control: production of article title (0) allowed
%Control: page (0) single
%Control: year (1) truncated
%Control: production of eprint (0) enabled
\begin{thebibliography}{136}%
\makeatletter
\providecommand \@ifxundefined [1]{%
 \@ifx{#1\undefined}
}%
\providecommand \@ifnum [1]{%
 \ifnum #1\expandafter \@firstoftwo
 \else \expandafter \@secondoftwo
 \fi
}%
\providecommand \@ifx [1]{%
 \ifx #1\expandafter \@firstoftwo
 \else \expandafter \@secondoftwo
 \fi
}%
\providecommand \natexlab [1]{#1}%
\providecommand \enquote  [1]{``#1''}%
\providecommand \bibnamefont  [1]{#1}%
\providecommand \bibfnamefont [1]{#1}%
\providecommand \citenamefont [1]{#1}%
\providecommand \href@noop [0]{\@secondoftwo}%
\providecommand \href [0]{\begingroup \@sanitize@url \@href}%
\providecommand \@href[1]{\@@startlink{#1}\@@href}%
\providecommand \@@href[1]{\endgroup#1\@@endlink}%
\providecommand \@sanitize@url [0]{\catcode `\\12\catcode `\$12\catcode
  `\&12\catcode `\#12\catcode `\^12\catcode `\_12\catcode `\%12\relax}%
\providecommand \@@startlink[1]{}%
\providecommand \@@endlink[0]{}%
\providecommand \url  [0]{\begingroup\@sanitize@url \@url }%
\providecommand \@url [1]{\endgroup\@href {#1}{\urlprefix }}%
\providecommand \urlprefix  [0]{URL }%
\providecommand \Eprint [0]{\href }%
\providecommand \doibase [0]{https://doi.org/}%
\providecommand \selectlanguage [0]{\@gobble}%
\providecommand \bibinfo  [0]{\@secondoftwo}%
\providecommand \bibfield  [0]{\@secondoftwo}%
\providecommand \translation [1]{[#1]}%
\providecommand \BibitemOpen [0]{}%
\providecommand \bibitemStop [0]{}%
\providecommand \bibitemNoStop [0]{.\EOS\space}%
\providecommand \EOS [0]{\spacefactor3000\relax}%
\providecommand \BibitemShut  [1]{\csname bibitem#1\endcsname}%
\let\auto@bib@innerbib\@empty
%</preamble>
\bibitem [{\citenamefont {Will}(2014)}]{Will:2014kxa}%
  \BibitemOpen
  \bibfield  {author} {\bibinfo {author} {\bibfnamefont {C.~M.}\ \bibnamefont
  {Will}},\ }\bibfield  {title} {\bibinfo {title} {{The Confrontation between
  General Relativity and Experiment}},\ }\href
  {https://doi.org/10.12942/lrr-2014-4} {\bibfield  {journal} {\bibinfo
  {journal} {Living Rev. Rel.}\ }\textbf {\bibinfo {volume} {17}},\ \bibinfo
  {pages} {4} (\bibinfo {year} {2014})},\ \Eprint
  {https://arxiv.org/abs/1403.7377} {arXiv:1403.7377 [gr-qc]} \BibitemShut
  {NoStop}%
\bibitem [{\citenamefont {Ishak}(2019)}]{Ishak:2018his}%
  \BibitemOpen
  \bibfield  {author} {\bibinfo {author} {\bibfnamefont {M.}~\bibnamefont
  {Ishak}},\ }\bibfield  {title} {\bibinfo {title} {{Testing General Relativity
  in Cosmology}},\ }\href {https://doi.org/10.1007/s41114-018-0017-4}
  {\bibfield  {journal} {\bibinfo  {journal} {Living Rev. Rel.}\ }\textbf
  {\bibinfo {volume} {22}},\ \bibinfo {pages} {1} (\bibinfo {year} {2019})},\
  \Eprint {https://arxiv.org/abs/1806.10122} {arXiv:1806.10122 [astro-ph.CO]}
  \BibitemShut {NoStop}%
\bibitem [{\citenamefont {Aasi}\ \emph
  {et~al.}(2015{\natexlab{a}})\citenamefont {Aasi} \emph
  {et~al.}}]{TheLIGOScientific:2014jea}%
  \BibitemOpen
  \bibfield  {author} {\bibinfo {author} {\bibfnamefont {J.}~\bibnamefont
  {Aasi}} \emph {et~al.} (\bibinfo {collaboration} {LIGO Scientific}),\
  }\bibfield  {title} {\bibinfo {title} {{Advanced LIGO}},\ }\href
  {https://doi.org/10.1088/0264-9381/32/7/074001} {\bibfield  {journal}
  {\bibinfo  {journal} {Class. Quant. Grav.}\ }\textbf {\bibinfo {volume}
  {32}},\ \bibinfo {pages} {074001} (\bibinfo {year} {2015}{\natexlab{a}})},\
  \Eprint {https://arxiv.org/abs/1411.4547} {arXiv:1411.4547 [gr-qc]}
  \BibitemShut {NoStop}%
\bibitem [{\citenamefont {Acernese}\ \emph {et~al.}(2015)\citenamefont
  {Acernese} \emph {et~al.}}]{TheVirgo:2014hva}%
  \BibitemOpen
  \bibfield  {author} {\bibinfo {author} {\bibfnamefont {F.}~\bibnamefont
  {Acernese}} \emph {et~al.} (\bibinfo {collaboration} {VIRGO}),\ }\bibfield
  {title} {\bibinfo {title} {{Advanced Virgo: a second-generation
  interferometric gravitational wave detector}},\ }\href
  {https://doi.org/10.1088/0264-9381/32/2/024001} {\bibfield  {journal}
  {\bibinfo  {journal} {Class. Quant. Grav.}\ }\textbf {\bibinfo {volume}
  {32}},\ \bibinfo {pages} {024001} (\bibinfo {year} {2015})},\ \Eprint
  {https://arxiv.org/abs/1408.3978} {arXiv:1408.3978 [gr-qc]} \BibitemShut
  {NoStop}%
%%CITATION = ARXIV:1408.3978;%%
\bibitem [{\citenamefont {Abbott}\ \emph {et~al.}(2016)\citenamefont {Abbott}
  \emph {et~al.}}]{TheLIGOScientific:2016src}%
  \BibitemOpen
  \bibfield  {author} {\bibinfo {author} {\bibfnamefont {B.~P.}\ \bibnamefont
  {Abbott}} \emph {et~al.} (\bibinfo {collaboration} {LIGO Scientific,
  Virgo}),\ }\bibfield  {title} {\bibinfo {title} {{Tests of general relativity
  with {GW}150914}},\ }\href {https://doi.org/10.1103/PhysRevLett.116.221101}
  {\bibfield  {journal} {\bibinfo  {journal} {Phys. Rev. Lett.}\ }\textbf
  {\bibinfo {volume} {116}},\ \bibinfo {pages} {221101} (\bibinfo {year}
  {2016})},\ \bibinfo {note} {[Erratum: Phys.Rev.Lett. 121, 129902 (2018)]},\
  \Eprint {https://arxiv.org/abs/1602.03841} {arXiv:1602.03841 [gr-qc]}
  \BibitemShut {NoStop}%
\bibitem [{\citenamefont {Abbott}\ \emph
  {et~al.}(2019{\natexlab{a}})\citenamefont {Abbott} \emph
  {et~al.}}]{LIGOScientific:2019fpa}%
  \BibitemOpen
  \bibfield  {author} {\bibinfo {author} {\bibfnamefont {B.~P.}\ \bibnamefont
  {Abbott}} \emph {et~al.} (\bibinfo {collaboration} {LIGO Scientific,
  Virgo}),\ }\bibfield  {title} {\bibinfo {title} {{Tests of General Relativity
  with the Binary Black Hole Signals from the {LIGO}-{V}irgo Catalog
  {GWTC}-1}},\ }\href {https://doi.org/10.1103/PhysRevD.100.104036} {\bibfield
  {journal} {\bibinfo  {journal} {Phys. Rev. D}\ }\textbf {\bibinfo {volume}
  {100}},\ \bibinfo {pages} {104036} (\bibinfo {year} {2019}{\natexlab{a}})},\
  \Eprint {https://arxiv.org/abs/1903.04467} {arXiv:1903.04467 [gr-qc]}
  \BibitemShut {NoStop}%
\bibitem [{\citenamefont {Abbott}\ \emph
  {et~al.}(2019{\natexlab{b}})\citenamefont {Abbott} \emph
  {et~al.}}]{LIGOScientific:2018dkp}%
  \BibitemOpen
  \bibfield  {author} {\bibinfo {author} {\bibfnamefont {B.~P.}\ \bibnamefont
  {Abbott}} \emph {et~al.} (\bibinfo {collaboration} {LIGO Scientific,
  Virgo}),\ }\bibfield  {title} {\bibinfo {title} {{Tests of General Relativity
  with GW170817}},\ }\href {https://doi.org/10.1103/PhysRevLett.123.011102}
  {\bibfield  {journal} {\bibinfo  {journal} {Phys. Rev. Lett.}\ }\textbf
  {\bibinfo {volume} {123}},\ \bibinfo {pages} {011102} (\bibinfo {year}
  {2019}{\natexlab{b}})},\ \Eprint {https://arxiv.org/abs/1811.00364}
  {arXiv:1811.00364 [gr-qc]} \BibitemShut {NoStop}%
\bibitem [{\citenamefont {Abbott}\ \emph
  {et~al.}(2021{\natexlab{a}})\citenamefont {Abbott} \emph
  {et~al.}}]{LIGOScientific:2020tif}%
  \BibitemOpen
  \bibfield  {author} {\bibinfo {author} {\bibfnamefont {R.}~\bibnamefont
  {Abbott}} \emph {et~al.} (\bibinfo {collaboration} {LIGO Scientific,
  Virgo}),\ }\bibfield  {title} {\bibinfo {title} {{Tests of general relativity
  with binary black holes from the second LIGO-Virgo gravitational-wave
  transient catalog}},\ }\href {https://doi.org/10.1103/PhysRevD.103.122002}
  {\bibfield  {journal} {\bibinfo  {journal} {Phys. Rev. D}\ }\textbf {\bibinfo
  {volume} {103}},\ \bibinfo {pages} {122002} (\bibinfo {year}
  {2021}{\natexlab{a}})},\ \Eprint {https://arxiv.org/abs/2010.14529}
  {arXiv:2010.14529 [gr-qc]} \BibitemShut {NoStop}%
\bibitem [{\citenamefont {Abbott}\ \emph
  {et~al.}(2021{\natexlab{b}})\citenamefont {Abbott} \emph
  {et~al.}}]{LIGOScientific:2021sio}%
  \BibitemOpen
  \bibfield  {author} {\bibinfo {author} {\bibfnamefont {R.}~\bibnamefont
  {Abbott}} \emph {et~al.} (\bibinfo {collaboration} {LIGO Scientific, VIRGO,
  KAGRA}),\ }\bibfield  {title} {\bibinfo {title} {{Tests of General Relativity
  with GWTC-3}},\ }\href@noop {} {\  (\bibinfo {year} {2021}{\natexlab{b}})},\
  \Eprint {https://arxiv.org/abs/2112.06861} {arXiv:2112.06861 [gr-qc]}
  \BibitemShut {NoStop}%
\bibitem [{\citenamefont {Cardoso}\ and\ \citenamefont
  {Pani}(2019)}]{Cardoso:2019rvt}%
  \BibitemOpen
  \bibfield  {author} {\bibinfo {author} {\bibfnamefont {V.}~\bibnamefont
  {Cardoso}}\ and\ \bibinfo {author} {\bibfnamefont {P.}~\bibnamefont {Pani}},\
  }\bibfield  {title} {\bibinfo {title} {{Testing the nature of dark compact
  objects: a status report}},\ }\href
  {https://doi.org/10.1007/s41114-019-0020-4} {\bibfield  {journal} {\bibinfo
  {journal} {Living Rev. Rel.}\ }\textbf {\bibinfo {volume} {22}},\ \bibinfo
  {pages} {4} (\bibinfo {year} {2019})},\ \Eprint
  {https://arxiv.org/abs/1904.05363} {arXiv:1904.05363 [gr-qc]} \BibitemShut
  {NoStop}%
\bibitem [{\citenamefont {Yunes}\ \emph {et~al.}(2016)\citenamefont {Yunes},
  \citenamefont {Yagi},\ and\ \citenamefont {Pretorius}}]{Yunes:2016jcc}%
  \BibitemOpen
  \bibfield  {author} {\bibinfo {author} {\bibfnamefont {N.}~\bibnamefont
  {Yunes}}, \bibinfo {author} {\bibfnamefont {K.}~\bibnamefont {Yagi}},\ and\
  \bibinfo {author} {\bibfnamefont {F.}~\bibnamefont {Pretorius}},\ }\bibfield
  {title} {\bibinfo {title} {{Theoretical Physics Implications of the Binary
  Black-Hole Mergers {GW}150914 and {GW}151226}},\ }\href
  {https://doi.org/10.1103/PhysRevD.94.084002} {\bibfield  {journal} {\bibinfo
  {journal} {Phys. Rev. D}\ }\textbf {\bibinfo {volume} {94}},\ \bibinfo
  {pages} {084002} (\bibinfo {year} {2016})},\ \Eprint
  {https://arxiv.org/abs/1603.08955} {arXiv:1603.08955 [gr-qc]} \BibitemShut
  {NoStop}%
\bibitem [{\citenamefont {Baker}\ \emph {et~al.}(2017)\citenamefont {Baker},
  \citenamefont {Bellini}, \citenamefont {Ferreira}, \citenamefont {Lagos},
  \citenamefont {Noller},\ and\ \citenamefont {Sawicki}}]{Baker:2017hug}%
  \BibitemOpen
  \bibfield  {author} {\bibinfo {author} {\bibfnamefont {T.}~\bibnamefont
  {Baker}}, \bibinfo {author} {\bibfnamefont {E.}~\bibnamefont {Bellini}},
  \bibinfo {author} {\bibfnamefont {P.~G.}\ \bibnamefont {Ferreira}}, \bibinfo
  {author} {\bibfnamefont {M.}~\bibnamefont {Lagos}}, \bibinfo {author}
  {\bibfnamefont {J.}~\bibnamefont {Noller}},\ and\ \bibinfo {author}
  {\bibfnamefont {I.}~\bibnamefont {Sawicki}},\ }\bibfield  {title} {\bibinfo
  {title} {{Strong constraints on cosmological gravity from {GW}170817 and
  {GRB} 170817{A}}},\ }\href {https://doi.org/10.1103/PhysRevLett.119.251301}
  {\bibfield  {journal} {\bibinfo  {journal} {Phys. Rev. Lett.}\ }\textbf
  {\bibinfo {volume} {119}},\ \bibinfo {pages} {251301} (\bibinfo {year}
  {2017})},\ \Eprint {https://arxiv.org/abs/1710.06394} {arXiv:1710.06394
  [astro-ph.CO]} \BibitemShut {NoStop}%
\bibitem [{\citenamefont {Lyu}\ \emph {et~al.}(2022{\natexlab{a}})\citenamefont
  {Lyu}, \citenamefont {Jiang},\ and\ \citenamefont {Yagi}}]{Lyu:2022gdr}%
  \BibitemOpen
  \bibfield  {author} {\bibinfo {author} {\bibfnamefont {Z.}~\bibnamefont
  {Lyu}}, \bibinfo {author} {\bibfnamefont {N.}~\bibnamefont {Jiang}},\ and\
  \bibinfo {author} {\bibfnamefont {K.}~\bibnamefont {Yagi}},\ }\bibfield
  {title} {\bibinfo {title} {{Constraints on
  {E}instein-dilation-{G}auss-{B}onnet gravity from black hole-neutron star
  gravitational wave events}},\ }\href
  {https://doi.org/10.1103/PhysRevD.105.064001} {\bibfield  {journal} {\bibinfo
   {journal} {Phys. Rev. D}\ }\textbf {\bibinfo {volume} {105}},\ \bibinfo
  {pages} {064001} (\bibinfo {year} {2022}{\natexlab{a}})},\ \Eprint
  {https://arxiv.org/abs/2201.02543} {arXiv:2201.02543 [gr-qc]} \BibitemShut
  {NoStop}%
\bibitem [{\citenamefont {Perkins}\ \emph {et~al.}(2021)\citenamefont
  {Perkins}, \citenamefont {Nair}, \citenamefont {Silva},\ and\ \citenamefont
  {Yunes}}]{Perkins:2021mhb}%
  \BibitemOpen
  \bibfield  {author} {\bibinfo {author} {\bibfnamefont {S.~E.}\ \bibnamefont
  {Perkins}}, \bibinfo {author} {\bibfnamefont {R.}~\bibnamefont {Nair}},
  \bibinfo {author} {\bibfnamefont {H.~O.}\ \bibnamefont {Silva}},\ and\
  \bibinfo {author} {\bibfnamefont {N.}~\bibnamefont {Yunes}},\ }\bibfield
  {title} {\bibinfo {title} {{Improved gravitational-wave constraints on
  higher-order curvature theories of gravity}},\ }\href
  {https://doi.org/10.1103/PhysRevD.104.024060} {\bibfield  {journal} {\bibinfo
   {journal} {Phys. Rev. D}\ }\textbf {\bibinfo {volume} {104}},\ \bibinfo
  {pages} {024060} (\bibinfo {year} {2021})},\ \Eprint
  {https://arxiv.org/abs/2104.11189} {arXiv:2104.11189 [gr-qc]} \BibitemShut
  {NoStop}%
\bibitem [{\citenamefont {Okounkova}\ \emph {et~al.}(2022)\citenamefont
  {Okounkova}, \citenamefont {Farr}, \citenamefont {Isi},\ and\ \citenamefont
  {Stein}}]{Okounkova:2021xjv}%
  \BibitemOpen
  \bibfield  {author} {\bibinfo {author} {\bibfnamefont {M.}~\bibnamefont
  {Okounkova}}, \bibinfo {author} {\bibfnamefont {W.~M.}\ \bibnamefont {Farr}},
  \bibinfo {author} {\bibfnamefont {M.}~\bibnamefont {Isi}},\ and\ \bibinfo
  {author} {\bibfnamefont {L.~C.}\ \bibnamefont {Stein}},\ }\bibfield  {title}
  {\bibinfo {title} {{Constraining gravitational wave amplitude birefringence
  and {C}hern-{S}imons gravity with {GWTC}-2}},\ }\href
  {https://doi.org/10.1103/PhysRevD.106.044067} {\bibfield  {journal} {\bibinfo
   {journal} {Phys. Rev. D}\ }\textbf {\bibinfo {volume} {106}},\ \bibinfo
  {pages} {044067} (\bibinfo {year} {2022})},\ \Eprint
  {https://arxiv.org/abs/2101.11153} {arXiv:2101.11153 [gr-qc]} \BibitemShut
  {NoStop}%
\bibitem [{\citenamefont {Weinberg}(2008)}]{Weinberg:2008hq}%
  \BibitemOpen
  \bibfield  {author} {\bibinfo {author} {\bibfnamefont {S.}~\bibnamefont
  {Weinberg}},\ }\bibfield  {title} {\bibinfo {title} {{Effective Field Theory
  for Inflation}},\ }\href {https://doi.org/10.1103/PhysRevD.77.123541}
  {\bibfield  {journal} {\bibinfo  {journal} {Phys. Rev. D}\ }\textbf {\bibinfo
  {volume} {77}},\ \bibinfo {pages} {123541} (\bibinfo {year} {2008})},\
  \Eprint {https://arxiv.org/abs/0804.4291} {arXiv:0804.4291 [hep-th]}
  \BibitemShut {NoStop}%
\bibitem [{\citenamefont {Kov\'acs}\ and\ \citenamefont
  {Reall}(2020)}]{Kovacs:2020pns}%
  \BibitemOpen
  \bibfield  {author} {\bibinfo {author} {\bibfnamefont {A.~D.}\ \bibnamefont
  {Kov\'acs}}\ and\ \bibinfo {author} {\bibfnamefont {H.~S.}\ \bibnamefont
  {Reall}},\ }\bibfield  {title} {\bibinfo {title} {{Well-Posed Formulation of
  Scalar-Tensor Effective Field Theory}},\ }\href
  {https://doi.org/10.1103/PhysRevLett.124.221101} {\bibfield  {journal}
  {\bibinfo  {journal} {Phys. Rev. Lett.}\ }\textbf {\bibinfo {volume} {124}},\
  \bibinfo {pages} {221101} (\bibinfo {year} {2020})},\ \Eprint
  {https://arxiv.org/abs/2003.04327} {arXiv:2003.04327 [gr-qc]} \BibitemShut
  {NoStop}%
\bibitem [{\citenamefont {Gross}\ and\ \citenamefont
  {Sloan}(1987)}]{Gross:1986mw}%
  \BibitemOpen
  \bibfield  {author} {\bibinfo {author} {\bibfnamefont {D.~J.}\ \bibnamefont
  {Gross}}\ and\ \bibinfo {author} {\bibfnamefont {J.~H.}\ \bibnamefont
  {Sloan}},\ }\bibfield  {title} {\bibinfo {title} {{The Quartic Effective
  Action for the Heterotic String}},\ }\href
  {https://doi.org/10.1016/0550-3213(87)90465-2} {\bibfield  {journal}
  {\bibinfo  {journal} {Nucl. Phys. B}\ }\textbf {\bibinfo {volume} {291}},\
  \bibinfo {pages} {41} (\bibinfo {year} {1987})}\BibitemShut {NoStop}%
\bibitem [{\citenamefont {Zwiebach}(1985)}]{Zwiebach:1985uq}%
  \BibitemOpen
  \bibfield  {author} {\bibinfo {author} {\bibfnamefont {B.}~\bibnamefont
  {Zwiebach}},\ }\bibfield  {title} {\bibinfo {title} {{Curvature Squared Terms
  and String Theories}},\ }\href {https://doi.org/10.1016/0370-2693(85)91616-8}
  {\bibfield  {journal} {\bibinfo  {journal} {Phys. Lett. B}\ }\textbf
  {\bibinfo {volume} {156}},\ \bibinfo {pages} {315} (\bibinfo {year}
  {1985})}\BibitemShut {NoStop}%
\bibitem [{\citenamefont {Cano}\ and\ \citenamefont
  {Ruip\'erez}(2022)}]{Cano:2021rey}%
  \BibitemOpen
  \bibfield  {author} {\bibinfo {author} {\bibfnamefont {P.~A.}\ \bibnamefont
  {Cano}}\ and\ \bibinfo {author} {\bibfnamefont {A.}~\bibnamefont
  {Ruip\'erez}},\ }\bibfield  {title} {\bibinfo {title} {{String gravity in
  {D}=4}},\ }\href {https://doi.org/10.1103/PhysRevD.105.044022} {\bibfield
  {journal} {\bibinfo  {journal} {Phys. Rev. D}\ }\textbf {\bibinfo {volume}
  {105}},\ \bibinfo {pages} {044022} (\bibinfo {year} {2022})},\ \Eprint
  {https://arxiv.org/abs/2111.04750} {arXiv:2111.04750 [hep-th]} \BibitemShut
  {NoStop}%
\bibitem [{\citenamefont {Kanti}\ \emph {et~al.}(1996)\citenamefont {Kanti},
  \citenamefont {Mavromatos}, \citenamefont {Rizos}, \citenamefont {Tamvakis},\
  and\ \citenamefont {Winstanley}}]{Kanti:1995vq}%
  \BibitemOpen
  \bibfield  {author} {\bibinfo {author} {\bibfnamefont {P.}~\bibnamefont
  {Kanti}}, \bibinfo {author} {\bibfnamefont {N.~E.}\ \bibnamefont
  {Mavromatos}}, \bibinfo {author} {\bibfnamefont {J.}~\bibnamefont {Rizos}},
  \bibinfo {author} {\bibfnamefont {K.}~\bibnamefont {Tamvakis}},\ and\
  \bibinfo {author} {\bibfnamefont {E.}~\bibnamefont {Winstanley}},\ }\bibfield
   {title} {\bibinfo {title} {{Dilatonic black holes in higher curvature string
  gravity}},\ }\href {https://doi.org/10.1103/PhysRevD.54.5049} {\bibfield
  {journal} {\bibinfo  {journal} {Phys. Rev. D}\ }\textbf {\bibinfo {volume}
  {54}},\ \bibinfo {pages} {5049} (\bibinfo {year} {1996})},\ \Eprint
  {https://arxiv.org/abs/hep-th/9511071} {arXiv:hep-th/9511071} \BibitemShut
  {NoStop}%
\bibitem [{\citenamefont {Sotiriou}\ and\ \citenamefont
  {Zhou}(2014)}]{Sotiriou:2014pfa}%
  \BibitemOpen
  \bibfield  {author} {\bibinfo {author} {\bibfnamefont {T.~P.}\ \bibnamefont
  {Sotiriou}}\ and\ \bibinfo {author} {\bibfnamefont {S.-Y.}\ \bibnamefont
  {Zhou}},\ }\bibfield  {title} {\bibinfo {title} {{Black hole hair in
  generalized scalar-tensor gravity: {A}n explicit example}},\ }\href
  {https://doi.org/10.1103/PhysRevD.90.124063} {\bibfield  {journal} {\bibinfo
  {journal} {Phys. Rev. D}\ }\textbf {\bibinfo {volume} {90}},\ \bibinfo
  {pages} {124063} (\bibinfo {year} {2014})},\ \Eprint
  {https://arxiv.org/abs/1408.1698} {arXiv:1408.1698 [gr-qc]} \BibitemShut
  {NoStop}%
\bibitem [{\citenamefont {Herdeiro}\ and\ \citenamefont
  {Radu}(2015)}]{Herdeiro:2015waa}%
  \BibitemOpen
  \bibfield  {author} {\bibinfo {author} {\bibfnamefont {C.~A.~R.}\
  \bibnamefont {Herdeiro}}\ and\ \bibinfo {author} {\bibfnamefont
  {E.}~\bibnamefont {Radu}},\ }\bibfield  {title} {\bibinfo {title}
  {{Asymptotically flat black holes with scalar hair: {A} review}},\ }\href
  {https://doi.org/10.1142/S0218271815420146} {\bibfield  {journal} {\bibinfo
  {journal} {Int. J. Mod. Phys. D}\ }\textbf {\bibinfo {volume} {24}},\
  \bibinfo {pages} {1542014} (\bibinfo {year} {2015})},\ \Eprint
  {https://arxiv.org/abs/1504.08209} {arXiv:1504.08209 [gr-qc]} \BibitemShut
  {NoStop}%
\bibitem [{\citenamefont {Yagi}\ \emph {et~al.}(2012)\citenamefont {Yagi},
  \citenamefont {Stein}, \citenamefont {Yunes},\ and\ \citenamefont
  {Tanaka}}]{Yagi:2011xp}%
  \BibitemOpen
  \bibfield  {author} {\bibinfo {author} {\bibfnamefont {K.}~\bibnamefont
  {Yagi}}, \bibinfo {author} {\bibfnamefont {L.~C.}\ \bibnamefont {Stein}},
  \bibinfo {author} {\bibfnamefont {N.}~\bibnamefont {Yunes}},\ and\ \bibinfo
  {author} {\bibfnamefont {T.}~\bibnamefont {Tanaka}},\ }\bibfield  {title}
  {\bibinfo {title} {{Post-Newtonian, Quasi-Circular Binary Inspirals in
  Quadratic Modified Gravity}},\ }\href
  {https://doi.org/10.1103/PhysRevD.85.064022} {\bibfield  {journal} {\bibinfo
  {journal} {Phys. Rev. D}\ }\textbf {\bibinfo {volume} {85}},\ \bibinfo
  {pages} {064022} (\bibinfo {year} {2012})},\ \bibinfo {note} {[Erratum:
  Phys.Rev.D 93, 029902 (2016)]},\ \Eprint {https://arxiv.org/abs/1110.5950}
  {arXiv:1110.5950 [gr-qc]} \BibitemShut {NoStop}%
\bibitem [{\citenamefont {Kobayashi}(2019)}]{Kobayashi_2019}%
  \BibitemOpen
  \bibfield  {author} {\bibinfo {author} {\bibfnamefont {T.}~\bibnamefont
  {Kobayashi}},\ }\bibfield  {title} {\bibinfo {title} {{Horndeski theory and
  beyond: {A} review}},\ }\href {https://doi.org/10.1088/1361-6633/ab2429}
  {\bibfield  {journal} {\bibinfo  {journal} {Rept. Prog. Phys.}\ }\textbf
  {\bibinfo {volume} {82}},\ \bibinfo {pages} {086901} (\bibinfo {year}
  {2019})},\ \Eprint {https://arxiv.org/abs/1901.07183} {arXiv:1901.07183
  [gr-qc]} \BibitemShut {NoStop}%
\bibitem [{\citenamefont {East}\ and\ \citenamefont
  {Ripley}(2021)}]{East_2021}%
  \BibitemOpen
  \bibfield  {author} {\bibinfo {author} {\bibfnamefont {W.~E.}\ \bibnamefont
  {East}}\ and\ \bibinfo {author} {\bibfnamefont {J.~L.}\ \bibnamefont
  {Ripley}},\ }\bibfield  {title} {\bibinfo {title} {Evolution of
  {E}instein-scalar-{G}auss-{B}onnet gravity using a modified harmonic
  formulation},\ }\bibfield  {journal} {\bibinfo  {journal} {Physical Review
  D}\ }\textbf {\bibinfo {volume} {103}},\ \href
  {https://doi.org/10.1103/physrevd.103.044040} {10.1103/physrevd.103.044040}
  (\bibinfo {year} {2021})\BibitemShut {NoStop}%
\bibitem [{\citenamefont {Arest\'e~Sal\'o}\ \emph {et~al.}(2022)\citenamefont
  {Arest\'e~Sal\'o}, \citenamefont {Clough},\ and\ \citenamefont
  {Figueras}}]{AresteSalo:2022hua}%
  \BibitemOpen
  \bibfield  {author} {\bibinfo {author} {\bibfnamefont {L.}~\bibnamefont
  {Arest\'e~Sal\'o}}, \bibinfo {author} {\bibfnamefont {K.}~\bibnamefont
  {Clough}},\ and\ \bibinfo {author} {\bibfnamefont {P.}~\bibnamefont
  {Figueras}},\ }\bibfield  {title} {\bibinfo {title} {{Well-posedness of the
  four-derivative scalar-tensor theory of gravity in singularity avoiding
  coordinates}},\ }\href@noop {} {\  (\bibinfo {year} {2022})},\ \Eprint
  {https://arxiv.org/abs/2208.14470} {arXiv:2208.14470 [gr-qc]} \BibitemShut
  {NoStop}%
\bibitem [{\citenamefont {East}\ and\ \citenamefont
  {Pretorius}(2022)}]{East:2022rqi}%
  \BibitemOpen
  \bibfield  {author} {\bibinfo {author} {\bibfnamefont {W.~E.}\ \bibnamefont
  {East}}\ and\ \bibinfo {author} {\bibfnamefont {F.}~\bibnamefont
  {Pretorius}},\ }\bibfield  {title} {\bibinfo {title} {{Binary neutron star
  mergers in Einstein-scalar-Gauss-Bonnet gravity}},\ }\href
  {https://doi.org/10.1103/PhysRevD.106.104055} {\bibfield  {journal} {\bibinfo
   {journal} {Phys. Rev. D}\ }\textbf {\bibinfo {volume} {106}},\ \bibinfo
  {pages} {104055} (\bibinfo {year} {2022})},\ \Eprint
  {https://arxiv.org/abs/2208.09488} {arXiv:2208.09488 [gr-qc]} \BibitemShut
  {NoStop}%
\bibitem [{\citenamefont {Ripley}(2022)}]{Ripley:2022cdh}%
  \BibitemOpen
  \bibfield  {author} {\bibinfo {author} {\bibfnamefont {J.~L.}\ \bibnamefont
  {Ripley}},\ }\bibfield  {title} {\bibinfo {title} {{Numerical relativity for
  {H}orndeski gravity}},\ }\href {https://doi.org/10.1142/S0218271822300178}
  {\bibfield  {journal} {\bibinfo  {journal} {Int. J. Mod. Phys. D}\ }\textbf
  {\bibinfo {volume} {31}},\ \bibinfo {pages} {2230017} (\bibinfo {year}
  {2022})},\ \Eprint {https://arxiv.org/abs/2207.13074} {arXiv:2207.13074
  [gr-qc]} \BibitemShut {NoStop}%
\bibitem [{\citenamefont {Ripley}\ and\ \citenamefont
  {Pretorius}(2019{\natexlab{a}})}]{Ripley:2019hxt}%
  \BibitemOpen
  \bibfield  {author} {\bibinfo {author} {\bibfnamefont {J.~L.}\ \bibnamefont
  {Ripley}}\ and\ \bibinfo {author} {\bibfnamefont {F.}~\bibnamefont
  {Pretorius}},\ }\bibfield  {title} {\bibinfo {title} {{Hyperbolicity in
  Spherical Gravitational Collapse in a Horndeski Theory}},\ }\href
  {https://doi.org/10.1103/PhysRevD.99.084014} {\bibfield  {journal} {\bibinfo
  {journal} {Phys. Rev. D}\ }\textbf {\bibinfo {volume} {99}},\ \bibinfo
  {pages} {084014} (\bibinfo {year} {2019}{\natexlab{a}})},\ \Eprint
  {https://arxiv.org/abs/1902.01468} {arXiv:1902.01468 [gr-qc]} \BibitemShut
  {NoStop}%
\bibitem [{\citenamefont {Ripley}\ and\ \citenamefont
  {Pretorius}(2019{\natexlab{b}})}]{Ripley:2019irj}%
  \BibitemOpen
  \bibfield  {author} {\bibinfo {author} {\bibfnamefont {J.~L.}\ \bibnamefont
  {Ripley}}\ and\ \bibinfo {author} {\bibfnamefont {F.}~\bibnamefont
  {Pretorius}},\ }\bibfield  {title} {\bibinfo {title} {{Gravitational collapse
  in Einstein dilaton-Gauss\textendash{}Bonnet gravity}},\ }\href
  {https://doi.org/10.1088/1361-6382/ab2416} {\bibfield  {journal} {\bibinfo
  {journal} {Class. Quant. Grav.}\ }\textbf {\bibinfo {volume} {36}},\ \bibinfo
  {pages} {134001} (\bibinfo {year} {2019}{\natexlab{b}})},\ \Eprint
  {https://arxiv.org/abs/1903.07543} {arXiv:1903.07543 [gr-qc]} \BibitemShut
  {NoStop}%
\bibitem [{\citenamefont {R.}\ \emph {et~al.}(2022)\citenamefont {R.},
  \citenamefont {Ripley},\ and\ \citenamefont {Yunes}}]{R:2022hlf}%
  \BibitemOpen
  \bibfield  {author} {\bibinfo {author} {\bibfnamefont {A.~H.~K.}\
  \bibnamefont {R.}}, \bibinfo {author} {\bibfnamefont {J.~L.}\ \bibnamefont
  {Ripley}},\ and\ \bibinfo {author} {\bibfnamefont {N.}~\bibnamefont
  {Yunes}},\ }\bibfield  {title} {\bibinfo {title} {{Where and why does
  Einstein-Scalar-Gauss-Bonnet theory break down?}},\ }\href@noop {} {\
  (\bibinfo {year} {2022})},\ \Eprint {https://arxiv.org/abs/2211.08477}
  {arXiv:2211.08477 [gr-qc]} \BibitemShut {NoStop}%
\bibitem [{\citenamefont {Yunes}\ and\ \citenamefont
  {Siemens}(2013)}]{Yunes:2013dva}%
  \BibitemOpen
  \bibfield  {author} {\bibinfo {author} {\bibfnamefont {N.}~\bibnamefont
  {Yunes}}\ and\ \bibinfo {author} {\bibfnamefont {X.}~\bibnamefont
  {Siemens}},\ }\bibfield  {title} {\bibinfo {title} {{Gravitational-Wave Tests
  of General Relativity with Ground-Based Detectors and Pulsar
  Timing-Arrays}},\ }\href {https://doi.org/10.12942/lrr-2013-9} {\bibfield
  {journal} {\bibinfo  {journal} {Living Rev. Rel.}\ }\textbf {\bibinfo
  {volume} {16}},\ \bibinfo {pages} {9} (\bibinfo {year} {2013})},\ \Eprint
  {https://arxiv.org/abs/1304.3473} {arXiv:1304.3473 [gr-qc]} \BibitemShut
  {NoStop}%
\bibitem [{\citenamefont {Berti}\ \emph
  {et~al.}(2018{\natexlab{a}})\citenamefont {Berti}, \citenamefont {Yagi},\
  and\ \citenamefont {Yunes}}]{Berti:2018cxi}%
  \BibitemOpen
  \bibfield  {author} {\bibinfo {author} {\bibfnamefont {E.}~\bibnamefont
  {Berti}}, \bibinfo {author} {\bibfnamefont {K.}~\bibnamefont {Yagi}},\ and\
  \bibinfo {author} {\bibfnamefont {N.}~\bibnamefont {Yunes}},\ }\bibfield
  {title} {\bibinfo {title} {{Extreme Gravity Tests with Gravitational Waves
  from Compact Binary Coalescences: {(I)} {I}nspiral-Merger}},\ }\href
  {https://doi.org/10.1007/s10714-018-2362-8} {\bibfield  {journal} {\bibinfo
  {journal} {Gen. Rel. Grav.}\ }\textbf {\bibinfo {volume} {50}},\ \bibinfo
  {pages} {46} (\bibinfo {year} {2018}{\natexlab{a}})},\ \Eprint
  {https://arxiv.org/abs/1801.03208} {arXiv:1801.03208 [gr-qc]} \BibitemShut
  {NoStop}%
\bibitem [{\citenamefont {Berti}\ \emph
  {et~al.}(2018{\natexlab{b}})\citenamefont {Berti}, \citenamefont {Yagi},
  \citenamefont {Yang},\ and\ \citenamefont {Yunes}}]{Berti:2018vdi}%
  \BibitemOpen
  \bibfield  {author} {\bibinfo {author} {\bibfnamefont {E.}~\bibnamefont
  {Berti}}, \bibinfo {author} {\bibfnamefont {K.}~\bibnamefont {Yagi}},
  \bibinfo {author} {\bibfnamefont {H.}~\bibnamefont {Yang}},\ and\ \bibinfo
  {author} {\bibfnamefont {N.}~\bibnamefont {Yunes}},\ }\bibfield  {title}
  {\bibinfo {title} {{Extreme Gravity Tests with Gravitational Waves from
  Compact Binary Coalescences: {(II} {R}ingdown}},\ }\href
  {https://doi.org/10.1007/s10714-018-2372-6} {\bibfield  {journal} {\bibinfo
  {journal} {Gen. Rel. Grav.}\ }\textbf {\bibinfo {volume} {50}},\ \bibinfo
  {pages} {49} (\bibinfo {year} {2018}{\natexlab{b}})},\ \Eprint
  {https://arxiv.org/abs/1801.03587} {arXiv:1801.03587 [gr-qc]} \BibitemShut
  {NoStop}%
\bibitem [{\citenamefont {Shiralilou}\ \emph
  {et~al.}(2021{\natexlab{a}})\citenamefont {Shiralilou}, \citenamefont
  {Hinderer}, \citenamefont {Nissanke}, \citenamefont {Ortiz},\ and\
  \citenamefont {Witek}}]{Shiralilou:2020gah}%
  \BibitemOpen
  \bibfield  {author} {\bibinfo {author} {\bibfnamefont {B.}~\bibnamefont
  {Shiralilou}}, \bibinfo {author} {\bibfnamefont {T.}~\bibnamefont
  {Hinderer}}, \bibinfo {author} {\bibfnamefont {S.}~\bibnamefont {Nissanke}},
  \bibinfo {author} {\bibfnamefont {N.}~\bibnamefont {Ortiz}},\ and\ \bibinfo
  {author} {\bibfnamefont {H.}~\bibnamefont {Witek}},\ }\bibfield  {title}
  {\bibinfo {title} {{Nonlinear curvature effects in gravitational waves from
  inspiralling black hole binaries}},\ }\href
  {https://doi.org/10.1103/PhysRevD.103.L121503} {\bibfield  {journal}
  {\bibinfo  {journal} {Phys. Rev. D}\ }\textbf {\bibinfo {volume} {103}},\
  \bibinfo {pages} {L121503} (\bibinfo {year} {2021}{\natexlab{a}})},\ \Eprint
  {https://arxiv.org/abs/2012.09162} {arXiv:2012.09162 [gr-qc]} \BibitemShut
  {NoStop}%
\bibitem [{\citenamefont {Wagle}\ \emph {et~al.}(2019)\citenamefont {Wagle},
  \citenamefont {Saffer},\ and\ \citenamefont
  {Yunes}}]{Wagle:2019polarisation}%
  \BibitemOpen
  \bibfield  {author} {\bibinfo {author} {\bibfnamefont {P.}~\bibnamefont
  {Wagle}}, \bibinfo {author} {\bibfnamefont {A.}~\bibnamefont {Saffer}},\ and\
  \bibinfo {author} {\bibfnamefont {N.}~\bibnamefont {Yunes}},\ }\bibfield
  {title} {\bibinfo {title} {Polarization modes of gravitational waves in
  quadratic gravity},\ }\bibfield  {journal} {\bibinfo  {journal} {Physical
  Review D}\ }\textbf {\bibinfo {volume} {100}},\ \href
  {https://doi.org/10.1103/physrevd.100.124007} {10.1103/physrevd.100.124007}
  (\bibinfo {year} {2019})\BibitemShut {NoStop}%
\bibitem [{\citenamefont {Isi}\ \emph {et~al.}(2015)\citenamefont {Isi},
  \citenamefont {Weinstein}, \citenamefont {Mead},\ and\ \citenamefont
  {Pitkin}}]{Isi:2015cva}%
  \BibitemOpen
  \bibfield  {author} {\bibinfo {author} {\bibfnamefont {M.}~\bibnamefont
  {Isi}}, \bibinfo {author} {\bibfnamefont {A.~J.}\ \bibnamefont {Weinstein}},
  \bibinfo {author} {\bibfnamefont {C.}~\bibnamefont {Mead}},\ and\ \bibinfo
  {author} {\bibfnamefont {M.}~\bibnamefont {Pitkin}},\ }\bibfield  {title}
  {\bibinfo {title} {{Detecting Beyond-{E}instein Polarizations of Continuous
  Gravitational Waves}},\ }\href {https://doi.org/10.1103/PhysRevD.91.082002}
  {\bibfield  {journal} {\bibinfo  {journal} {Phys. Rev. D}\ }\textbf {\bibinfo
  {volume} {91}},\ \bibinfo {pages} {082002} (\bibinfo {year} {2015})},\
  \Eprint {https://arxiv.org/abs/1502.00333} {arXiv:1502.00333 [gr-qc]}
  \BibitemShut {NoStop}%
\bibitem [{\citenamefont {Isi}\ \emph {et~al.}(2017)\citenamefont {Isi},
  \citenamefont {Pitkin},\ and\ \citenamefont {Weinstein}}]{Isi:2017equ}%
  \BibitemOpen
  \bibfield  {author} {\bibinfo {author} {\bibfnamefont {M.}~\bibnamefont
  {Isi}}, \bibinfo {author} {\bibfnamefont {M.}~\bibnamefont {Pitkin}},\ and\
  \bibinfo {author} {\bibfnamefont {A.~J.}\ \bibnamefont {Weinstein}},\
  }\bibfield  {title} {\bibinfo {title} {{Probing Dynamical Gravity with the
  Polarization of Continuous Gravitational Waves}},\ }\href
  {https://doi.org/10.1103/PhysRevD.96.042001} {\bibfield  {journal} {\bibinfo
  {journal} {Phys. Rev. D}\ }\textbf {\bibinfo {volume} {96}},\ \bibinfo
  {pages} {042001} (\bibinfo {year} {2017})},\ \Eprint
  {https://arxiv.org/abs/1703.07530} {arXiv:1703.07530 [gr-qc]} \BibitemShut
  {NoStop}%
\bibitem [{\citenamefont {Isi}\ and\ \citenamefont
  {Weinstein}(2017)}]{Isi:2017fbj}%
  \BibitemOpen
  \bibfield  {author} {\bibinfo {author} {\bibfnamefont {M.}~\bibnamefont
  {Isi}}\ and\ \bibinfo {author} {\bibfnamefont {A.~J.}\ \bibnamefont
  {Weinstein}},\ }\bibfield  {title} {\bibinfo {title} {{Probing gravitational
  wave polarizations with signals from compact binary coalescences}},\
  }\href@noop {} {\  (\bibinfo {year} {2017})},\ \Eprint
  {https://arxiv.org/abs/1710.03794} {arXiv:1710.03794 [gr-qc]} \BibitemShut
  {NoStop}%
\bibitem [{\citenamefont {Isi}(2022)}]{Isi:2022mbx}%
  \BibitemOpen
  \bibfield  {author} {\bibinfo {author} {\bibfnamefont {M.}~\bibnamefont
  {Isi}},\ }\bibfield  {title} {\bibinfo {title} {{Parametrizing
  gravitational-wave polarizations}},\ }\href@noop {} {\  (\bibinfo {year}
  {2022})},\ \Eprint {https://arxiv.org/abs/2208.03372} {arXiv:2208.03372
  [gr-qc]} \BibitemShut {NoStop}%
\bibitem [{\citenamefont {Isi}\ and\ \citenamefont
  {Stein}(2018)}]{Isi:2018miq}%
  \BibitemOpen
  \bibfield  {author} {\bibinfo {author} {\bibfnamefont {M.}~\bibnamefont
  {Isi}}\ and\ \bibinfo {author} {\bibfnamefont {L.~C.}\ \bibnamefont
  {Stein}},\ }\bibfield  {title} {\bibinfo {title} {{Measuring stochastic
  gravitational-wave energy beyond general relativity}},\ }\href
  {https://doi.org/10.1103/PhysRevD.98.104025} {\bibfield  {journal} {\bibinfo
  {journal} {Phys. Rev. D}\ }\textbf {\bibinfo {volume} {98}},\ \bibinfo
  {pages} {104025} (\bibinfo {year} {2018})},\ \Eprint
  {https://arxiv.org/abs/1807.02123} {arXiv:1807.02123 [gr-qc]} \BibitemShut
  {NoStop}%
\bibitem [{\citenamefont {Joyce}\ \emph {et~al.}(2015)\citenamefont {Joyce},
  \citenamefont {Jain}, \citenamefont {Khoury},\ and\ \citenamefont
  {Trodden}}]{Joyce:2014kja}%
  \BibitemOpen
  \bibfield  {author} {\bibinfo {author} {\bibfnamefont {A.}~\bibnamefont
  {Joyce}}, \bibinfo {author} {\bibfnamefont {B.}~\bibnamefont {Jain}},
  \bibinfo {author} {\bibfnamefont {J.}~\bibnamefont {Khoury}},\ and\ \bibinfo
  {author} {\bibfnamefont {M.}~\bibnamefont {Trodden}},\ }\bibfield  {title}
  {\bibinfo {title} {{Beyond the Cosmological {S}tandard {M}odel}},\ }\href
  {https://doi.org/10.1016/j.physrep.2014.12.002} {\bibfield  {journal}
  {\bibinfo  {journal} {Phys. Rept.}\ }\textbf {\bibinfo {volume} {568}},\
  \bibinfo {pages} {1} (\bibinfo {year} {2015})},\ \Eprint
  {https://arxiv.org/abs/1407.0059} {arXiv:1407.0059 [astro-ph.CO]}
  \BibitemShut {NoStop}%
\bibitem [{\citenamefont {Berti}\ \emph {et~al.}(2015)\citenamefont {Berti}
  \emph {et~al.}}]{Berti_2015}%
  \BibitemOpen
  \bibfield  {author} {\bibinfo {author} {\bibfnamefont {E.}~\bibnamefont
  {Berti}} \emph {et~al.},\ }\bibfield  {title} {\bibinfo {title} {{Testing
  General Relativity with Present and Future Astrophysical Observations}},\
  }\href {https://doi.org/10.1088/0264-9381/32/24/243001} {\bibfield  {journal}
  {\bibinfo  {journal} {Class. Quant. Grav.}\ }\textbf {\bibinfo {volume}
  {32}},\ \bibinfo {pages} {243001} (\bibinfo {year} {2015})},\ \Eprint
  {https://arxiv.org/abs/1501.07274} {arXiv:1501.07274 [gr-qc]} \BibitemShut
  {NoStop}%
\bibitem [{\citenamefont {Eardley}\ \emph {et~al.}(1973)\citenamefont
  {Eardley}, \citenamefont {Lee},\ and\ \citenamefont
  {Lightman}}]{1973PhRvL..30..884E}%
  \BibitemOpen
  \bibfield  {author} {\bibinfo {author} {\bibfnamefont {D.~M.}\ \bibnamefont
  {Eardley}}, \bibinfo {author} {\bibfnamefont {D.~L.}\ \bibnamefont {Lee}},\
  and\ \bibinfo {author} {\bibfnamefont {A.~P.}\ \bibnamefont {Lightman}},\
  }\bibfield  {title} {\bibinfo {title} {{Gravitational-wave observations as a
  tool for testing relativistic gravity}},\ }\href
  {https://doi.org/10.1103/PhysRevD.8.3308} {\bibfield  {journal} {\bibinfo
  {journal} {Phys. Rev. D}\ }\textbf {\bibinfo {volume} {8}},\ \bibinfo {pages}
  {3308} (\bibinfo {year} {1973})}\BibitemShut {NoStop}%
\bibitem [{\citenamefont {Chatziioannou}\ \emph {et~al.}(2012)\citenamefont
  {Chatziioannou}, \citenamefont {Yunes},\ and\ \citenamefont
  {Cornish}}]{Chatziioannou:2012rf}%
  \BibitemOpen
  \bibfield  {author} {\bibinfo {author} {\bibfnamefont {K.}~\bibnamefont
  {Chatziioannou}}, \bibinfo {author} {\bibfnamefont {N.}~\bibnamefont
  {Yunes}},\ and\ \bibinfo {author} {\bibfnamefont {N.}~\bibnamefont
  {Cornish}},\ }\bibfield  {title} {\bibinfo {title} {{Model-Independent Test
  of General Relativity: {A}n Extended post-Einsteinian Framework with Complete
  Polarization Content}},\ }\href {https://doi.org/10.1103/PhysRevD.86.022004}
  {\bibfield  {journal} {\bibinfo  {journal} {Phys. Rev. D}\ }\textbf {\bibinfo
  {volume} {86}},\ \bibinfo {pages} {022004} (\bibinfo {year} {2012})},\
  \bibinfo {note} {[Erratum: Phys.Rev.D 95, 129901 (2017)]},\ \Eprint
  {https://arxiv.org/abs/1204.2585} {arXiv:1204.2585 [gr-qc]} \BibitemShut
  {NoStop}%
\bibitem [{\citenamefont {Hou}\ \emph {et~al.}(2018)\citenamefont {Hou},
  \citenamefont {Gong},\ and\ \citenamefont {Liu}}]{Hou:2018hordenski}%
  \BibitemOpen
  \bibfield  {author} {\bibinfo {author} {\bibfnamefont {S.}~\bibnamefont
  {Hou}}, \bibinfo {author} {\bibfnamefont {Y.}~\bibnamefont {Gong}},\ and\
  \bibinfo {author} {\bibfnamefont {Y.}~\bibnamefont {Liu}},\ }\bibfield
  {title} {\bibinfo {title} {Polarizations of gravitational waves in
  {H}orndeski theory},\ }\bibfield  {journal} {\bibinfo  {journal} {The
  European Physical Journal C}\ }\textbf {\bibinfo {volume} {78}},\ \href
  {https://doi.org/10.1140/epjc/s10052-018-5869-y}
  {10.1140/epjc/s10052-018-5869-y} (\bibinfo {year} {2018})\BibitemShut
  {NoStop}%
\bibitem [{\citenamefont {Lee}(1974)}]{PhysRevD.10.2374}%
  \BibitemOpen
  \bibfield  {author} {\bibinfo {author} {\bibfnamefont {D.~L.}\ \bibnamefont
  {Lee}},\ }\bibfield  {title} {\bibinfo {title} {Conservation laws,
  gravitational waves, and mass losses in the {D}icke-{B}rans-{J}ordan theory
  of gravity},\ }\href {https://doi.org/10.1103/PhysRevD.10.2374} {\bibfield
  {journal} {\bibinfo  {journal} {Phys. Rev. D}\ }\textbf {\bibinfo {volume}
  {10}},\ \bibinfo {pages} {2374} (\bibinfo {year} {1974})}\BibitemShut
  {NoStop}%
\bibitem [{\citenamefont {Okounkova}\ \emph {et~al.}(2017)\citenamefont
  {Okounkova}, \citenamefont {Stein}, \citenamefont {Scheel},\ and\
  \citenamefont {Hemberger}}]{Okounkova:2017yby}%
  \BibitemOpen
  \bibfield  {author} {\bibinfo {author} {\bibfnamefont {M.}~\bibnamefont
  {Okounkova}}, \bibinfo {author} {\bibfnamefont {L.~C.}\ \bibnamefont
  {Stein}}, \bibinfo {author} {\bibfnamefont {M.~A.}\ \bibnamefont {Scheel}},\
  and\ \bibinfo {author} {\bibfnamefont {D.~A.}\ \bibnamefont {Hemberger}},\
  }\bibfield  {title} {\bibinfo {title} {{Numerical binary black hole mergers
  in dynamical {C}hern-{S}imons gravity: {S}calar field}},\ }\href
  {https://doi.org/10.1103/PhysRevD.96.044020} {\bibfield  {journal} {\bibinfo
  {journal} {Phys. Rev. D}\ }\textbf {\bibinfo {volume} {96}},\ \bibinfo
  {pages} {044020} (\bibinfo {year} {2017})},\ \Eprint
  {https://arxiv.org/abs/1705.07924} {arXiv:1705.07924 [gr-qc]} \BibitemShut
  {NoStop}%
\bibitem [{\citenamefont {Witek}\ \emph
  {et~al.}(2019{\natexlab{a}})\citenamefont {Witek}, \citenamefont {Gualtieri},
  \citenamefont {Pani},\ and\ \citenamefont {Sotiriou}}]{Witek:2018dmd}%
  \BibitemOpen
  \bibfield  {author} {\bibinfo {author} {\bibfnamefont {H.}~\bibnamefont
  {Witek}}, \bibinfo {author} {\bibfnamefont {L.}~\bibnamefont {Gualtieri}},
  \bibinfo {author} {\bibfnamefont {P.}~\bibnamefont {Pani}},\ and\ \bibinfo
  {author} {\bibfnamefont {T.~P.}\ \bibnamefont {Sotiriou}},\ }\bibfield
  {title} {\bibinfo {title} {{Black holes and binary mergers in scalar
  {G}auss-{B}onnet gravity: {S}calar field dynamics}},\ }\href
  {https://doi.org/10.1103/PhysRevD.99.064035} {\bibfield  {journal} {\bibinfo
  {journal} {Phys. Rev. D}\ }\textbf {\bibinfo {volume} {99}},\ \bibinfo
  {pages} {064035} (\bibinfo {year} {2019}{\natexlab{a}})},\ \Eprint
  {https://arxiv.org/abs/1810.05177} {arXiv:1810.05177 [gr-qc]} \BibitemShut
  {NoStop}%
\bibitem [{\citenamefont {Hawking}(1972)}]{cmp/1103857885}%
  \BibitemOpen
  \bibfield  {author} {\bibinfo {author} {\bibfnamefont {S.~W.}\ \bibnamefont
  {Hawking}},\ }\bibfield  {title} {\bibinfo {title} {{Black holes in the
  {B}rans-{D}icke theory of gravitation}},\ }\href
  {https://doi.org/cmp/1103857885} {\bibfield  {journal} {\bibinfo  {journal}
  {Communications in Mathematical Physics}\ }\textbf {\bibinfo {volume} {25}},\
  \bibinfo {pages} {167 } (\bibinfo {year} {1972})}\BibitemShut {NoStop}%
\bibitem [{\citenamefont {Abbott}\ \emph
  {et~al.}(2017{\natexlab{a}})\citenamefont {Abbott} \emph
  {et~al.}}]{LIGOScientific:2017zic}%
  \BibitemOpen
  \bibfield  {author} {\bibinfo {author} {\bibfnamefont {B.~P.}\ \bibnamefont
  {Abbott}} \emph {et~al.} (\bibinfo {collaboration} {LIGO Scientific, Virgo,
  Fermi-GBM, INTEGRAL}),\ }\bibfield  {title} {\bibinfo {title} {{Gravitational
  Waves and Gamma-rays from a Binary Neutron Star Merger: GW170817 and GRB
  170817A}},\ }\href {https://doi.org/10.3847/2041-8213/aa920c} {\bibfield
  {journal} {\bibinfo  {journal} {Astrophys. J. Lett.}\ }\textbf {\bibinfo
  {volume} {848}},\ \bibinfo {pages} {L13} (\bibinfo {year}
  {2017}{\natexlab{a}})},\ \Eprint {https://arxiv.org/abs/1710.05834}
  {arXiv:1710.05834 [astro-ph.HE]} \BibitemShut {NoStop}%
\bibitem [{\citenamefont {Okounkova}(2020)}]{Okounkova:2020rqw}%
  \BibitemOpen
  \bibfield  {author} {\bibinfo {author} {\bibfnamefont {M.}~\bibnamefont
  {Okounkova}},\ }\bibfield  {title} {\bibinfo {title} {{Numerical relativity
  simulation of {GW}150914 in {E}instein dilaton {G}auss-{B}onnet gravity}},\
  }\href {https://doi.org/10.1103/PhysRevD.102.084046} {\bibfield  {journal}
  {\bibinfo  {journal} {Phys. Rev. D}\ }\textbf {\bibinfo {volume} {102}},\
  \bibinfo {pages} {084046} (\bibinfo {year} {2020})},\ \Eprint
  {https://arxiv.org/abs/2001.03571} {arXiv:2001.03571 [gr-qc]} \BibitemShut
  {NoStop}%
\bibitem [{\citenamefont {Corman}\ \emph {et~al.}(2022)\citenamefont {Corman},
  \citenamefont {Ripley},\ and\ \citenamefont {East}}]{Corman:2022xqg}%
  \BibitemOpen
  \bibfield  {author} {\bibinfo {author} {\bibfnamefont {M.}~\bibnamefont
  {Corman}}, \bibinfo {author} {\bibfnamefont {J.~L.}\ \bibnamefont {Ripley}},\
  and\ \bibinfo {author} {\bibfnamefont {W.~E.}\ \bibnamefont {East}},\
  }\bibfield  {title} {\bibinfo {title} {{Nonlinear studies of binary black
  hole mergers in {E}instein-scalar-{G}auss-{B}onnet gravity}},\ }\href@noop {}
  {\  (\bibinfo {year} {2022})},\ \Eprint {https://arxiv.org/abs/2210.09235}
  {arXiv:2210.09235 [gr-qc]} \BibitemShut {NoStop}%
\bibitem [{\citenamefont {Ripley}\ and\ \citenamefont
  {Pretorius}(2020)}]{Ripley:2019aqj}%
  \BibitemOpen
  \bibfield  {author} {\bibinfo {author} {\bibfnamefont {J.~L.}\ \bibnamefont
  {Ripley}}\ and\ \bibinfo {author} {\bibfnamefont {F.}~\bibnamefont
  {Pretorius}},\ }\bibfield  {title} {\bibinfo {title} {{Scalarized Black Hole
  dynamics in {E}instein dilaton {G}auss-{B}onnet Gravity}},\ }\href
  {https://doi.org/10.1103/PhysRevD.101.044015} {\bibfield  {journal} {\bibinfo
   {journal} {Phys. Rev. D}\ }\textbf {\bibinfo {volume} {101}},\ \bibinfo
  {pages} {044015} (\bibinfo {year} {2020})},\ \Eprint
  {https://arxiv.org/abs/1911.11027} {arXiv:1911.11027 [gr-qc]} \BibitemShut
  {NoStop}%
\bibitem [{\citenamefont {Lyu}\ \emph {et~al.}(2022{\natexlab{b}})\citenamefont
  {Lyu}, \citenamefont {Jiang},\ and\ \citenamefont {Yagi}}]{Lyu_2022}%
  \BibitemOpen
  \bibfield  {author} {\bibinfo {author} {\bibfnamefont {Z.}~\bibnamefont
  {Lyu}}, \bibinfo {author} {\bibfnamefont {N.}~\bibnamefont {Jiang}},\ and\
  \bibinfo {author} {\bibfnamefont {K.}~\bibnamefont {Yagi}},\ }\bibfield
  {title} {\bibinfo {title} {Constraints on einstein-dilation-gauss-bonnet
  gravity from black hole-neutron star gravitational wave events},\ }\bibfield
  {journal} {\bibinfo  {journal} {Physical Review D}\ }\textbf {\bibinfo
  {volume} {105}},\ \href {https://doi.org/10.1103/physrevd.105.064001}
  {10.1103/physrevd.105.064001} (\bibinfo {year}
  {2022}{\natexlab{b}})\BibitemShut {NoStop}%
\bibitem [{\citenamefont {Amendola}\ \emph {et~al.}(2007)\citenamefont
  {Amendola}, \citenamefont {Charmousis},\ and\ \citenamefont
  {Davis}}]{Amendola:2007ni}%
  \BibitemOpen
  \bibfield  {author} {\bibinfo {author} {\bibfnamefont {L.}~\bibnamefont
  {Amendola}}, \bibinfo {author} {\bibfnamefont {C.}~\bibnamefont
  {Charmousis}},\ and\ \bibinfo {author} {\bibfnamefont {S.~C.}\ \bibnamefont
  {Davis}},\ }\bibfield  {title} {\bibinfo {title} {{Solar System Constraints
  on Gauss-Bonnet Mediated Dark Energy}},\ }\href
  {https://doi.org/10.1088/1475-7516/2007/10/004} {\bibfield  {journal}
  {\bibinfo  {journal} {JCAP}\ }\textbf {\bibinfo {volume} {10}},\ \bibinfo
  {pages} {004}},\ \Eprint {https://arxiv.org/abs/0704.0175} {arXiv:0704.0175
  [astro-ph]} \BibitemShut {NoStop}%
\bibitem [{\citenamefont {{Reasenberg}}\ \emph {et~al.}(1979)\citenamefont
  {{Reasenberg}}, \citenamefont {{Shapiro}}, \citenamefont {{MacNeil}},
  \citenamefont {{Goldstein}}, \citenamefont {{Breidenthal}}, \citenamefont
  {{Brenkle}}, \citenamefont {{Cain}}, \citenamefont {{Kaufman}}, \citenamefont
  {{Komarek}},\ and\ \citenamefont {{Zygielbaum}}}]{1979ApJ...234L.219R}%
  \BibitemOpen
  \bibfield  {author} {\bibinfo {author} {\bibfnamefont {R.~D.}\ \bibnamefont
  {{Reasenberg}}}, \bibinfo {author} {\bibfnamefont {I.~I.}\ \bibnamefont
  {{Shapiro}}}, \bibinfo {author} {\bibfnamefont {P.~E.}\ \bibnamefont
  {{MacNeil}}}, \bibinfo {author} {\bibfnamefont {R.~B.}\ \bibnamefont
  {{Goldstein}}}, \bibinfo {author} {\bibfnamefont {J.~C.}\ \bibnamefont
  {{Breidenthal}}}, \bibinfo {author} {\bibfnamefont {J.~P.}\ \bibnamefont
  {{Brenkle}}}, \bibinfo {author} {\bibfnamefont {D.~L.}\ \bibnamefont
  {{Cain}}}, \bibinfo {author} {\bibfnamefont {T.~M.}\ \bibnamefont
  {{Kaufman}}}, \bibinfo {author} {\bibfnamefont {T.~A.}\ \bibnamefont
  {{Komarek}}},\ and\ \bibinfo {author} {\bibfnamefont {A.~I.}\ \bibnamefont
  {{Zygielbaum}}},\ }\bibfield  {title} {\bibinfo {title} {{Viking relativity
  experiment - Verification of signal retardation by solar gravity}},\ }\href
  {https://doi.org/10.1086/183144} {\bibfield  {journal} {\bibinfo  {journal}
  {ApJl}\ }\textbf {\bibinfo {volume} {234}},\ \bibinfo {pages} {L219}
  (\bibinfo {year} {1979})}\BibitemShut {NoStop}%
\bibitem [{\citenamefont {Bertotti}\ \emph {et~al.}(2003)\citenamefont
  {Bertotti}, \citenamefont {Iess},\ and\ \citenamefont
  {Tortora}}]{Bertotti_2022}%
  \BibitemOpen
  \bibfield  {author} {\bibinfo {author} {\bibfnamefont {B.}~\bibnamefont
  {Bertotti}}, \bibinfo {author} {\bibfnamefont {L.}~\bibnamefont {Iess}},\
  and\ \bibinfo {author} {\bibfnamefont {P.}~\bibnamefont {Tortora}},\
  }\bibfield  {title} {\bibinfo {title} {{A test of general relativity using
  radio links with the {C}assini spacecraft}},\ }\href
  {https://doi.org/10.1038/nature01997} {\bibfield  {journal} {\bibinfo
  {journal} {Nature}\ }\textbf {\bibinfo {volume} {425}},\ \bibinfo {pages}
  {374} (\bibinfo {year} {2003})}\BibitemShut {NoStop}%
\bibitem [{\citenamefont {Wang}\ \emph {et~al.}(2021)\citenamefont {Wang},
  \citenamefont {Tang}, \citenamefont {Li}, \citenamefont {Han},\ and\
  \citenamefont {Fan}}]{Wang:2021jfc}%
  \BibitemOpen
  \bibfield  {author} {\bibinfo {author} {\bibfnamefont {H.-T.}\ \bibnamefont
  {Wang}}, \bibinfo {author} {\bibfnamefont {S.-P.}\ \bibnamefont {Tang}},
  \bibinfo {author} {\bibfnamefont {P.-C.}\ \bibnamefont {Li}}, \bibinfo
  {author} {\bibfnamefont {M.-Z.}\ \bibnamefont {Han}},\ and\ \bibinfo {author}
  {\bibfnamefont {Y.-Z.}\ \bibnamefont {Fan}},\ }\bibfield  {title} {\bibinfo
  {title} {{Tight constraints on Einstein-dilation-Gauss-Bonnet gravity from
  GW190412 and GW190814}},\ }\href
  {https://doi.org/10.1103/PhysRevD.104.024015} {\bibfield  {journal} {\bibinfo
   {journal} {Phys. Rev. D}\ }\textbf {\bibinfo {volume} {104}},\ \bibinfo
  {pages} {024015} (\bibinfo {year} {2021})},\ \Eprint
  {https://arxiv.org/abs/2104.07590} {arXiv:2104.07590 [gr-qc]} \BibitemShut
  {NoStop}%
\bibitem [{\citenamefont {Tahura}\ \emph {et~al.}(2019)\citenamefont {Tahura},
  \citenamefont {Yagi},\ and\ \citenamefont {Carson}}]{Shammi_2019}%
  \BibitemOpen
  \bibfield  {author} {\bibinfo {author} {\bibfnamefont {S.}~\bibnamefont
  {Tahura}}, \bibinfo {author} {\bibfnamefont {K.}~\bibnamefont {Yagi}},\ and\
  \bibinfo {author} {\bibfnamefont {Z.}~\bibnamefont {Carson}},\ }\bibfield
  {title} {\bibinfo {title} {Testing gravity with gravitational waves from
  binary black hole mergers: {C}ontributions from amplitude corrections},\
  }\href {https://doi.org/10.1103/PhysRevD.100.104001} {\bibfield  {journal}
  {\bibinfo  {journal} {Phys. Rev. D}\ }\textbf {\bibinfo {volume} {100}},\
  \bibinfo {pages} {104001} (\bibinfo {year} {2019})}\BibitemShut {NoStop}%
\bibitem [{\citenamefont {Carullo}(2021)}]{Carullo:2021dui}%
  \BibitemOpen
  \bibfield  {author} {\bibinfo {author} {\bibfnamefont {G.}~\bibnamefont
  {Carullo}},\ }\bibfield  {title} {\bibinfo {title} {{Enhancing modified
  gravity detection from gravitational-wave observations using the parametrized
  ringdown spin expansion coeffcients formalism}},\ }\href
  {https://doi.org/10.1103/PhysRevD.103.124043} {\bibfield  {journal} {\bibinfo
   {journal} {Phys. Rev. D}\ }\textbf {\bibinfo {volume} {103}},\ \bibinfo
  {pages} {124043} (\bibinfo {year} {2021})},\ \Eprint
  {https://arxiv.org/abs/2102.05939} {arXiv:2102.05939 [gr-qc]} \BibitemShut
  {NoStop}%
\bibitem [{\citenamefont {Silva}\ \emph {et~al.}(2022)\citenamefont {Silva},
  \citenamefont {Ghosh},\ and\ \citenamefont {Buonanno}}]{Silva:2022srr}%
  \BibitemOpen
  \bibfield  {author} {\bibinfo {author} {\bibfnamefont {H.~O.}\ \bibnamefont
  {Silva}}, \bibinfo {author} {\bibfnamefont {A.}~\bibnamefont {Ghosh}},\ and\
  \bibinfo {author} {\bibfnamefont {A.}~\bibnamefont {Buonanno}},\ }\bibfield
  {title} {\bibinfo {title} {{Black-hole ringdown as a probe of
  higher-curvature gravity theories}},\ }\href@noop {} {\  (\bibinfo {year}
  {2022})},\ \Eprint {https://arxiv.org/abs/2205.05132} {arXiv:2205.05132
  [gr-qc]} \BibitemShut {NoStop}%
\bibitem [{\citenamefont {Yagi}(2012)}]{Yagi_2012}%
  \BibitemOpen
  \bibfield  {author} {\bibinfo {author} {\bibfnamefont {K.}~\bibnamefont
  {Yagi}},\ }\bibfield  {title} {\bibinfo {title} {New constraint on scalar
  {G}auss-{B}onnet gravity and a possible explanation for the excess of the
  orbital decay rate in a low-mass {X}-ray binary},\ }\href
  {https://doi.org/10.1103/PhysRevD.86.081504} {\bibfield  {journal} {\bibinfo
  {journal} {Phys. Rev. D}\ }\textbf {\bibinfo {volume} {86}},\ \bibinfo
  {pages} {081504} (\bibinfo {year} {2012})}\BibitemShut {NoStop}%
\bibitem [{\citenamefont {Stein}\ and\ \citenamefont
  {Yagi}(2014)}]{Stein_2014}%
  \BibitemOpen
  \bibfield  {author} {\bibinfo {author} {\bibfnamefont {L.~C.}\ \bibnamefont
  {Stein}}\ and\ \bibinfo {author} {\bibfnamefont {K.}~\bibnamefont {Yagi}},\
  }\bibfield  {title} {\bibinfo {title} {Parametrizing and constraining scalar
  corrections to general relativity},\ }\href
  {https://doi.org/10.1103/PhysRevD.89.044026} {\bibfield  {journal} {\bibinfo
  {journal} {Phys. Rev. D}\ }\textbf {\bibinfo {volume} {89}},\ \bibinfo
  {pages} {044026} (\bibinfo {year} {2014})}\BibitemShut {NoStop}%
\bibitem [{\citenamefont {Pani}\ \emph {et~al.}(2011)\citenamefont {Pani},
  \citenamefont {Berti}, \citenamefont {Cardoso},\ and\ \citenamefont
  {Read}}]{Pani_2011}%
  \BibitemOpen
  \bibfield  {author} {\bibinfo {author} {\bibfnamefont {P.}~\bibnamefont
  {Pani}}, \bibinfo {author} {\bibfnamefont {E.}~\bibnamefont {Berti}},
  \bibinfo {author} {\bibfnamefont {V.}~\bibnamefont {Cardoso}},\ and\ \bibinfo
  {author} {\bibfnamefont {J.}~\bibnamefont {Read}},\ }\bibfield  {title}
  {\bibinfo {title} {Compact stars in alternative theories of gravity:
  {E}instein-{D}ilaton-{G}auss-{B}onnet gravity},\ }\href
  {https://doi.org/10.1103/PhysRevD.84.104035} {\bibfield  {journal} {\bibinfo
  {journal} {Phys. Rev. D}\ }\textbf {\bibinfo {volume} {84}},\ \bibinfo
  {pages} {104035} (\bibinfo {year} {2011})}\BibitemShut {NoStop}%
\bibitem [{\citenamefont {Witek}\ \emph
  {et~al.}(2019{\natexlab{b}})\citenamefont {Witek}, \citenamefont {Gualtieri},
  \citenamefont {Pani},\ and\ \citenamefont {Sotiriou}}]{Witek_2019}%
  \BibitemOpen
  \bibfield  {author} {\bibinfo {author} {\bibfnamefont {H.}~\bibnamefont
  {Witek}}, \bibinfo {author} {\bibfnamefont {L.}~\bibnamefont {Gualtieri}},
  \bibinfo {author} {\bibfnamefont {P.}~\bibnamefont {Pani}},\ and\ \bibinfo
  {author} {\bibfnamefont {T.~P.}\ \bibnamefont {Sotiriou}},\ }\bibfield
  {title} {\bibinfo {title} {Black holes and binary mergers in scalar
  {G}auss-{B}onnet gravity: {S}calar field dynamics},\ }\href
  {https://doi.org/10.1103/PhysRevD.99.064035} {\bibfield  {journal} {\bibinfo
  {journal} {Phys. Rev. D}\ }\textbf {\bibinfo {volume} {99}},\ \bibinfo
  {pages} {064035} (\bibinfo {year} {2019}{\natexlab{b}})}\BibitemShut
  {NoStop}%
\bibitem [{\citenamefont {Damour}\ and\ \citenamefont
  {Esposito-Far\`ese}(1993)}]{PhysRevLett.70.2220}%
  \BibitemOpen
  \bibfield  {author} {\bibinfo {author} {\bibfnamefont {T.}~\bibnamefont
  {Damour}}\ and\ \bibinfo {author} {\bibfnamefont {G.}~\bibnamefont
  {Esposito-Far\`ese}},\ }\bibfield  {title} {\bibinfo {title} {Nonperturbative
  strong-field effects in tensor-scalar theories of gravitation},\ }\href
  {https://doi.org/10.1103/PhysRevLett.70.2220} {\bibfield  {journal} {\bibinfo
   {journal} {Phys. Rev. Lett.}\ }\textbf {\bibinfo {volume} {70}},\ \bibinfo
  {pages} {2220} (\bibinfo {year} {1993})}\BibitemShut {NoStop}%
\bibitem [{\citenamefont {Shapiro}\ \emph {et~al.}(2004)\citenamefont
  {Shapiro}, \citenamefont {Davis}, \citenamefont {Lebach},\ and\ \citenamefont
  {Gregory}}]{Shapiro:2004zz}%
  \BibitemOpen
  \bibfield  {author} {\bibinfo {author} {\bibfnamefont {S.~S.}\ \bibnamefont
  {Shapiro}}, \bibinfo {author} {\bibfnamefont {J.~L.}\ \bibnamefont {Davis}},
  \bibinfo {author} {\bibfnamefont {D.~E.}\ \bibnamefont {Lebach}},\ and\
  \bibinfo {author} {\bibfnamefont {J.~S.}\ \bibnamefont {Gregory}},\
  }\bibfield  {title} {\bibinfo {title} {{Measurement of the Solar
  Gravitational Deflection of Radio Waves using Geodetic Very-Long-Baseline
  Interferometry Data, 1979-1999}},\ }\href
  {https://doi.org/10.1103/PhysRevLett.92.121101} {\bibfield  {journal}
  {\bibinfo  {journal} {Phys. Rev. Lett.}\ }\textbf {\bibinfo {volume} {92}},\
  \bibinfo {pages} {121101} (\bibinfo {year} {2004})}\BibitemShut {NoStop}%
\bibitem [{\citenamefont {Andrade}\ \emph {et~al.}(2021)\citenamefont
  {Andrade}, \citenamefont {Salo}, \citenamefont {Aurrekoetxea}, \citenamefont
  {Bamber}, \citenamefont {Clough}, \citenamefont {Croft}, \citenamefont
  {de~Jong}, \citenamefont {Drew}, \citenamefont {Duran}, \citenamefont
  {Ferreira}, \citenamefont {Figueras}, \citenamefont {Finkel}, \citenamefont
  {Fran\c{c}a}, \citenamefont {Ge}, \citenamefont {Gu}, \citenamefont {Helfer},
  \citenamefont {Jäykkä}, \citenamefont {Joana}, \citenamefont {Kunesch},
  \citenamefont {Kornet}, \citenamefont {Lim}, \citenamefont {Muia},
  \citenamefont {Nazari}, \citenamefont {Radia}, \citenamefont {Ripley},
  \citenamefont {Shellard}, \citenamefont {Sperhake}, \citenamefont {Traykova},
  \citenamefont {Tunyasuvunakool}, \citenamefont {Wang}, \citenamefont
  {Widdicombe},\ and\ \citenamefont {Wong}}]{GRChombo2021}%
  \BibitemOpen
  \bibfield  {author} {\bibinfo {author} {\bibfnamefont {T.}~\bibnamefont
  {Andrade}}, \bibinfo {author} {\bibfnamefont {L.~A.}\ \bibnamefont {Salo}},
  \bibinfo {author} {\bibfnamefont {J.~C.}\ \bibnamefont {Aurrekoetxea}},
  \bibinfo {author} {\bibfnamefont {J.}~\bibnamefont {Bamber}}, \bibinfo
  {author} {\bibfnamefont {K.}~\bibnamefont {Clough}}, \bibinfo {author}
  {\bibfnamefont {R.}~\bibnamefont {Croft}}, \bibinfo {author} {\bibfnamefont
  {E.}~\bibnamefont {de~Jong}}, \bibinfo {author} {\bibfnamefont
  {A.}~\bibnamefont {Drew}}, \bibinfo {author} {\bibfnamefont {A.}~\bibnamefont
  {Duran}}, \bibinfo {author} {\bibfnamefont {P.~G.}\ \bibnamefont {Ferreira}},
  \bibinfo {author} {\bibfnamefont {P.}~\bibnamefont {Figueras}}, \bibinfo
  {author} {\bibfnamefont {H.}~\bibnamefont {Finkel}}, \bibinfo {author}
  {\bibfnamefont {T.}~\bibnamefont {Fran\c{c}a}}, \bibinfo {author}
  {\bibfnamefont {B.-X.}\ \bibnamefont {Ge}}, \bibinfo {author} {\bibfnamefont
  {C.}~\bibnamefont {Gu}}, \bibinfo {author} {\bibfnamefont {T.}~\bibnamefont
  {Helfer}}, \bibinfo {author} {\bibfnamefont {J.}~\bibnamefont {Jäykkä}},
  \bibinfo {author} {\bibfnamefont {C.}~\bibnamefont {Joana}}, \bibinfo
  {author} {\bibfnamefont {M.}~\bibnamefont {Kunesch}}, \bibinfo {author}
  {\bibfnamefont {K.}~\bibnamefont {Kornet}}, \bibinfo {author} {\bibfnamefont
  {E.~A.}\ \bibnamefont {Lim}}, \bibinfo {author} {\bibfnamefont
  {F.}~\bibnamefont {Muia}}, \bibinfo {author} {\bibfnamefont {Z.}~\bibnamefont
  {Nazari}}, \bibinfo {author} {\bibfnamefont {M.}~\bibnamefont {Radia}},
  \bibinfo {author} {\bibfnamefont {J.}~\bibnamefont {Ripley}}, \bibinfo
  {author} {\bibfnamefont {P.}~\bibnamefont {Shellard}}, \bibinfo {author}
  {\bibfnamefont {U.}~\bibnamefont {Sperhake}}, \bibinfo {author}
  {\bibfnamefont {D.}~\bibnamefont {Traykova}}, \bibinfo {author}
  {\bibfnamefont {S.}~\bibnamefont {Tunyasuvunakool}}, \bibinfo {author}
  {\bibfnamefont {Z.}~\bibnamefont {Wang}}, \bibinfo {author} {\bibfnamefont
  {J.~Y.}\ \bibnamefont {Widdicombe}},\ and\ \bibinfo {author} {\bibfnamefont
  {K.}~\bibnamefont {Wong}},\ }\bibfield  {title} {\bibinfo {title}
  {{GRC}hombo: {A}n adaptable numerical relativity code for fundamental
  physics},\ }\href {https://doi.org/10.21105/joss.03703} {\bibfield  {journal}
  {\bibinfo  {journal} {Journal of Open Source Software}\ }\textbf {\bibinfo
  {volume} {6}},\ \bibinfo {pages} {3703} (\bibinfo {year} {2021})}\BibitemShut
  {NoStop}%
\bibitem [{\citenamefont {Clough}\ \emph {et~al.}(2015)\citenamefont {Clough},
  \citenamefont {Figueras}, \citenamefont {Finkel}, \citenamefont {Kunesch},
  \citenamefont {Lim},\ and\ \citenamefont {Tunyasuvunakool}}]{Clough:2015sqa}%
  \BibitemOpen
  \bibfield  {author} {\bibinfo {author} {\bibfnamefont {K.}~\bibnamefont
  {Clough}}, \bibinfo {author} {\bibfnamefont {P.}~\bibnamefont {Figueras}},
  \bibinfo {author} {\bibfnamefont {H.}~\bibnamefont {Finkel}}, \bibinfo
  {author} {\bibfnamefont {M.}~\bibnamefont {Kunesch}}, \bibinfo {author}
  {\bibfnamefont {E.~A.}\ \bibnamefont {Lim}},\ and\ \bibinfo {author}
  {\bibfnamefont {S.}~\bibnamefont {Tunyasuvunakool}},\ }\bibfield  {title}
  {\bibinfo {title} {{{GRC}hombo : {N}umerical Relativity with Adaptive Mesh
  Refinement}},\ }\href {https://doi.org/10.1088/0264-9381/32/24/245011}
  {\bibfield  {journal} {\bibinfo  {journal} {Class. Quant. Grav.}\ }\textbf
  {\bibinfo {volume} {32}},\ \bibinfo {pages} {245011} (\bibinfo {year}
  {2015})},\ \Eprint {https://arxiv.org/abs/1503.03436} {arXiv:1503.03436
  [gr-qc]} \BibitemShut {NoStop}%
\bibitem [{\citenamefont {Radia}\ \emph {et~al.}(2022)\citenamefont {Radia},
  \citenamefont {Sperhake}, \citenamefont {Drew}, \citenamefont {Clough},
  \citenamefont {Figueras}, \citenamefont {Lim}, \citenamefont {Ripley},
  \citenamefont {Aurrekoetxea}, \citenamefont {Fran\c{c}a},\ and\ \citenamefont
  {Helfer}}]{Radia_2022}%
  \BibitemOpen
  \bibfield  {author} {\bibinfo {author} {\bibfnamefont {M.}~\bibnamefont
  {Radia}}, \bibinfo {author} {\bibfnamefont {U.}~\bibnamefont {Sperhake}},
  \bibinfo {author} {\bibfnamefont {A.}~\bibnamefont {Drew}}, \bibinfo {author}
  {\bibfnamefont {K.}~\bibnamefont {Clough}}, \bibinfo {author} {\bibfnamefont
  {P.}~\bibnamefont {Figueras}}, \bibinfo {author} {\bibfnamefont {E.~A.}\
  \bibnamefont {Lim}}, \bibinfo {author} {\bibfnamefont {J.~L.}\ \bibnamefont
  {Ripley}}, \bibinfo {author} {\bibfnamefont {J.~C.}\ \bibnamefont
  {Aurrekoetxea}}, \bibinfo {author} {\bibfnamefont {T.}~\bibnamefont
  {Fran\c{c}a}},\ and\ \bibinfo {author} {\bibfnamefont {T.}~\bibnamefont
  {Helfer}},\ }\bibfield  {title} {\bibinfo {title} {{Lessons for adaptive mesh
  refinement in numerical relativity}},\ }\href
  {https://doi.org/10.1088/1361-6382/ac6fa9} {\bibfield  {journal} {\bibinfo
  {journal} {Class. Quant. Grav.}\ }\textbf {\bibinfo {volume} {39}},\ \bibinfo
  {pages} {135006} (\bibinfo {year} {2022})},\ \Eprint
  {https://arxiv.org/abs/2112.10567} {arXiv:2112.10567 [gr-qc]} \BibitemShut
  {NoStop}%
\bibitem [{\citenamefont {et~al.}(2019)}]{chombo}%
  \BibitemOpen
  \bibfield  {author} {\bibinfo {author} {\bibfnamefont {M.~A.}\ \bibnamefont
  {et~al.}},\ }\href@noop {} {\bibinfo {title} {Tech. {R}eport {N}o.
  {LBNL}-6616{E}}} (\bibinfo {year} {2019})\BibitemShut {NoStop}%
\bibitem [{\citenamefont {Alic}\ \emph {et~al.}(2012)\citenamefont {Alic},
  \citenamefont {Bona-Casas}, \citenamefont {Bona}, \citenamefont {Rezzolla},\
  and\ \citenamefont {Palenzuela}}]{ccz4}%
  \BibitemOpen
  \bibfield  {author} {\bibinfo {author} {\bibfnamefont {D.}~\bibnamefont
  {Alic}}, \bibinfo {author} {\bibfnamefont {C.}~\bibnamefont {Bona-Casas}},
  \bibinfo {author} {\bibfnamefont {C.}~\bibnamefont {Bona}}, \bibinfo {author}
  {\bibfnamefont {L.}~\bibnamefont {Rezzolla}},\ and\ \bibinfo {author}
  {\bibfnamefont {C.}~\bibnamefont {Palenzuela}},\ }\bibfield  {title}
  {\bibinfo {title} {Conformal and covariant formulation of the {Z}4 system
  with constraint-violation damping},\ }\href
  {https://doi.org/10.1103/PhysRevD.85.064040} {\bibfield  {journal} {\bibinfo
  {journal} {Phys. Rev. D}\ }\textbf {\bibinfo {volume} {85}},\ \bibinfo
  {pages} {064040} (\bibinfo {year} {2012})}\BibitemShut {NoStop}%
\bibitem [{\citenamefont {Campanelli}\ \emph {et~al.}(2006)\citenamefont
  {Campanelli}, \citenamefont {Lousto}, \citenamefont {Marronetti},\ and\
  \citenamefont {Zlochower}}]{movingpuncture1}%
  \BibitemOpen
  \bibfield  {author} {\bibinfo {author} {\bibfnamefont {M.}~\bibnamefont
  {Campanelli}}, \bibinfo {author} {\bibfnamefont {C.~O.}\ \bibnamefont
  {Lousto}}, \bibinfo {author} {\bibfnamefont {P.}~\bibnamefont {Marronetti}},\
  and\ \bibinfo {author} {\bibfnamefont {Y.}~\bibnamefont {Zlochower}},\
  }\bibfield  {title} {\bibinfo {title} {{Accurate evolutions of orbiting
  black-hole binaries without excision}},\ }\href
  {https://doi.org/10.1103/PhysRevLett.96.111101} {\bibfield  {journal}
  {\bibinfo  {journal} {Phys. Rev. Lett.}\ }\textbf {\bibinfo {volume} {96}},\
  \bibinfo {pages} {111101} (\bibinfo {year} {2006})},\ \Eprint
  {https://arxiv.org/abs/gr-qc/0511048} {arXiv:gr-qc/0511048} \BibitemShut
  {NoStop}%
\bibitem [{\citenamefont {Baker}\ \emph {et~al.}(2006)\citenamefont {Baker},
  \citenamefont {Centrella}, \citenamefont {Choi}, \citenamefont {Koppitz},\
  and\ \citenamefont {van Meter}}]{movingpuncture2}%
  \BibitemOpen
  \bibfield  {author} {\bibinfo {author} {\bibfnamefont {J.~G.}\ \bibnamefont
  {Baker}}, \bibinfo {author} {\bibfnamefont {J.}~\bibnamefont {Centrella}},
  \bibinfo {author} {\bibfnamefont {D.-I.}\ \bibnamefont {Choi}}, \bibinfo
  {author} {\bibfnamefont {M.}~\bibnamefont {Koppitz}},\ and\ \bibinfo {author}
  {\bibfnamefont {J.}~\bibnamefont {van Meter}},\ }\bibfield  {title} {\bibinfo
  {title} {{Gravitational wave extraction from an inspiraling configuration of
  merging black holes}},\ }\href
  {https://doi.org/10.1103/PhysRevLett.96.111102} {\bibfield  {journal}
  {\bibinfo  {journal} {Phys. Rev. Lett.}\ }\textbf {\bibinfo {volume} {96}},\
  \bibinfo {pages} {111102} (\bibinfo {year} {2006})},\ \Eprint
  {https://arxiv.org/abs/gr-qc/0511103} {arXiv:gr-qc/0511103} \BibitemShut
  {NoStop}%
\bibitem [{\citenamefont {Alic}\ \emph {et~al.}(2013)\citenamefont {Alic},
  \citenamefont {Kastaun},\ and\ \citenamefont {Rezzolla}}]{Z4eqs}%
  \BibitemOpen
  \bibfield  {author} {\bibinfo {author} {\bibfnamefont {D.}~\bibnamefont
  {Alic}}, \bibinfo {author} {\bibfnamefont {W.}~\bibnamefont {Kastaun}},\ and\
  \bibinfo {author} {\bibfnamefont {L.}~\bibnamefont {Rezzolla}},\ }\bibfield
  {title} {\bibinfo {title} {Constraint damping of the conformal and covariant
  formulation of the {Z}4 system in simulations of binary neutron stars},\
  }\href {https://doi.org/10.1103/PhysRevD.88.064049} {\bibfield  {journal}
  {\bibinfo  {journal} {Phys. Rev. D}\ }\textbf {\bibinfo {volume} {88}},\
  \bibinfo {pages} {064049} (\bibinfo {year} {2013})}\BibitemShut {NoStop}%
\bibitem [{\citenamefont {Baumgarte}\ and\ \citenamefont
  {Shapiro}(1998)}]{Baumgarte_1998}%
  \BibitemOpen
  \bibfield  {author} {\bibinfo {author} {\bibfnamefont {T.~W.}\ \bibnamefont
  {Baumgarte}}\ and\ \bibinfo {author} {\bibfnamefont {S.~L.}\ \bibnamefont
  {Shapiro}},\ }\bibfield  {title} {\bibinfo {title} {{On the numerical
  integration of {E}instein's field equations}},\ }\href
  {https://doi.org/10.1103/PhysRevD.59.024007} {\bibfield  {journal} {\bibinfo
  {journal} {Phys. Rev. D}\ }\textbf {\bibinfo {volume} {59}},\ \bibinfo
  {pages} {024007} (\bibinfo {year} {1998})},\ \Eprint
  {https://arxiv.org/abs/gr-qc/9810065} {arXiv:gr-qc/9810065} \BibitemShut
  {NoStop}%
\bibitem [{\citenamefont {Shibata}\ and\ \citenamefont
  {Nakamura}(1995)}]{Shibata:1995we}%
  \BibitemOpen
  \bibfield  {author} {\bibinfo {author} {\bibfnamefont {M.}~\bibnamefont
  {Shibata}}\ and\ \bibinfo {author} {\bibfnamefont {T.}~\bibnamefont
  {Nakamura}},\ }\bibfield  {title} {\bibinfo {title} {Evolution of
  three-dimensional gravitational waves: {H}armonic slicing case},\ }\href
  {https://doi.org/10.1103/PhysRevD.52.5428} {\bibfield  {journal} {\bibinfo
  {journal} {Phys. Rev. D}\ }\textbf {\bibinfo {volume} {52}},\ \bibinfo
  {pages} {5428} (\bibinfo {year} {1995})}\BibitemShut {NoStop}%
\bibitem [{\citenamefont {Nakamura}\ \emph {et~al.}(1987)\citenamefont
  {Nakamura}, \citenamefont {Oohara},\ and\ \citenamefont
  {Kojima}}]{Nakamura:1987zz}%
  \BibitemOpen
  \bibfield  {author} {\bibinfo {author} {\bibfnamefont {T.}~\bibnamefont
  {Nakamura}}, \bibinfo {author} {\bibfnamefont {K.}~\bibnamefont {Oohara}},\
  and\ \bibinfo {author} {\bibfnamefont {Y.}~\bibnamefont {Kojima}},\
  }\bibfield  {title} {\bibinfo {title} {{General Relativistic Collapse To
  Black Holes And Gravitational Waves From Black Holes}},\ }\href
  {https://doi.org/10.1143/PTPS.90.1} {\bibfield  {journal} {\bibinfo
  {journal} {Prog. Theor. Phys. Suppl.}\ }\textbf {\bibinfo {volume} {90}},\
  \bibinfo {pages} {1} (\bibinfo {year} {1987})}\BibitemShut {NoStop}%
\bibitem [{\citenamefont {Brandt}\ and\ \citenamefont
  {Br\"ugmann}(1997)}]{puncture}%
  \BibitemOpen
  \bibfield  {author} {\bibinfo {author} {\bibfnamefont {S.}~\bibnamefont
  {Brandt}}\ and\ \bibinfo {author} {\bibfnamefont {B.}~\bibnamefont
  {Br\"ugmann}},\ }\bibfield  {title} {\bibinfo {title} {A simple construction
  of initial data for multiple black holes},\ }\href
  {https://doi.org/10.1103/PhysRevLett.78.3606} {\bibfield  {journal} {\bibinfo
   {journal} {Phys. Rev. Lett.}\ }\textbf {\bibinfo {volume} {78}},\ \bibinfo
  {pages} {3606} (\bibinfo {year} {1997})}\BibitemShut {NoStop}%
\bibitem [{\citenamefont {Bowen}\ and\ \citenamefont {York}(1980)}]{bowenyork}%
  \BibitemOpen
  \bibfield  {author} {\bibinfo {author} {\bibfnamefont {J.~M.}\ \bibnamefont
  {Bowen}}\ and\ \bibinfo {author} {\bibfnamefont {J.~W.}\ \bibnamefont
  {York}},\ }\bibfield  {title} {\bibinfo {title} {Time-asymmetric initial data
  for black holes and black-hole collisions},\ }\href
  {https://doi.org/10.1103/PhysRevD.21.2047} {\bibfield  {journal} {\bibinfo
  {journal} {Phys. Rev. D}\ }\textbf {\bibinfo {volume} {21}},\ \bibinfo
  {pages} {2047} (\bibinfo {year} {1980})}\BibitemShut {NoStop}%
\bibitem [{\citenamefont {Radia}\ \emph {et~al.}(2021)\citenamefont {Radia},
  \citenamefont {Sperhake}, \citenamefont {Berti},\ and\ \citenamefont
  {Croft}}]{mirenanomalies}%
  \BibitemOpen
  \bibfield  {author} {\bibinfo {author} {\bibfnamefont {M.}~\bibnamefont
  {Radia}}, \bibinfo {author} {\bibfnamefont {U.}~\bibnamefont {Sperhake}},
  \bibinfo {author} {\bibfnamefont {E.}~\bibnamefont {Berti}},\ and\ \bibinfo
  {author} {\bibfnamefont {R.}~\bibnamefont {Croft}},\ }\bibfield  {title}
  {\bibinfo {title} {Anomalies in the gravitational recoil of eccentric
  black-hole mergers with unequal mass ratios},\ }\href
  {https://doi.org/10.1103/PhysRevD.103.104006} {\bibfield  {journal} {\bibinfo
   {journal} {Phys. Rev. D}\ }\textbf {\bibinfo {volume} {103}},\ \bibinfo
  {pages} {104006} (\bibinfo {year} {2021})}\BibitemShut {NoStop}%
\bibitem [{\citenamefont {Ansorg}\ \emph {et~al.}(2004)\citenamefont {Ansorg},
  \citenamefont {Bruegmann},\ and\ \citenamefont {Tichy}}]{spectralmarcus}%
  \BibitemOpen
  \bibfield  {author} {\bibinfo {author} {\bibfnamefont {M.}~\bibnamefont
  {Ansorg}}, \bibinfo {author} {\bibfnamefont {B.}~\bibnamefont {Bruegmann}},\
  and\ \bibinfo {author} {\bibfnamefont {W.}~\bibnamefont {Tichy}},\ }\bibfield
   {title} {\bibinfo {title} {{A Single-domain spectral method for black hole
  puncture data}},\ }\href {https://doi.org/10.1103/PhysRevD.70.064011}
  {\bibfield  {journal} {\bibinfo  {journal} {Phys. Rev. D}\ }\textbf {\bibinfo
  {volume} {70}},\ \bibinfo {pages} {064011} (\bibinfo {year} {2004})},\
  \Eprint {https://arxiv.org/abs/gr-qc/0404056} {arXiv:gr-qc/0404056}
  \BibitemShut {NoStop}%
\bibitem [{\citenamefont {Mroue}\ \emph {et~al.}(2010)\citenamefont {Mroue},
  \citenamefont {Pfeiffer}, \citenamefont {Kidder},\ and\ \citenamefont
  {Teukolsky}}]{Mroue:2010re}%
  \BibitemOpen
  \bibfield  {author} {\bibinfo {author} {\bibfnamefont {A.~H.}\ \bibnamefont
  {Mroue}}, \bibinfo {author} {\bibfnamefont {H.~P.}\ \bibnamefont {Pfeiffer}},
  \bibinfo {author} {\bibfnamefont {L.~E.}\ \bibnamefont {Kidder}},\ and\
  \bibinfo {author} {\bibfnamefont {S.~A.}\ \bibnamefont {Teukolsky}},\
  }\bibfield  {title} {\bibinfo {title} {{Measuring orbital eccentricity and
  periastron advance in quasi-circular black hole simulations}},\ }\href
  {https://doi.org/10.1103/PhysRevD.82.124016} {\bibfield  {journal} {\bibinfo
  {journal} {Phys. Rev. D}\ }\textbf {\bibinfo {volume} {82}},\ \bibinfo
  {pages} {124016} (\bibinfo {year} {2010})},\ \Eprint
  {https://arxiv.org/abs/1004.4697} {arXiv:1004.4697 [gr-qc]} \BibitemShut
  {NoStop}%
\bibitem [{\citenamefont {Newman}\ and\ \citenamefont
  {Penrose}(1962)}]{npformalism}%
  \BibitemOpen
  \bibfield  {author} {\bibinfo {author} {\bibfnamefont {E.}~\bibnamefont
  {Newman}}\ and\ \bibinfo {author} {\bibfnamefont {R.}~\bibnamefont
  {Penrose}},\ }\bibfield  {title} {\bibinfo {title} {{An Approach to
  gravitational radiation by a method of spin coefficients}},\ }\href
  {https://doi.org/10.1063/1.1724257} {\bibfield  {journal} {\bibinfo
  {journal} {J. Math. Phys.}\ }\textbf {\bibinfo {volume} {3}},\ \bibinfo
  {pages} {566} (\bibinfo {year} {1962})}\BibitemShut {NoStop}%
\bibitem [{\citenamefont {Bishop}\ and\ \citenamefont
  {Rezzolla}(2016)}]{Bishop:2016lgv}%
  \BibitemOpen
  \bibfield  {author} {\bibinfo {author} {\bibfnamefont {N.~T.}\ \bibnamefont
  {Bishop}}\ and\ \bibinfo {author} {\bibfnamefont {L.}~\bibnamefont
  {Rezzolla}},\ }\bibfield  {title} {\bibinfo {title} {{Extraction of
  Gravitational Waves in Numerical Relativity}},\ }\href
  {https://doi.org/10.1007/s41114-016-0001-9} {\bibfield  {journal} {\bibinfo
  {journal} {Living Rev. Rel.}\ }\textbf {\bibinfo {volume} {19}},\ \bibinfo
  {pages} {2} (\bibinfo {year} {2016})},\ \Eprint
  {https://arxiv.org/abs/1606.02532} {arXiv:1606.02532 [gr-qc]} \BibitemShut
  {NoStop}%
\bibitem [{\citenamefont {Bruegmann}\ \emph {et~al.}(2008)\citenamefont
  {Bruegmann}, \citenamefont {Gonzalez}, \citenamefont {Hannam}, \citenamefont
  {Husa}, \citenamefont {Sperhake},\ and\ \citenamefont
  {Tichy}}]{ulipunctures}%
  \BibitemOpen
  \bibfield  {author} {\bibinfo {author} {\bibfnamefont {B.}~\bibnamefont
  {Bruegmann}}, \bibinfo {author} {\bibfnamefont {J.~A.}\ \bibnamefont
  {Gonzalez}}, \bibinfo {author} {\bibfnamefont {M.}~\bibnamefont {Hannam}},
  \bibinfo {author} {\bibfnamefont {S.}~\bibnamefont {Husa}}, \bibinfo {author}
  {\bibfnamefont {U.}~\bibnamefont {Sperhake}},\ and\ \bibinfo {author}
  {\bibfnamefont {W.}~\bibnamefont {Tichy}},\ }\bibfield  {title} {\bibinfo
  {title} {{Calibration of Moving Puncture Simulations}},\ }\href
  {https://doi.org/10.1103/PhysRevD.77.024027} {\bibfield  {journal} {\bibinfo
  {journal} {Phys. Rev. D}\ }\textbf {\bibinfo {volume} {77}},\ \bibinfo
  {pages} {024027} (\bibinfo {year} {2008})},\ \Eprint
  {https://arxiv.org/abs/gr-qc/0610128} {arXiv:gr-qc/0610128} \BibitemShut
  {NoStop}%
\bibitem [{\citenamefont {Wiaux}\ \emph {et~al.}(2007)\citenamefont {Wiaux},
  \citenamefont {Jacques},\ and\ \citenamefont {Vandergheynst}}]{spinweighted}%
  \BibitemOpen
  \bibfield  {author} {\bibinfo {author} {\bibfnamefont {Y.}~\bibnamefont
  {Wiaux}}, \bibinfo {author} {\bibfnamefont {L.}~\bibnamefont {Jacques}},\
  and\ \bibinfo {author} {\bibfnamefont {P.}~\bibnamefont {Vandergheynst}},\
  }\bibfield  {title} {\bibinfo {title} {{Fast spin +-2 spherical harmonics
  transforms}},\ }\href {https://doi.org/10.1016/j.jcp.2007.07.005} {\bibfield
  {journal} {\bibinfo  {journal} {J. Comput. Phys.}\ }\textbf {\bibinfo
  {volume} {226}},\ \bibinfo {pages} {2359} (\bibinfo {year} {2007})},\ \Eprint
  {https://arxiv.org/abs/astro-ph/0508514} {arXiv:astro-ph/0508514}
  \BibitemShut {NoStop}%
\bibitem [{\citenamefont {Witek}\ \emph {et~al.}(2010)\citenamefont {Witek},
  \citenamefont {Cardoso}, \citenamefont {Herdeiro}, \citenamefont {Nerozzi},
  \citenamefont {Sperhake},\ and\ \citenamefont {Zilhao}}]{Witek_2010}%
  \BibitemOpen
  \bibfield  {author} {\bibinfo {author} {\bibfnamefont {H.}~\bibnamefont
  {Witek}}, \bibinfo {author} {\bibfnamefont {V.}~\bibnamefont {Cardoso}},
  \bibinfo {author} {\bibfnamefont {C.}~\bibnamefont {Herdeiro}}, \bibinfo
  {author} {\bibfnamefont {A.}~\bibnamefont {Nerozzi}}, \bibinfo {author}
  {\bibfnamefont {U.}~\bibnamefont {Sperhake}},\ and\ \bibinfo {author}
  {\bibfnamefont {M.}~\bibnamefont {Zilhao}},\ }\bibfield  {title} {\bibinfo
  {title} {{Black holes in a box: {T}owards the numerical evolution of black
  holes in {A}d{S}}},\ }\href {https://doi.org/10.1103/PhysRevD.82.104037}
  {\bibfield  {journal} {\bibinfo  {journal} {Phys. Rev. D}\ }\textbf {\bibinfo
  {volume} {82}},\ \bibinfo {pages} {104037} (\bibinfo {year} {2010})},\
  \Eprint {https://arxiv.org/abs/1004.4633} {arXiv:1004.4633 [hep-th]}
  \BibitemShut {NoStop}%
\bibitem [{\citenamefont {Campanelli}\ and\ \citenamefont
  {Lousto}(1999)}]{PhysRevD.59.124022}%
  \BibitemOpen
  \bibfield  {author} {\bibinfo {author} {\bibfnamefont {M.}~\bibnamefont
  {Campanelli}}\ and\ \bibinfo {author} {\bibfnamefont {C.~O.}\ \bibnamefont
  {Lousto}},\ }\bibfield  {title} {\bibinfo {title} {Second order gauge
  invariant gravitational perturbations of a {K}err black hole},\ }\href
  {https://doi.org/10.1103/PhysRevD.59.124022} {\bibfield  {journal} {\bibinfo
  {journal} {Phys. Rev. D}\ }\textbf {\bibinfo {volume} {59}},\ \bibinfo
  {pages} {124022} (\bibinfo {year} {1999})}\BibitemShut {NoStop}%
\bibitem [{\citenamefont {Lousto}\ and\ \citenamefont
  {Zlochower}(2007)}]{PhysRevD.76.041502}%
  \BibitemOpen
  \bibfield  {author} {\bibinfo {author} {\bibfnamefont {C.~O.}\ \bibnamefont
  {Lousto}}\ and\ \bibinfo {author} {\bibfnamefont {Y.}~\bibnamefont
  {Zlochower}},\ }\bibfield  {title} {\bibinfo {title} {Practical formula for
  the radiated angular momentum},\ }\href
  {https://doi.org/10.1103/PhysRevD.76.041502} {\bibfield  {journal} {\bibinfo
  {journal} {Phys. Rev. D}\ }\textbf {\bibinfo {volume} {76}},\ \bibinfo
  {pages} {041502} (\bibinfo {year} {2007})}\BibitemShut {NoStop}%
\bibitem [{\citenamefont {Shiralilou}\ \emph
  {et~al.}(2021{\natexlab{b}})\citenamefont {Shiralilou}, \citenamefont
  {Hinderer}, \citenamefont {Nissanke}, \citenamefont {Ortiz},\ and\
  \citenamefont {Witek}}]{Shiralilou:2021mfl}%
  \BibitemOpen
  \bibfield  {author} {\bibinfo {author} {\bibfnamefont {B.}~\bibnamefont
  {Shiralilou}}, \bibinfo {author} {\bibfnamefont {T.}~\bibnamefont
  {Hinderer}}, \bibinfo {author} {\bibfnamefont {S.}~\bibnamefont {Nissanke}},
  \bibinfo {author} {\bibfnamefont {N.}~\bibnamefont {Ortiz}},\ and\ \bibinfo
  {author} {\bibfnamefont {H.}~\bibnamefont {Witek}},\ }\bibfield  {title}
  {\bibinfo {title} {{Post-Newtonian Gravitational and Scalar Waves in
  Scalar-{G}auss-{B}onnet Gravity}},\ }\href@noop {} {\  (\bibinfo {year}
  {2021}{\natexlab{b}})},\ \Eprint {https://arxiv.org/abs/2105.13972}
  {arXiv:2105.13972 [gr-qc]} \BibitemShut {NoStop}%
\bibitem [{\citenamefont {Berti}\ \emph {et~al.}(2007)\citenamefont {Berti},
  \citenamefont {Cardoso}, \citenamefont {Gonzalez}, \citenamefont {Sperhake},
  \citenamefont {Hannam}, \citenamefont {Husa},\ and\ \citenamefont
  {Bruegmann}}]{Berti:2007fi}%
  \BibitemOpen
  \bibfield  {author} {\bibinfo {author} {\bibfnamefont {E.}~\bibnamefont
  {Berti}}, \bibinfo {author} {\bibfnamefont {V.}~\bibnamefont {Cardoso}},
  \bibinfo {author} {\bibfnamefont {J.~A.}\ \bibnamefont {Gonzalez}}, \bibinfo
  {author} {\bibfnamefont {U.}~\bibnamefont {Sperhake}}, \bibinfo {author}
  {\bibfnamefont {M.}~\bibnamefont {Hannam}}, \bibinfo {author} {\bibfnamefont
  {S.}~\bibnamefont {Husa}},\ and\ \bibinfo {author} {\bibfnamefont
  {B.}~\bibnamefont {Bruegmann}},\ }\bibfield  {title} {\bibinfo {title}
  {{Inspiral, merger and ringdown of unequal mass black hole binaries: {A}
  Multipolar analysis}},\ }\href {https://doi.org/10.1103/PhysRevD.76.064034}
  {\bibfield  {journal} {\bibinfo  {journal} {Phys. Rev. D}\ }\textbf {\bibinfo
  {volume} {76}},\ \bibinfo {pages} {064034} (\bibinfo {year} {2007})},\
  \Eprint {https://arxiv.org/abs/gr-qc/0703053} {arXiv:gr-qc/0703053}
  \BibitemShut {NoStop}%
\bibitem [{\citenamefont {Press}(1971)}]{1971ApJ...170L.105P}%
  \BibitemOpen
  \bibfield  {author} {\bibinfo {author} {\bibfnamefont {W.~H.}\ \bibnamefont
  {Press}},\ }\bibfield  {title} {\bibinfo {title} {{Long Wave Trains of
  Gravitational Waves from a Vibrating Black Hole}},\ }\href
  {https://doi.org/10.1086/180849} {\bibfield  {journal} {\bibinfo  {journal}
  {Astrophys. J. Lett.}\ }\textbf {\bibinfo {volume} {170}},\ \bibinfo {pages}
  {L105} (\bibinfo {year} {1971})}\BibitemShut {NoStop}%
\bibitem [{\citenamefont {Berti}\ \emph {et~al.}(2009)\citenamefont {Berti},
  \citenamefont {Cardoso},\ and\ \citenamefont {Starinets}}]{Berti:2009kk}%
  \BibitemOpen
  \bibfield  {author} {\bibinfo {author} {\bibfnamefont {E.}~\bibnamefont
  {Berti}}, \bibinfo {author} {\bibfnamefont {V.}~\bibnamefont {Cardoso}},\
  and\ \bibinfo {author} {\bibfnamefont {A.~O.}\ \bibnamefont {Starinets}},\
  }\bibfield  {title} {\bibinfo {title} {{Quasinormal modes of black holes and
  black branes}},\ }\href {https://doi.org/10.1088/0264-9381/26/16/163001}
  {\bibfield  {journal} {\bibinfo  {journal} {Class. Quant. Grav.}\ }\textbf
  {\bibinfo {volume} {26}},\ \bibinfo {pages} {163001} (\bibinfo {year}
  {2009})},\ \Eprint {https://arxiv.org/abs/0905.2975} {arXiv:0905.2975
  [gr-qc]} \BibitemShut {NoStop}%
\bibitem [{\citenamefont {Teukolsky}(1973)}]{teukolskyformalism}%
  \BibitemOpen
  \bibfield  {author} {\bibinfo {author} {\bibfnamefont {S.~A.}\ \bibnamefont
  {Teukolsky}},\ }\bibfield  {title} {\bibinfo {title} {{Perturbations of a
  rotating black hole. 1. {F}undamental equations for gravitational
  electromagnetic and neutrino field perturbations}},\ }\href
  {https://doi.org/10.1086/152444} {\bibfield  {journal} {\bibinfo  {journal}
  {Astrophys. J.}\ }\textbf {\bibinfo {volume} {185}},\ \bibinfo {pages} {635}
  (\bibinfo {year} {1973})}\BibitemShut {NoStop}%
\bibitem [{\citenamefont {Konoplya}\ and\ \citenamefont
  {Zinhailo}(2020)}]{Konoplya_2020}%
  \BibitemOpen
  \bibfield  {author} {\bibinfo {author} {\bibfnamefont {R.~A.}\ \bibnamefont
  {Konoplya}}\ and\ \bibinfo {author} {\bibfnamefont {A.~F.}\ \bibnamefont
  {Zinhailo}},\ }\bibfield  {title} {\bibinfo {title} {{Quasinormal modes,
  stability and shadows of a black hole in the 4{D}
  {E}instein\textendash{}{G}auss\textendash{}{B}onnet gravity}},\ }\href
  {https://doi.org/10.1140/epjc/s10052-020-08639-8} {\bibfield  {journal}
  {\bibinfo  {journal} {Eur. Phys. J. C}\ }\textbf {\bibinfo {volume} {80}},\
  \bibinfo {pages} {1049} (\bibinfo {year} {2020})},\ \Eprint
  {https://arxiv.org/abs/2003.01188} {arXiv:2003.01188 [gr-qc]} \BibitemShut
  {NoStop}%
\bibitem [{\citenamefont {Bryant}\ \emph {et~al.}(2021)\citenamefont {Bryant},
  \citenamefont {Silva}, \citenamefont {Yagi},\ and\ \citenamefont
  {Glampedakis}}]{Bryant:2021xdh}%
  \BibitemOpen
  \bibfield  {author} {\bibinfo {author} {\bibfnamefont {A.}~\bibnamefont
  {Bryant}}, \bibinfo {author} {\bibfnamefont {H.~O.}\ \bibnamefont {Silva}},
  \bibinfo {author} {\bibfnamefont {K.}~\bibnamefont {Yagi}},\ and\ \bibinfo
  {author} {\bibfnamefont {K.}~\bibnamefont {Glampedakis}},\ }\bibfield
  {title} {\bibinfo {title} {{Eikonal quasinormal modes of black holes beyond
  general relativity. {III}. {S}calar {G}auss-{B}onnet gravity}},\ }\href
  {https://doi.org/10.1103/PhysRevD.104.044051} {\bibfield  {journal} {\bibinfo
   {journal} {Phys. Rev. D}\ }\textbf {\bibinfo {volume} {104}},\ \bibinfo
  {pages} {044051} (\bibinfo {year} {2021})},\ \Eprint
  {https://arxiv.org/abs/2106.09657} {arXiv:2106.09657 [gr-qc]} \BibitemShut
  {NoStop}%
\bibitem [{\citenamefont {Pierini}\ and\ \citenamefont
  {Gualtieri}(2021)}]{Pierini:2021qnms}%
  \BibitemOpen
  \bibfield  {author} {\bibinfo {author} {\bibfnamefont {L.}~\bibnamefont
  {Pierini}}\ and\ \bibinfo {author} {\bibfnamefont {L.}~\bibnamefont
  {Gualtieri}},\ }\bibfield  {title} {\bibinfo {title} {Quasinormal modes of
  rotating black holes in {E}instein-dilaton {G}auss-{B}onnet gravity: {T}he
  first order in rotation},\ }\bibfield  {journal} {\bibinfo  {journal}
  {Physical Review D}\ }\textbf {\bibinfo {volume} {103}},\ \href
  {https://doi.org/10.1103/physrevd.103.124017} {10.1103/physrevd.103.124017}
  (\bibinfo {year} {2021})\BibitemShut {NoStop}%
\bibitem [{\citenamefont {Pierini}\ and\ \citenamefont
  {Gualtieri}(2022)}]{Pierini:2022eim}%
  \BibitemOpen
  \bibfield  {author} {\bibinfo {author} {\bibfnamefont {L.}~\bibnamefont
  {Pierini}}\ and\ \bibinfo {author} {\bibfnamefont {L.}~\bibnamefont
  {Gualtieri}},\ }\bibfield  {title} {\bibinfo {title} {{Quasi-normal modes of
  rotating black holes in {E}instein-dilaton {G}auss-{B}onnet gravity: {T}he
  second order in rotation}},\ }\href@noop {} {\  (\bibinfo {year} {2022})},\
  \Eprint {https://arxiv.org/abs/2207.11267} {arXiv:2207.11267 [gr-qc]}
  \BibitemShut {NoStop}%
\bibitem [{\citenamefont {Li}\ \emph {et~al.}(2022)\citenamefont {Li},
  \citenamefont {Wagle}, \citenamefont {Chen},\ and\ \citenamefont
  {Yunes}}]{Li:2022pcy}%
  \BibitemOpen
  \bibfield  {author} {\bibinfo {author} {\bibfnamefont {D.}~\bibnamefont
  {Li}}, \bibinfo {author} {\bibfnamefont {P.}~\bibnamefont {Wagle}}, \bibinfo
  {author} {\bibfnamefont {Y.}~\bibnamefont {Chen}},\ and\ \bibinfo {author}
  {\bibfnamefont {N.}~\bibnamefont {Yunes}},\ }\bibfield  {title} {\bibinfo
  {title} {{Perturbations of spinning black holes beyond General Relativity:
  {M}odified Teukolsky equation}},\ }\href@noop {} {\  (\bibinfo {year}
  {2022})},\ \Eprint {https://arxiv.org/abs/2206.10652} {arXiv:2206.10652
  [gr-qc]} \BibitemShut {NoStop}%
\bibitem [{\citenamefont {Okounkova}\ \emph {et~al.}(2019)\citenamefont
  {Okounkova}, \citenamefont {Stein}, \citenamefont {Scheel},\ and\
  \citenamefont {Teukolsky}}]{Okunkova:2019NRsecondorder}%
  \BibitemOpen
  \bibfield  {author} {\bibinfo {author} {\bibfnamefont {M.}~\bibnamefont
  {Okounkova}}, \bibinfo {author} {\bibfnamefont {L.~C.}\ \bibnamefont
  {Stein}}, \bibinfo {author} {\bibfnamefont {M.~A.}\ \bibnamefont {Scheel}},\
  and\ \bibinfo {author} {\bibfnamefont {S.~A.}\ \bibnamefont {Teukolsky}},\
  }\bibfield  {title} {\bibinfo {title} {{Numerical binary black hole
  collisions in dynamical {C}hern-{S}imons gravity}},\ }\href
  {https://doi.org/10.1103/PhysRevD.100.104026} {\bibfield  {journal} {\bibinfo
   {journal} {Phys. Rev. D}\ }\textbf {\bibinfo {volume} {100}},\ \bibinfo
  {pages} {104026} (\bibinfo {year} {2019})},\ \Eprint
  {https://arxiv.org/abs/1906.08789} {arXiv:1906.08789 [gr-qc]} \BibitemShut
  {NoStop}%
\bibitem [{\citenamefont {{Foreman-Mackey}}\ \emph {et~al.}(2013)\citenamefont
  {{Foreman-Mackey}}, \citenamefont {{Hogg}}, \citenamefont {{Lang}},\ and\
  \citenamefont {{Goodman}}}]{emcee}%
  \BibitemOpen
  \bibfield  {author} {\bibinfo {author} {\bibfnamefont {D.}~\bibnamefont
  {{Foreman-Mackey}}}, \bibinfo {author} {\bibfnamefont {D.~W.}\ \bibnamefont
  {{Hogg}}}, \bibinfo {author} {\bibfnamefont {D.}~\bibnamefont {{Lang}}},\
  and\ \bibinfo {author} {\bibfnamefont {J.}~\bibnamefont {{Goodman}}},\
  }\bibfield  {title} {\bibinfo {title} {emcee: {T}he {MCMC} {H}ammer},\ }\href
  {https://doi.org/10.1086/670067} {\bibfield  {journal} {\bibinfo  {journal}
  {PASP}\ }\textbf {\bibinfo {volume} {125}},\ \bibinfo {pages} {306} (\bibinfo
  {year} {2013})},\ \Eprint {https://arxiv.org/abs/1202.3665} {1202.3665}
  \BibitemShut {NoStop}%
\bibitem [{\citenamefont {Stein}(2019)}]{Stein:2019mop}%
  \BibitemOpen
  \bibfield  {author} {\bibinfo {author} {\bibfnamefont {L.~C.}\ \bibnamefont
  {Stein}},\ }\bibfield  {title} {\bibinfo {title} {{qnm: {A} {P}ython package
  for calculating {K}err quasinormal modes, separation constants, and
  spherical-spheroidal mixing coefficients}},\ }\href
  {https://doi.org/10.21105/joss.01683} {\bibfield  {journal} {\bibinfo
  {journal} {J. Open Source Softw.}\ }\textbf {\bibinfo {volume} {4}},\
  \bibinfo {pages} {1683} (\bibinfo {year} {2019})},\ \Eprint
  {https://arxiv.org/abs/1908.10377} {arXiv:1908.10377 [gr-qc]} \BibitemShut
  {NoStop}%
%%CITATION = ARXIV:1908.10377;%%
\bibitem [{\citenamefont {Bhagwat}\ \emph {et~al.}(2018)\citenamefont
  {Bhagwat}, \citenamefont {Okounkova}, \citenamefont {Ballmer}, \citenamefont
  {Brown}, \citenamefont {Giesler}, \citenamefont {Scheel},\ and\ \citenamefont
  {Teukolsky}}]{Bhagwat:2017tkm}%
  \BibitemOpen
  \bibfield  {author} {\bibinfo {author} {\bibfnamefont {S.}~\bibnamefont
  {Bhagwat}}, \bibinfo {author} {\bibfnamefont {M.}~\bibnamefont {Okounkova}},
  \bibinfo {author} {\bibfnamefont {S.~W.}\ \bibnamefont {Ballmer}}, \bibinfo
  {author} {\bibfnamefont {D.~A.}\ \bibnamefont {Brown}}, \bibinfo {author}
  {\bibfnamefont {M.}~\bibnamefont {Giesler}}, \bibinfo {author} {\bibfnamefont
  {M.~A.}\ \bibnamefont {Scheel}},\ and\ \bibinfo {author} {\bibfnamefont
  {S.~A.}\ \bibnamefont {Teukolsky}},\ }\bibfield  {title} {\bibinfo {title}
  {{On choosing the start time of binary black hole ringdowns}},\ }\href
  {https://doi.org/10.1103/PhysRevD.97.104065} {\bibfield  {journal} {\bibinfo
  {journal} {Phys. Rev. D}\ }\textbf {\bibinfo {volume} {97}},\ \bibinfo
  {pages} {104065} (\bibinfo {year} {2018})},\ \Eprint
  {https://arxiv.org/abs/1711.00926} {arXiv:1711.00926 [gr-qc]} \BibitemShut
  {NoStop}%
\bibitem [{\citenamefont {Capano}\ \emph {et~al.}(2021)\citenamefont {Capano},
  \citenamefont {Cabero}, \citenamefont {Westerweck}, \citenamefont {Abedi},
  \citenamefont {Kastha}, \citenamefont {Nitz}, \citenamefont {Wang},
  \citenamefont {Nielsen},\ and\ \citenamefont {Krishnan}}]{Capano:2021etf}%
  \BibitemOpen
  \bibfield  {author} {\bibinfo {author} {\bibfnamefont {C.~D.}\ \bibnamefont
  {Capano}}, \bibinfo {author} {\bibfnamefont {M.}~\bibnamefont {Cabero}},
  \bibinfo {author} {\bibfnamefont {J.}~\bibnamefont {Westerweck}}, \bibinfo
  {author} {\bibfnamefont {J.}~\bibnamefont {Abedi}}, \bibinfo {author}
  {\bibfnamefont {S.}~\bibnamefont {Kastha}}, \bibinfo {author} {\bibfnamefont
  {A.~H.}\ \bibnamefont {Nitz}}, \bibinfo {author} {\bibfnamefont {Y.-F.}\
  \bibnamefont {Wang}}, \bibinfo {author} {\bibfnamefont {A.~B.}\ \bibnamefont
  {Nielsen}},\ and\ \bibinfo {author} {\bibfnamefont {B.}~\bibnamefont
  {Krishnan}},\ }\bibfield  {title} {\bibinfo {title} {{Observation of a
  multimode quasi-normal spectrum from a perturbed black hole}},\ }\href@noop
  {} {\  (\bibinfo {year} {2021})},\ \Eprint {https://arxiv.org/abs/2105.05238}
  {arXiv:2105.05238 [gr-qc]} \BibitemShut {NoStop}%
\bibitem [{\citenamefont {London}\ \emph {et~al.}(2014)\citenamefont {London},
  \citenamefont {Shoemaker},\ and\ \citenamefont {Healy}}]{London:2014cma}%
  \BibitemOpen
  \bibfield  {author} {\bibinfo {author} {\bibfnamefont {L.}~\bibnamefont
  {London}}, \bibinfo {author} {\bibfnamefont {D.}~\bibnamefont {Shoemaker}},\
  and\ \bibinfo {author} {\bibfnamefont {J.}~\bibnamefont {Healy}},\ }\bibfield
   {title} {\bibinfo {title} {{Modeling ringdown: Beyond the fundamental
  quasinormal modes}},\ }\href {https://doi.org/10.1103/PhysRevD.90.124032}
  {\bibfield  {journal} {\bibinfo  {journal} {Phys. Rev. D}\ }\textbf {\bibinfo
  {volume} {90}},\ \bibinfo {pages} {124032} (\bibinfo {year} {2014})},\
  \bibinfo {note} {[Erratum: Phys.Rev.D 94, 069902 (2016)]},\ \Eprint
  {https://arxiv.org/abs/1404.3197} {arXiv:1404.3197 [gr-qc]} \BibitemShut
  {NoStop}%
\bibitem [{\citenamefont {Giesler}\ \emph {et~al.}(2019)\citenamefont
  {Giesler}, \citenamefont {Isi}, \citenamefont {Scheel},\ and\ \citenamefont
  {Teukolsky}}]{Giesler_2019}%
  \BibitemOpen
  \bibfield  {author} {\bibinfo {author} {\bibfnamefont {M.}~\bibnamefont
  {Giesler}}, \bibinfo {author} {\bibfnamefont {M.}~\bibnamefont {Isi}},
  \bibinfo {author} {\bibfnamefont {M.~A.}\ \bibnamefont {Scheel}},\ and\
  \bibinfo {author} {\bibfnamefont {S.}~\bibnamefont {Teukolsky}},\ }\bibfield
  {title} {\bibinfo {title} {{Black Hole Ringdown: {T}he Importance of
  Overtones}},\ }\href {https://doi.org/10.1103/PhysRevX.9.041060} {\bibfield
  {journal} {\bibinfo  {journal} {Phys. Rev. X}\ }\textbf {\bibinfo {volume}
  {9}},\ \bibinfo {pages} {041060} (\bibinfo {year} {2019})},\ \Eprint
  {https://arxiv.org/abs/1903.08284} {arXiv:1903.08284 [gr-qc]} \BibitemShut
  {NoStop}%
\bibitem [{\citenamefont {Jim\'enez~Forteza}\ \emph {et~al.}(2020)\citenamefont
  {Jim\'enez~Forteza}, \citenamefont {Bhagwat}, \citenamefont {Pani},\ and\
  \citenamefont {Ferrari}}]{JimenezForteza:2020cve}%
  \BibitemOpen
  \bibfield  {author} {\bibinfo {author} {\bibfnamefont {X.}~\bibnamefont
  {Jim\'enez~Forteza}}, \bibinfo {author} {\bibfnamefont {S.}~\bibnamefont
  {Bhagwat}}, \bibinfo {author} {\bibfnamefont {P.}~\bibnamefont {Pani}},\ and\
  \bibinfo {author} {\bibfnamefont {V.}~\bibnamefont {Ferrari}},\ }\bibfield
  {title} {\bibinfo {title} {{Spectroscopy of binary black hole ringdown using
  overtones and angular modes}},\ }\href
  {https://doi.org/10.1103/PhysRevD.102.044053} {\bibfield  {journal} {\bibinfo
   {journal} {Phys. Rev. D}\ }\textbf {\bibinfo {volume} {102}},\ \bibinfo
  {pages} {044053} (\bibinfo {year} {2020})},\ \Eprint
  {https://arxiv.org/abs/2005.03260} {arXiv:2005.03260 [gr-qc]} \BibitemShut
  {NoStop}%
\bibitem [{\citenamefont {Maga\~na Zertuche}\ \emph {et~al.}(2022)\citenamefont
  {Maga\~na Zertuche} \emph {et~al.}}]{MaganaZertuche:2021syq}%
  \BibitemOpen
  \bibfield  {author} {\bibinfo {author} {\bibfnamefont {L.}~\bibnamefont
  {Maga\~na Zertuche}} \emph {et~al.},\ }\bibfield  {title} {\bibinfo {title}
  {{High Precision Ringdown Modeling: {M}ultimode Fits and {BMS} Frames}},\
  }\href {https://doi.org/10.1103/PhysRevD.105.104015} {\bibfield  {journal}
  {\bibinfo  {journal} {Phys. Rev. D}\ }\textbf {\bibinfo {volume} {105}},\
  \bibinfo {pages} {104015} (\bibinfo {year} {2022})},\ \Eprint
  {https://arxiv.org/abs/2110.15922} {arXiv:2110.15922 [gr-qc]} \BibitemShut
  {NoStop}%
\bibitem [{\citenamefont {Isi}\ and\ \citenamefont {Farr}(2021)}]{Isi:2021iql}%
  \BibitemOpen
  \bibfield  {author} {\bibinfo {author} {\bibfnamefont {M.}~\bibnamefont
  {Isi}}\ and\ \bibinfo {author} {\bibfnamefont {W.~M.}\ \bibnamefont {Farr}},\
  }\bibfield  {title} {\bibinfo {title} {{Analyzing black-hole ringdowns}},\
  }\href@noop {} {\  (\bibinfo {year} {2021})},\ \Eprint
  {https://arxiv.org/abs/2107.05609} {arXiv:2107.05609 [gr-qc]} \BibitemShut
  {NoStop}%
\bibitem [{\citenamefont {Cotesta}\ \emph {et~al.}(2022)\citenamefont
  {Cotesta}, \citenamefont {Carullo}, \citenamefont {Berti},\ and\
  \citenamefont {Cardoso}}]{Cotesta:2022pci}%
  \BibitemOpen
  \bibfield  {author} {\bibinfo {author} {\bibfnamefont {R.}~\bibnamefont
  {Cotesta}}, \bibinfo {author} {\bibfnamefont {G.}~\bibnamefont {Carullo}},
  \bibinfo {author} {\bibfnamefont {E.}~\bibnamefont {Berti}},\ and\ \bibinfo
  {author} {\bibfnamefont {V.}~\bibnamefont {Cardoso}},\ }\bibfield  {title}
  {\bibinfo {title} {{Analysis of Ringdown Overtones in {GW}150914}},\ }\href
  {https://doi.org/10.1103/PhysRevLett.129.111102} {\bibfield  {journal}
  {\bibinfo  {journal} {Phys. Rev. Lett.}\ }\textbf {\bibinfo {volume} {129}},\
  \bibinfo {pages} {111102} (\bibinfo {year} {2022})},\ \Eprint
  {https://arxiv.org/abs/2201.00822} {arXiv:2201.00822 [gr-qc]} \BibitemShut
  {NoStop}%
\bibitem [{\citenamefont {Cheung}\ \emph {et~al.}(2022)\citenamefont {Cheung}
  \emph {et~al.}}]{Cheung:2022rbm}%
  \BibitemOpen
  \bibfield  {author} {\bibinfo {author} {\bibfnamefont {M.~H.-Y.}\
  \bibnamefont {Cheung}} \emph {et~al.},\ }\bibfield  {title} {\bibinfo {title}
  {{Nonlinear effects in black hole ringdown}},\ }\href@noop {} {\  (\bibinfo
  {year} {2022})},\ \Eprint {https://arxiv.org/abs/2208.07374}
  {arXiv:2208.07374 [gr-qc]} \BibitemShut {NoStop}%
\bibitem [{\citenamefont {Mitman}\ \emph {et~al.}(2022)\citenamefont {Mitman}
  \emph {et~al.}}]{Mitman:2022qdl}%
  \BibitemOpen
  \bibfield  {author} {\bibinfo {author} {\bibfnamefont {K.}~\bibnamefont
  {Mitman}} \emph {et~al.},\ }\bibfield  {title} {\bibinfo {title}
  {{Nonlinearities in black hole ringdowns}},\ }\href@noop {} {\  (\bibinfo
  {year} {2022})},\ \Eprint {https://arxiv.org/abs/2208.07380}
  {arXiv:2208.07380 [gr-qc]} \BibitemShut {NoStop}%
\bibitem [{\citenamefont {Aasi}\ \emph
  {et~al.}(2015{\natexlab{b}})\citenamefont {Aasi} \emph
  {et~al.}}]{LIGOScientific:2014pky}%
  \BibitemOpen
  \bibfield  {author} {\bibinfo {author} {\bibfnamefont {J.}~\bibnamefont
  {Aasi}} \emph {et~al.} (\bibinfo {collaboration} {LIGO Scientific}),\
  }\bibfield  {title} {\bibinfo {title} {{Advanced LIGO}},\ }\href
  {https://doi.org/10.1088/0264-9381/32/7/074001} {\bibfield  {journal}
  {\bibinfo  {journal} {Class. Quant. Grav.}\ }\textbf {\bibinfo {volume}
  {32}},\ \bibinfo {pages} {074001} (\bibinfo {year} {2015}{\natexlab{b}})},\
  \Eprint {https://arxiv.org/abs/1411.4547} {arXiv:1411.4547 [gr-qc]}
  \BibitemShut {NoStop}%
\bibitem [{\citenamefont {Abbott}\ \emph {et~al.}(2018)\citenamefont {Abbott}
  \emph {et~al.}}]{KAGRA:2013rdx}%
  \BibitemOpen
  \bibfield  {author} {\bibinfo {author} {\bibfnamefont {B.~P.}\ \bibnamefont
  {Abbott}} \emph {et~al.} (\bibinfo {collaboration} {KAGRA, LIGO Scientific,
  Virgo, VIRGO}),\ }\bibfield  {title} {\bibinfo {title} {{Prospects for
  observing and localizing gravitational-wave transients with {A}dvanced
  {LIGO}, {A}dvanced {V}irgo and {KAGRA}}},\ }\href
  {https://doi.org/10.1007/s41114-020-00026-9} {\bibfield  {journal} {\bibinfo
  {journal} {Living Rev. Rel.}\ }\textbf {\bibinfo {volume} {21}},\ \bibinfo
  {pages} {3} (\bibinfo {year} {2018})},\ \Eprint
  {https://arxiv.org/abs/1304.0670} {arXiv:1304.0670 [gr-qc]} \BibitemShut
  {NoStop}%
\bibitem [{\citenamefont {Punturo}\ \emph {et~al.}(2010)\citenamefont {Punturo}
  \emph {et~al.}}]{Punturo_2010}%
  \BibitemOpen
  \bibfield  {author} {\bibinfo {author} {\bibfnamefont {M.}~\bibnamefont
  {Punturo}} \emph {et~al.},\ }\bibfield  {title} {\bibinfo {title} {{The
  {E}instein {T}elescope: {A} third-generation gravitational wave
  observatory}},\ }\href {https://doi.org/10.1088/0264-9381/27/19/194002}
  {\bibfield  {journal} {\bibinfo  {journal} {Class. Quant. Grav.}\ }\textbf
  {\bibinfo {volume} {27}},\ \bibinfo {pages} {194002} (\bibinfo {year}
  {2010})}\BibitemShut {NoStop}%
\bibitem [{\citenamefont {Hild}\ \emph {et~al.}(2011)\citenamefont {Hild} \emph
  {et~al.}}]{Hild:2010id}%
  \BibitemOpen
  \bibfield  {author} {\bibinfo {author} {\bibfnamefont {S.}~\bibnamefont
  {Hild}} \emph {et~al.},\ }\bibfield  {title} {\bibinfo {title} {{Sensitivity
  Studies for Third-Generation Gravitational Wave Observatories}},\ }\href
  {https://doi.org/10.1088/0264-9381/28/9/094013} {\bibfield  {journal}
  {\bibinfo  {journal} {Class. Quant. Grav.}\ }\textbf {\bibinfo {volume}
  {28}},\ \bibinfo {pages} {094013} (\bibinfo {year} {2011})},\ \Eprint
  {https://arxiv.org/abs/1012.0908} {arXiv:1012.0908 [gr-qc]} \BibitemShut
  {NoStop}%
\bibitem [{\citenamefont {Abbott}\ \emph
  {et~al.}(2017{\natexlab{b}})\citenamefont {Abbott} \emph
  {et~al.}}]{Abbott_2017}%
  \BibitemOpen
  \bibfield  {author} {\bibinfo {author} {\bibfnamefont {B.~P.}\ \bibnamefont
  {Abbott}} \emph {et~al.} (\bibinfo {collaboration} {LIGO Scientific,
  Virgo}),\ }\bibfield  {title} {\bibinfo {title} {{Search for intermediate
  mass black hole binaries in the first observing run of {A}dvanced `{LIGO}}},\
  }\href {https://doi.org/10.1103/PhysRevD.96.022001} {\bibfield  {journal}
  {\bibinfo  {journal} {Phys. Rev. D}\ }\textbf {\bibinfo {volume} {96}},\
  \bibinfo {pages} {022001} (\bibinfo {year} {2017}{\natexlab{b}})},\ \Eprint
  {https://arxiv.org/abs/1704.04628} {arXiv:1704.04628 [gr-qc]} \BibitemShut
  {NoStop}%
\bibitem [{\citenamefont {Iacovelli}\ \emph {et~al.}(2022)\citenamefont
  {Iacovelli}, \citenamefont {Mancarella}, \citenamefont {Foffa},\ and\
  \citenamefont {Maggiore}}]{Iacovelli:2022bbs}%
  \BibitemOpen
  \bibfield  {author} {\bibinfo {author} {\bibfnamefont {F.}~\bibnamefont
  {Iacovelli}}, \bibinfo {author} {\bibfnamefont {M.}~\bibnamefont
  {Mancarella}}, \bibinfo {author} {\bibfnamefont {S.}~\bibnamefont {Foffa}},\
  and\ \bibinfo {author} {\bibfnamefont {M.}~\bibnamefont {Maggiore}},\
  }\bibfield  {title} {\bibinfo {title} {{Forecasting the detection
  capabilities of third-generation gravitational-wave detectors using
  $\texttt{GWFAST}$}},\ }\href@noop {} {\  (\bibinfo {year} {2022})},\ \Eprint
  {https://arxiv.org/abs/2207.02771} {arXiv:2207.02771 [gr-qc]} \BibitemShut
  {NoStop}%
\bibitem [{\citenamefont {Abbott}\ and\ \citenamefont
  {et.al.}(2016)}]{PhysRevLett.116.241103}%
  \BibitemOpen
  \bibfield  {author} {\bibinfo {author} {\bibfnamefont {B.~P.}\ \bibnamefont
  {Abbott}}\ and\ \bibinfo {author} {\bibnamefont {et.al.}} (\bibinfo
  {collaboration} {LIGO Scientific Collaboration and Virgo Collaboration}),\
  }\bibfield  {title} {\bibinfo {title} {{GW}151226:{O}bservation of
  gravitational waves from a 22-solar-mass binary black hole coalescence},\
  }\href {https://doi.org/10.1103/PhysRevLett.116.241103} {\bibfield  {journal}
  {\bibinfo  {journal} {Phys. Rev. Lett.}\ }\textbf {\bibinfo {volume} {116}},\
  \bibinfo {pages} {241103} (\bibinfo {year} {2016})}\BibitemShut {NoStop}%
\bibitem [{\citenamefont {Abbott}\ \emph
  {et~al.}(2021{\natexlab{c}})\citenamefont {Abbott} \emph
  {et~al.}}]{LIGOScientific:2020ibl}%
  \BibitemOpen
  \bibfield  {author} {\bibinfo {author} {\bibfnamefont {R.}~\bibnamefont
  {Abbott}} \emph {et~al.} (\bibinfo {collaboration} {LIGO Scientific,
  Virgo}),\ }\bibfield  {title} {\bibinfo {title} {{{GWTC}-2: {C}ompact Binary
  Coalescences Observed by {LIGO} and {V}irgo During the First Half of the
  Third Observing Run}},\ }\href {https://doi.org/10.1103/PhysRevX.11.021053}
  {\bibfield  {journal} {\bibinfo  {journal} {Phys. Rev. X}\ }\textbf {\bibinfo
  {volume} {11}},\ \bibinfo {pages} {021053} (\bibinfo {year}
  {2021}{\natexlab{c}})},\ \Eprint {https://arxiv.org/abs/2010.14527}
  {arXiv:2010.14527 [gr-qc]} \BibitemShut {NoStop}%
\bibitem [{\citenamefont {Carullo}\ \emph {et~al.}(2019)\citenamefont
  {Carullo}, \citenamefont {Del~Pozzo},\ and\ \citenamefont
  {Veitch}}]{Carullo_2019}%
  \BibitemOpen
  \bibfield  {author} {\bibinfo {author} {\bibfnamefont {G.}~\bibnamefont
  {Carullo}}, \bibinfo {author} {\bibfnamefont {W.}~\bibnamefont {Del~Pozzo}},\
  and\ \bibinfo {author} {\bibfnamefont {J.}~\bibnamefont {Veitch}},\
  }\bibfield  {title} {\bibinfo {title} {{Observational Black Hole
  Spectroscopy: {A} time-domain multimode analysis of {GW}150914}},\ }\href
  {https://doi.org/10.1103/PhysRevD.99.123029} {\bibfield  {journal} {\bibinfo
  {journal} {Phys. Rev. D}\ }\textbf {\bibinfo {volume} {99}},\ \bibinfo
  {pages} {123029} (\bibinfo {year} {2019})},\ \bibinfo {note} {[Erratum:
  Phys.Rev.D 100, 089903 (2019)]},\ \Eprint {https://arxiv.org/abs/1902.07527}
  {arXiv:1902.07527 [gr-qc]} \BibitemShut {NoStop}%
\bibitem [{\citenamefont {Isi}\ \emph {et~al.}(2019)\citenamefont {Isi},
  \citenamefont {Giesler}, \citenamefont {Farr}, \citenamefont {Scheel},\ and\
  \citenamefont {Teukolsky}}]{Isi_2019}%
  \BibitemOpen
  \bibfield  {author} {\bibinfo {author} {\bibfnamefont {M.}~\bibnamefont
  {Isi}}, \bibinfo {author} {\bibfnamefont {M.}~\bibnamefont {Giesler}},
  \bibinfo {author} {\bibfnamefont {W.~M.}\ \bibnamefont {Farr}}, \bibinfo
  {author} {\bibfnamefont {M.~A.}\ \bibnamefont {Scheel}},\ and\ \bibinfo
  {author} {\bibfnamefont {S.~A.}\ \bibnamefont {Teukolsky}},\ }\bibfield
  {title} {\bibinfo {title} {{Testing the no-hair theorem with {GW}150914}},\
  }\href {https://doi.org/10.1103/PhysRevLett.123.111102} {\bibfield  {journal}
  {\bibinfo  {journal} {Phys. Rev. Lett.}\ }\textbf {\bibinfo {volume} {123}},\
  \bibinfo {pages} {111102} (\bibinfo {year} {2019})},\ \Eprint
  {https://arxiv.org/abs/1905.00869} {arXiv:1905.00869 [gr-qc]} \BibitemShut
  {NoStop}%
\bibitem [{\citenamefont {{Skilling}}(2004)}]{2004AIPC..735..395S}%
  \BibitemOpen
  \bibfield  {author} {\bibinfo {author} {\bibfnamefont {J.}~\bibnamefont
  {{Skilling}}},\ }\bibfield  {title} {\bibinfo {title} {{Nested Sampling}},\
  }in\ \href {https://doi.org/10.1063/1.1835238} {\emph {\bibinfo {booktitle}
  {Bayesian Inference and Maximum Entropy Methods in Science and Engineering:
  24th International Workshop on Bayesian Inference and Maximum Entropy Methods
  in Science and Engineering}}},\ \bibinfo {series} {American Institute of
  Physics Conference Series}, Vol.\ \bibinfo {volume} {735},\ \bibinfo {editor}
  {edited by\ \bibinfo {editor} {\bibfnamefont {R.}~\bibnamefont {{Fischer}}},
  \bibinfo {editor} {\bibfnamefont {R.}~\bibnamefont {{Preuss}}},\ and\
  \bibinfo {editor} {\bibfnamefont {U.~V.}\ \bibnamefont {{Toussaint}}}}\
  (\bibinfo {year} {2004})\ pp.\ \bibinfo {pages} {395--405}\BibitemShut
  {NoStop}%
\bibitem [{\citenamefont {Pozzo}\ and\ \citenamefont {Veitch}(2022)}]{cpnest}%
  \BibitemOpen
  \bibfield  {author} {\bibinfo {author} {\bibfnamefont {W.~D.}\ \bibnamefont
  {Pozzo}}\ and\ \bibinfo {author} {\bibfnamefont {J.}~\bibnamefont {Veitch}},\
  }\href@noop {} {\bibinfo {title} {{CPN}est: {A}n efficient {PYTHON}
  parallelizable nested sampling algorithm}},\ \bibinfo {howpublished}
  {Astrophysics Source Code Library, record ascl:2205.021} (\bibinfo {year}
  {2022}),\ \Eprint {https://arxiv.org/abs/2205.021} {ascl:2205.021}
  \BibitemShut {NoStop}%
\bibitem [{\citenamefont {Schmidt}\ \emph {et~al.}(2017)\citenamefont
  {Schmidt}, \citenamefont {Harry},\ and\ \citenamefont
  {Pfeiffer}}]{https://doi.org/10.48550/arxiv.1703.01076}%
  \BibitemOpen
  \bibfield  {author} {\bibinfo {author} {\bibfnamefont {P.}~\bibnamefont
  {Schmidt}}, \bibinfo {author} {\bibfnamefont {I.~W.}\ \bibnamefont {Harry}},\
  and\ \bibinfo {author} {\bibfnamefont {H.~P.}\ \bibnamefont {Pfeiffer}},\
  }\bibfield  {title} {\bibinfo {title} {{Numerical Relativity Injection
  Infrastructure}},\ }\href@noop {} {\  (\bibinfo {year} {2017})},\ \Eprint
  {https://arxiv.org/abs/1703.01076} {arXiv:1703.01076 [gr-qc]} \BibitemShut
  {NoStop}%
\bibitem [{\citenamefont {{LIGO Scientific Collaboration}}(2018)}]{lalsuite}%
  \BibitemOpen
  \bibfield  {author} {\bibinfo {author} {\bibnamefont {{LIGO Scientific
  Collaboration}}},\ }\href {https://doi.org/10.7935/GT1W-FZ16} {\bibinfo
  {title} {{LIGO} {A}lgorithm {L}ibrary - {LALS}uite}},\ \bibinfo
  {howpublished} {free software (GPL)} (\bibinfo {year} {2018})\BibitemShut
  {NoStop}%
\bibitem [{\citenamefont {Borhanian}\ and\ \citenamefont
  {Sathyaprakash}(2022)}]{Borhanian:2022czq}%
  \BibitemOpen
  \bibfield  {author} {\bibinfo {author} {\bibfnamefont {S.}~\bibnamefont
  {Borhanian}}\ and\ \bibinfo {author} {\bibfnamefont {B.~S.}\ \bibnamefont
  {Sathyaprakash}},\ }\bibfield  {title} {\bibinfo {title} {{Listening to the
  Universe with Next Generation Ground-Based Gravitational-Wave Detectors}},\
  }\href@noop {} {\  (\bibinfo {year} {2022})},\ \Eprint
  {https://arxiv.org/abs/2202.11048} {arXiv:2202.11048 [gr-qc]} \BibitemShut
  {NoStop}%
\bibitem [{\citenamefont {Zimmerman}\ \emph {et~al.}(2019)\citenamefont
  {Zimmerman}, \citenamefont {Haster},\ and\ \citenamefont
  {Chatziioannou}}]{PhysRevD.99.124044}%
  \BibitemOpen
  \bibfield  {author} {\bibinfo {author} {\bibfnamefont {A.}~\bibnamefont
  {Zimmerman}}, \bibinfo {author} {\bibfnamefont {C.-J.}\ \bibnamefont
  {Haster}},\ and\ \bibinfo {author} {\bibfnamefont {K.}~\bibnamefont
  {Chatziioannou}},\ }\bibfield  {title} {\bibinfo {title} {On combining
  information from multiple gravitational wave sources},\ }\href
  {https://doi.org/10.1103/PhysRevD.99.124044} {\bibfield  {journal} {\bibinfo
  {journal} {Phys. Rev. D}\ }\textbf {\bibinfo {volume} {99}},\ \bibinfo
  {pages} {124044} (\bibinfo {year} {2019})}\BibitemShut {NoStop}%
\bibitem [{\citenamefont {Amaro-Seoane}\ \emph {et~al.}(2017)\citenamefont
  {Amaro-Seoane} \emph {et~al.}}]{https://doi.org/10.48550/arxiv.1702.00786}%
  \BibitemOpen
  \bibfield  {author} {\bibinfo {author} {\bibfnamefont {P.}~\bibnamefont
  {Amaro-Seoane}} \emph {et~al.} (\bibinfo {collaboration} {LISA}),\ }\bibfield
   {title} {\bibinfo {title} {{Laser Interferometer Space Antenna}},\
  }\href@noop {} {\  (\bibinfo {year} {2017})},\ \Eprint
  {https://arxiv.org/abs/1702.00786} {arXiv:1702.00786 [astro-ph.IM]}
  \BibitemShut {NoStop}%
\bibitem [{\citenamefont {Benkel}\ \emph {et~al.}(2016)\citenamefont {Benkel},
  \citenamefont {Sotiriou},\ and\ \citenamefont {Witek}}]{Benkel:2016kcq}%
  \BibitemOpen
  \bibfield  {author} {\bibinfo {author} {\bibfnamefont {R.}~\bibnamefont
  {Benkel}}, \bibinfo {author} {\bibfnamefont {T.~P.}\ \bibnamefont
  {Sotiriou}},\ and\ \bibinfo {author} {\bibfnamefont {H.}~\bibnamefont
  {Witek}},\ }\bibfield  {title} {\bibinfo {title} {{Dynamical scalar hair
  formation around a {S}chwarzschild black hole}},\ }\href
  {https://doi.org/10.1103/PhysRevD.94.121503} {\bibfield  {journal} {\bibinfo
  {journal} {Phys. Rev. D}\ }\textbf {\bibinfo {volume} {94}},\ \bibinfo
  {pages} {121503} (\bibinfo {year} {2016})},\ \Eprint
  {https://arxiv.org/abs/1612.08184} {arXiv:1612.08184 [gr-qc]} \BibitemShut
  {NoStop}%
\bibitem [{\citenamefont {Silva}\ \emph {et~al.}(2021)\citenamefont {Silva},
  \citenamefont {Witek}, \citenamefont {Elley},\ and\ \citenamefont
  {Yunes}}]{Silva:2020omi}%
  \BibitemOpen
  \bibfield  {author} {\bibinfo {author} {\bibfnamefont {H.~O.}\ \bibnamefont
  {Silva}}, \bibinfo {author} {\bibfnamefont {H.}~\bibnamefont {Witek}},
  \bibinfo {author} {\bibfnamefont {M.}~\bibnamefont {Elley}},\ and\ \bibinfo
  {author} {\bibfnamefont {N.}~\bibnamefont {Yunes}},\ }\bibfield  {title}
  {\bibinfo {title} {{Dynamical Descalarization in Binary Black Hole
  Mergers}},\ }\href {https://doi.org/10.1103/PhysRevLett.127.031101}
  {\bibfield  {journal} {\bibinfo  {journal} {Phys. Rev. Lett.}\ }\textbf
  {\bibinfo {volume} {127}},\ \bibinfo {pages} {031101} (\bibinfo {year}
  {2021})},\ \Eprint {https://arxiv.org/abs/2012.10436} {arXiv:2012.10436
  [gr-qc]} \BibitemShut {NoStop}%
\bibitem [{\citenamefont {Kamaretsos}\ \emph {et~al.}(2012)\citenamefont
  {Kamaretsos}, \citenamefont {Hannam},\ and\ \citenamefont
  {Sathyaprakash}}]{Kamaretsos:2012bs}%
  \BibitemOpen
  \bibfield  {author} {\bibinfo {author} {\bibfnamefont {I.}~\bibnamefont
  {Kamaretsos}}, \bibinfo {author} {\bibfnamefont {M.}~\bibnamefont {Hannam}},\
  and\ \bibinfo {author} {\bibfnamefont {B.}~\bibnamefont {Sathyaprakash}},\
  }\bibfield  {title} {\bibinfo {title} {{Is black-hole ringdown a memory of
  its progenitor?}},\ }\href {https://doi.org/10.1103/PhysRevLett.109.141102}
  {\bibfield  {journal} {\bibinfo  {journal} {Phys. Rev. Lett.}\ }\textbf
  {\bibinfo {volume} {109}},\ \bibinfo {pages} {141102} (\bibinfo {year}
  {2012})},\ \Eprint {https://arxiv.org/abs/1207.0399} {arXiv:1207.0399
  [gr-qc]} \BibitemShut {NoStop}%
\bibitem [{\citenamefont {London}(2020)}]{London:2018gaq}%
  \BibitemOpen
  \bibfield  {author} {\bibinfo {author} {\bibfnamefont {L.~T.}\ \bibnamefont
  {London}},\ }\bibfield  {title} {\bibinfo {title} {{Modeling ringdown. {II}.
  {A}ligned-spin binary black holes, implications for data analysis and
  fundamental theory}},\ }\href {https://doi.org/10.1103/PhysRevD.102.084052}
  {\bibfield  {journal} {\bibinfo  {journal} {Phys. Rev. D}\ }\textbf {\bibinfo
  {volume} {102}},\ \bibinfo {pages} {084052} (\bibinfo {year} {2020})},\
  \Eprint {https://arxiv.org/abs/1801.08208} {arXiv:1801.08208 [gr-qc]}
  \BibitemShut {NoStop}%
\end{thebibliography}%
%==============================================================================
\end{document}